\documentclass[frontimage]{jkpaper}
\usepackage{tikz}
\usepackage{bigints}
\usepackage{biblatex}
\usepackage{aas_macros}

\theoremstyle{plain}

\theoremstyle{definition}

\newcommand{\OIST}{\raisebox{-0.08em}{\includegraphics[height=0.8em]{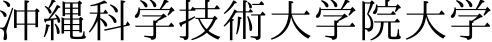}}}

\title{Emergent classical gauge symmetry from quantum entanglement}
\author{Josh Kirklin}
\institution{Okinawa Institute of Science and Technology {\em(}\OIST{\em)},\texorpdfstring{\\}{ } 1919-1 Tancha, Onna-son, Kunigami-gun, Okinawa, Japan 904-0495}
\email{\emaillink{joshua.kirklin@oist.jp}}

\abstr{
    We describe explicitly how entanglement between quantum mechanical subsystems can lead to emergent gauge symmetry in a classical limit. We first provide a precise characterisation of when it is consistent to treat a quantum subsystem classically in such a limit, namely: in any quantum state corresponding to a definite classical state in the classical limit, the reduced density matrix of the subsystem must be approximately proportional to a projection operator, and the projection operators for different classical subsystem states must obey an approximate mutual orthogonality condition. These are strong constraints on the entanglement structure of classical states. They generically give rise to fundamentally non-local classical degrees of freedom, which may nevertheless be accounted for using a completely local kinematical description, if one gauges this description in the right way. The mechanism we describe is very general, but for concreteness we exhibit a toy example involving three entangled spins at high angular momentum, and we also describe a significant group-theoretic generalisation of this toy example. Finally, we give evidence that this phenomenon plays a role in the emergence of bulk diffeomorphism invariance in gravity.
}

\bibliography{refs}

\begin{document}
\maketitleandtoc

\section{Introduction}
\label{Section: Introduction}

Our most well-established modern physical theories involve fundamentally non-local degrees of freedom, which gauge symmetry allows us to describe using purely local mathematical structures. The nature of this is well understood in the case of field theories such as the Standard Model, where it is possible for the gauge symmetry and its associated non-local degrees of freedom to be fundamentally present at the quantum level. It is remarkable that this works so well~\cite{Harvey:2005it}. 

Compared to the Standard Model, the status of gravitational gauge symmetry (a.k.a.\ diffeomorphism invariance) is undoubtably murky. We know how it works in the classical limit, but there is no firm consensus for its role in the quantum theory. However, a clue may come from AdS/CFT, where quantum gravity is a CFT on the boundary of a gravitational `bulk' spacetime~\cite{AdSCFT1,AdSCFT2}. The bulk description is only relevant when one approaches a certain limit in the CFT (which is essentially a classical or semiclassical limit in the bulk). Thus, bulk diffeomorphism invariance would appear not to be a fundamental part of the quantum theory, only emerging in a limiting regime~\cite{Harlow:2015lma,Witten:2017hdv}. Since the bulk spacetime is believed to be a reflection of the entanglement structure in the quantum state~\cite{Maldacena:2001kr,VanRaamsdonk:2010pw,Faulkner:2013ica,Swingle:2014uza,Jacobson:2015hqa,Verlinde:2016toy}, classical gravitational gauge symmetry can be said to emerge from quantum entanglement.

It is reasonable to speculate that this phenomenon happens more generally than just in gravity. Indeed, gauge symmetry and entanglement are both reflections of \emph{non-locality}, the former because it allows us to describe non-local degrees of freedom using local mathematical structures, as we have already mentioned, and the latter because entanglement is the non-local distribution of quantum information between subsystems. If this information is well-behaved in a classical limit, then one might interpret it as describing the state of emergent non-local classical degrees of freedom. If this can be captured using the language of gauge symmetry, then there is an emergent classical gauge symmetry, arising from quantum entanglement.

The purpose of this paper is to confirm that this does happen, and to describe exactly how it works, as well as give some examples. Our results are precise and elementary, but also sufficiently general that we believe they must apply in some way to the gravitational case. It would also not be surprising if they apply to other known examples of emergent gauge symmetry (see for example~\cite{Polyakov:1987ez,Kitaev:1997wr,Wen:2003yv,Lee:2010fy,Bass:2021wxv,Qi:2022lbd}).

As is often true, the progress we present here stems mainly from simply asking the right questions. For us, these are:
\begin{enumerate}
    \item What does it mean for a physical system to have local structure?
    \item How does classical physics emerge from a classical limit of quantum physics?
    \item When do a quantum system and its classical limit share the same local structure?
    \item If they do share the same local structure, do they have the same kinds of fundamentally non-local degrees of freedom? In other words, do they have the same amount of gauge symmetry?
\end{enumerate}
We will spend a small portion of time describing partial answers to the first two questions. Basically, local structure means that one can divide a system into a consistent set of subsystems. Classical physics emerges when operators approximately commute, and when transition probabilities take on approximately classical properties. These answers are more or less intuitive and widely understood, but the effort is worthwhile because establishing these preliminaries is key to finding answers to the third and fourth questions. 

Since in this paper we are interested only in \emph{emergent} gauge symmetry, we will for simplicity assume that there is no pre-existing gauge symmetry at the quantum level (but it is likely that our results generalise in a simple way to the case where there is already some quantum gauge symmetry). Then we find:
\begin{quote}
    A quantum system and its classical limit share the same local structure if and only if in any classical state the reduced density matrix of any local subsystem is approximately proportional to one of a set of mutually orthogonal projection operators.
\end{quote}
The key observation that allows up to obtain this result is that a \emph{quantum} subsystem can only be viewed as a \emph{classical} subsystem if it is `classically resolvable', i.e.\ if there is some set of classical measurements which allow us to fully determine the state of the subsystem.

When the projection operators are rank 1, there is no entanglement between the various subsystems. Then we can treat each subsystem classically in the expected way, and there are no fundamentally non-local degrees of freedom at either the quantum or classical levels, answering the fourth question for this case. Somewhat more interesting is the other case, where the projection operators are of rank higher than 1. Physically speaking, this means that the subsystems can be understood as sharing sets of maximally entangled degrees of freedom. As we will show, the classical observables restricted to each subsystem cannot depend on these shared degrees of freedom --- but the classical observables of the total system can. Thus, at the classical level there are fundamentally non-local degrees of freedom, in contrast with the quantum theory. We will show how this structure can be described using a classically emergent gauge symmetry. Thus, the answer to the fourth question is:
\begin{quote}
    Suppose a quantum system and its classical limit share the same local structure. If all of the subsystems are unentangled from each other in any classical state, then the quantum system and its classical limit have the same amount of gauge symmetry. Otherwise, there is a classically emergent gauge symmetry.
\end{quote}
We will describe properties of this emergent gauge symmetry in this paper.

One key upshot of our results is that one does not necessarily need to use operator constraints to quantise a classical theory with gauge symmetry. Thus, for example, even though a classical system may be constructed using the `fusion product' described in~\cite{Donnelly:2016auv}, the corresponding quantum system need not be constructed using the `entangling product' also described in that paper.

Another important takeaway concerns the nature of multipartite entanglement in the classical limit. In general, multipartite entanglement is notoriously difficult to analyse and characterise, in contrast to the bipartite case~\cite{Multipartite}. What we show demonstrates that such an analysis simplifies significantly in the classical regime; indeed, it demonstrates that all the properties of multipartite entanglement in this regime may be understood using the familiar and comparatively much simpler language of gauge symmetry.

The paper proceeds as follows. In Section~\ref{Section: Preliminaries}, we will review some relevant properties of local structure and subsystems, and the meaning of the classical limit. Then, in Section~\ref{Section: Local structures in the classical limit} we describe what happens to local structures in the classical limit, explain what it means for a local structure to be `classically resolvable', and describe the implications of classical resolvability for reduced states. Classically resolvable subsystems can be entangled with each other; in Section~\ref{Section: Gauge symmetry from entanglement} we show that this entanglement leads to the emergence of fundamentally non-local classical degrees of freedom, and we describe the emergent gauge symmetry that can be used to account for them. A toy example is given in Section~\ref{Section: Example} of a simple system exhibiting the mechanism we have described: a set of three entangled spins at large angular momentum. Then, in Section~\ref{Section: group}, we describe a vast group-theoretic generalisation of this example. The phenomenon we describe has many properties reminiscent of gravity, some of which we set out in Section~\ref{Section: Gravitational properties}. We conclude the paper in Section~\ref{Section: Conclusion} with some open questions. We also provide an appendix giving some important details on Schur's lemma for the unfamiliar reader.

\section{Preliminaries}
\label{Section: Preliminaries}

\subsection{Local structure as a decomposition into subsystems}

Consider a physical system with a notion of \emph{locality}. The key feature of such systems is that they are composite, i.e.\ they are divisible into subsystems. In the context of this paper, this is all that we will explicitly require (though in different contexts it might be useful to add other requirements). Let us give two examples. In a theory of fields on a spacetime, we may consider the values of the fields in a spatial subregion; this is a subsystem of the full set of degrees of freedom over all of space. In a gas of particles we may consider the motion of some subset of the particles; this is a subsystem of the gas.

Subsystems obey certain axiomatic properties we will not describe in full here, since it is mostly intuitively obvious what they should be. Let us at this stage just mention two. First, subsystems may be contained within other subsystems. Second, any set of subsystems may be viewed collectively as a single subsystem, which we call their `union'. The union of a set of subsystems contains all of those subsystems.

We will call the set $\mathscr{S}$ of all subsystems of a given physical system its `local structure'. Depending on the context, one can use different local structures for the same physical system. For example, in a continuum field theory, one may wish to  impose a lower limit on the size of spatial subregions under consideration. Different choices of lower limit will give different sets of subsystems, i.e.\ different local structures. Importantly, only certain local structures of a quantum system will be consistent with its classical limit, as will be described in greater detail in Section~\ref{Section: Local structures in the classical limit}.

Suppose $\{s_i\in\mathscr{S}\mid i=1,\dots,n\}$ is some collection of subsystems of a physical system. If these subsystems are mutually disjoint, and if their union is the entire system, then we will call the collection a `subdivision' of the system. By studying the subdivisions of a physical system, we can better understand its local structure.

Each subsystem is associated with a set of observables for the degrees of freedom it contains. These are the observables `local to' that subsystem. If the set of observables local to a union of subsystems contains observables which cannot be formed out of the observables local to each individual subsystem, then clearly there must be \emph{non-local} degrees of freedom. 

In this paper, we are interested in two different types of physical system: quantum systems, and classical systems. Each has its own kind of observable. 

\subsubsection{Quantum systems with local structure}

Let us first address the quantum case. Suppose a quantum system with Hilbert space $\mathcal{H}$ has a local structure $\mathscr{S}$, and let $s\in\mathscr{S}$ be a subsystem. Then the set of observables local to $s$ is a von Neumann algebra $\mathcal{A}_s$ of operators acting on $\mathcal{H}$. In general, $\mathcal{A}_s$ can be any such algebra. This includes algebras with non-trivial center, which are relevant when there is a gauge symmetry present in the quantum theory, and Type~\textls[-50]{II} and Type~\textls[-50]{III} algebras, which are relevant in certain settings with an infinite-dimensional Hilbert space, such as QFT~\cite{Witten:2018zxz}.

Unless stated otherwise, from now on we will for simplicity assume that $\mathcal{A}_s$ is a `factor', meaning it has trivial center (since we are only interested in emergent gauge symmetries), and that $\mathcal{H}$ is finite-dimensional\footnote{On the other hand we will allow the dimension of $\mathcal{H}$ to be arbitrarily large in the classical limit. So even though $\dim(\mathcal{H})$ is always finite for any fixed value of the parameter $\chi$ defining the classical limit $\chi\to 0$ (see Section~\ref{Subsection: classical limit}), we can still have $\dim(\mathcal{H})\to\infty$ as $\chi\to 0$.} (as it will make the analysis more conceptually straightforward). Thus, $\mathcal{A}_s$ is a Type~I factor, and so may always be written in the form
\begin{equation}
    \mathcal{A}_s = \mathcal{B}(\mathcal{H}_s) \otimes \mathds{1}_{\bar{s}},
    \label{Equation: subsystem algebra}
\end{equation}
for some factorisation of the system Hilbert space 
\begin{equation}
    \mathcal{H}= \mathcal{H}_s \otimes \mathcal{H}_{\bar{s}},
    \label{Equation: Hilbert space factorise}
\end{equation}
with $\mathcal{B}(\mathcal{H}_s)$ being the algebra of operators acting on $\mathcal{H}_s$, and $\mathds{1}_{\bar{s}}$ being the identity acting on $\mathcal{H}_{\bar{s}}$. We may think of $\mathcal{H}_s$ as the Hilbert space of subsystem $s$, and $\mathcal{H}_{\bar{s}}$ as the Hilbert space of its complement.

If $\{s_i\in\mathscr{S}\mid i=1,\dots,n\}$ is a subdivision of the quantum system, then we can straightforwardly generalise this to write the algebra of operators of subsystem $s_i$ as
\begin{equation}
    \mathcal{A}_i = \mathds{1}_{1}\otimes\dots\otimes\mathds{1}_{i-1}\otimes\mathcal{B}(\mathcal{H}_i)\otimes\mathds{1}_{i+1}\otimes\dots\otimes\mathds{1}_{n},
\end{equation}
where
\begin{equation}
    \mathcal{H} = \mathcal{H}_1\otimes\dots\otimes\mathcal{H}_i\otimes\dots\otimes\mathcal{H}_n,
    \label{Equation: quantum tensor factorisation}
\end{equation}
and $\mathcal{H}_i$ is the Hilbert space of subsystem $s_i$. This tensor product structure means that there are no non-local constraints on the quantum state of the system, which reflects the lack of a fundamental gauge symmetry.

\subsubsection{Classical systems with local structure}
\label{Subsubsection: classical local structure}

The classical case is slightly different. For a classical system with local structure $\mathscr{S}$, each subsystem $s\in\mathscr{S}$ comes equipped with a space $\mathcal{N}_s$ of possible classical states, and the observables of that subsystem are functions on $\mathcal{N}_s$. In the absence of gauge symmetry, the space of states $\mathcal{N}$ for the total system may then be written as
\begin{equation}
    \mathcal{N}=\mathcal{N}_s\times\mathcal{N}_{\bar{s}}
\end{equation}
where $\mathcal{N}_{\bar{s}}$ is the space of states of the complement of $s$, mirroring the local decomposition structure of~\eqref{Equation: Hilbert space factorise}. However, unlike quantum systems, in this paper we do want to allow the classical systems we consider to have gauge symmetry, in which case this simple decomposition does not work.

When gauge symmetry is present, to get something resembling a local decomposition we have to augment each subsystem with some extra degrees of freedom. This involves replacing the space $\mathcal{N}_s$ of possible states for the original degrees of freedom in subsystem $s$, which we will call `physical states' for subsystem $s$, with a space $\mathcal{N}^{\text{kin.}}_s$ and a surjective map $R_s:\mathcal{N}^{\text{kin.}}_s\to\mathcal{N}_s$. The elements of $\mathcal{N}^{\text{kin.}}_s$ are known as `kinematical states' for subsystem $s$, and are states of both its original degrees of freedom and the new extra degrees of freedom. The map $R_s$ simply picks out the physical state corresponding to a given kinematical state. We also do the same thing for the complement of $s$, i.e.\ we construct a space $\mathcal{N}_{\bar{s}}^{\text{kin.}}$ of kinematical states in the complement, and a surjective map $R_{\bar{s}}:\mathcal{N}_{\bar{s}}^{\text{kin.}}\to\mathcal{N}_{\bar{s}}$ picking out the physical state of the complement.

We can then construct a space of kinematical states for the total system via
\begin{equation}
    \mathcal{N}^{\text{kin.}} = \mathcal{N}^{\text{kin.}}_s\times\mathcal{N}^{\text{kin.}}_{\bar{s}}.
\end{equation}
This is the desired local decomposition. However, $\mathcal{N}^{\text{kin.}}$ is not equivalent to $\mathcal{N}$ (the space of physical states for the full system), because of the presence of the extra degrees of freedom. To get $\mathcal{N}$, we have to impose some constraints, and then perform gauge reduction. To give a bit more detail, the constraints relate in a non-local way the kinematical state in subsystem $s$ with the kinematical state in its complement. The set
\begin{equation}
    \overline{\mathcal{N}^{\text{kin.}}} \subset \mathcal{N}^{\text{kin.}}
\end{equation}
of states obeying these constraints is sometimes known as the `constraint surface'. The constraints are chosen such that there exists a surjective map 
\begin{equation}
    R:\overline{\mathcal{N}^{\text{kin.}}}\to \mathcal{N}.
\end{equation}
Often there is a group (known as the gauge group) which acts on kinematical states in the constraint surface. The elements of the group are the gauge transformations, and $R$ performs a quotient with respect to this action. In other words, the physical states are equivalence classes of kinematical states modulo gauge transformations. Acting with $R$ is known as `gauge reduction'.

If $\{s_i\in\mathscr{S}\mid i=1,\dots,n\}$ is a subdivision of the classical system, then the above generalises in the following way. Suppose $\mathcal{N}_i$ is the physical state space of subsystem $i$. Then for each $i$ we have a kinematical state space $\mathcal{N}_i^{\text{kin.}}$ and a map $R_i:\mathcal{N}_i^{\text{kin.}}\to\mathcal{N}_i$. Then the kinematical state space for the total system is
\begin{equation}
    \mathcal{N}^{\text{kin.}} = \mathcal{N}_1^{\text{kin.}}\times \dots\times\mathcal{N}_i^{\text{kin.}}\times\dots\times \mathcal{N}_n^{\text{kin.}},
\end{equation}
and the gauge reduction map is $R:\overline{\mathcal{N}^{\text{kin.}}}\to \mathcal{N}$, where $\overline{\mathcal{N}^{\text{kin.}}} \subset \mathcal{N}^{\text{kin.}}$ is some constraint surface in $\mathcal{N}^{\text{kin.}}$. Note that the forms of the kinematical state spaces $\mathcal{N}_i^{\text{kin.}}$, $\mathcal{N}^{\text{kin.}}$, constraint surface $\overline{\mathcal{N}^{\text{kin.}}}$, and gauge reduction map $R$ can depend on which subdivision $\{s_i\in\mathscr{S}\mid i=1,\dots,n\}$ we are considering. However, the physical space of states $\mathcal{N}$ must be independent of the subdivision.

In the setting of field theory, with subsystems corresponding to spatial subregions, the extra degrees of freedom one includes in the kinematical state space are sometimes known as `edge modes'~\cite{Donnelly:2016auv,Donnelly:2016qqt,Geiller:2017xad,Speranza:2017gxd,Harlow:2019yfa,Donnelly:2020xgu,Carrozza:2021gju}.\footnote{Edge modes are sometimes promoted to physical degrees of freedom, but we will not consider that possibility in this paper.} We will suggestively use this terminology more generally in this paper, i.e.\ we will refer to the extra degrees of freedom as `edge modes' even in the non-field-theoretic context.

\subsection{Classical physics in quantum systems}
\label{Subsection: classical limit}

Classical physics suffices to describe many physical observations to a high degree of precision, despite the fact that the wider world is more accurately modelled by quantum theory, so there must be some way to `approximately embed' the classical picture within the quantum one. One manner in which this can be made precise (although certainly not the only one) is as follows.

Suppose we wish to embed some classical degrees of freedom, the possible states of which are the elements of a set $\mathcal{N}$, inside of a quantum system whose states are elements of a Hilbert space $\mathcal{H}$. In the classical picture, for each $x\in\mathcal{N}$ we can always answer with certainty the question: is $x$ the current state of the classical degrees of freedom? For this to also be approximately true within the quantum picture, there must be some operator acting on $\mathcal{H}$ which we can measure to answer this question. We can choose for this operator to have eigenvalue 1 when the answer is ``yes'', and eigenvalue 0 when the answer is ``no'' --- so it is a projection operator. Thus, for each classical state $x\in\mathcal{N}$, there must  a projection operator $\hat \pi(x)=\hat \pi(x)^2=\hat \pi(x)^\dagger$ acting on $\mathcal{H}$. By measuring $\hat \pi(x)$, we can decide with high precision whether the classical state is $x$.\footnote{To keep things simple we avoid the use of projection-valued measures (PVM). But a more precise treatment would probably involve them.}

The projection operators should have certain special properties. Since classical degrees of freedom cannot be in more than one state at a time, the projection operators must be approximately orthogonal:
\begin{equation}
    \hat \pi(x)\hat \pi(y) \approx \delta_{xy}\hat \pi(x), \text{ for all }x,y\in\mathcal{N},
    \label{Equation: approximately orthogonal}
\end{equation}
where $\delta_{xy}=1$ if $x=y$ and $\delta_{xy}=0$ otherwise. On the other hand, since the classical degrees of freedom must be in \emph{some} state in $\mathcal{N}$, the union of the images of the projection operators must span $\mathcal{H}$, which immediately implies that there is some measure $\mu$ on $\mathcal{N}$ obeying
\begin{equation}
    \int_{\mathcal{N}} \dd{\mu(x)}\frac{\hat \pi(x)}{N(x)} \approx \frac{\mathds{1}}{N}.
    \label{Equation: approximate resolution}
\end{equation}
Thus, the projection operators furnish an approximate resolution of the identity. Here, $N(x)$ is the rank of $\hat\pi(x)$, $N=\dim(\mathcal{H})$, and $\mathds{1}$ is the identity acting on $\mathcal{H}$. The constant factors are chosen such that the measure $\mu$ is (approximately) normalised, as can be verified by taking a trace of~\eqref{Equation: approximate resolution}. Acting with $\hat\pi(y)$ on both sides of~\eqref{Equation: approximate resolution} yields
\begin{equation}
    \hat\pi(y) \approx 
    \int_{\mathcal{N}} \dd{\mu(x)}\frac{N}{N(x)}\,\hat \pi(x) \hat \pi(y).
\end{equation}
Since~\eqref{Equation: approximately orthogonal} implies that this integral is dominated by contributions at $x=y$, we can write\footnote{If $\mathcal{N}$ is a continuous space, then the consistency of~\eqref{Equation: approximately orthogonal} and~\eqref{Equation: approximate delta} requires $N/N(x)\to\infty$ in the classical limit.}
\begin{equation}
    \frac{N}{N(x)}\,\hat\pi(x)\hat\pi(y) \approx \delta_{\mu}(x,y)\hat\pi(x),
    \label{Equation: approximate delta}
\end{equation}
where $\delta_{\mu}(x,y)$ is a delta function for the measure $\mu$.

Classical observables are functions $A(x)$ of the classical state $x\in\mathcal{N}$, and can be translated into quantum operators via
\begin{equation}
    \hat{A} \approx \int_{\mathcal{N}}\dd{\mu(x)}\,\frac{N}{N(x)}\,\hat\pi(x)\,A(x).
    \label{Equation: approximate classical operator}
\end{equation}
We refer to such operators as `classical operators'.
By~\eqref{Equation: approximate delta}, we have
\begin{equation}
    \hat{A}\,\hat\pi(x) \approx A(x)\,\hat\pi(x) \approx \hat\pi(x)\,\hat{A},
\end{equation}
so projecting onto the classical state $x$ and then measuring $\hat{A}$ (or vice versa) gives the expected classical answer $A(x)$. Also, for two classical observables $A_1(x),A_2(x)$ with corresponding classical operators $\hat{A}_1,\hat{A}_2$,~\eqref{Equation: approximate delta} implies that
\begin{equation}
    \hat{A}_1\hat{A}_2 \approx \int_{\mathcal{N}}\dd{\mu(x)}\,\frac{N}{N(x)}\,\hat\pi(x)\,A_1(x)A_2(x).
\end{equation}
The right-hand side is the operator corresponding to the classical observable $A_1(x)A_2(x)$. Thus, these operators (approximately) reproduce the commutative algebra of functions on $\mathcal{N}$.

We have been vague about what we mean by approximate equality; let us partially remediate this. Any of the approximate equalities appearing in this paper should be understood to indicate equality in a \emph{classical limit}. More precisely, the quantum theory depends on some parameter $\chi$ (which could be Planck's constant $\hbar$, or Newton's constant $G$, etc.) which can be taken to be arbitrarily small, and $a\approx b$ is shorthand for $\lim_{\chi\to 0} a = b$, where this limit is taken with respect to some topology on whatever space contains $a$ and $b$. For the purposes of this paper, it does not matter too much what this topology is. Indeed, it can depend on what kind of classical limit one is considering, and on what kind of quantum theory one started with. All that matters is that, for a given classical limit of a given quantum theory, we use a single self-consistent set of such topologies. Throughout the paper, we will be somewhat cavalier about the precise nature of these topologies, leaving a more rigorous treatment to later work.

A privileged role is played by quantum states for which the classical degrees of freedom are in a definite classical state, in the classical limit.\footnote{The exact reason for only considering states of this form can vary from theory to theory, and is unimportant for the purposes of this paper. Perhaps in the classical limit they are dynamically favoured, or they statistically dominate over other states.}\textsuperscript{,}\footnote{They are sometimes called `coherent states'. However, this terminology is also sometimes reserved for states which are also associated with the action of a Lie group in a certain way. We will describe examples of this in Sections~\ref{Section: Example} and~\ref{Section: group}.} For these quantum states $\ket{\psi}$, there must be an $x\in\mathcal{N}$ such that
\begin{equation}
    \hat\pi(y)\ket{\psi} \approx
    \begin{cases}
        \ket{\psi} & \text{if } x=y,\\
        0 & \text{otherwise}.
    \end{cases}
    \label{Equation: definite classical state}
\end{equation}
In this case, we say $x$ is the classical state of $\ket{\psi}$. In states of this form, classical operators act approximately as multiplication by the classical observables upon which they are based:
\begin{equation}
    \hat{A}\ket{\psi} \approx A(x) \ket{\psi}.
\end{equation}
Using~\eqref{Equation: approximate resolution}, more general states (i.e.\ those $\ket{\psi}$ which do not necessarily obey~\eqref{Equation: definite classical state}) can always be written as a superposition of such states:
\begin{equation}
    \ket{\psi} = \int_{\mathcal{N}}\dd{\mu(x)}\,\frac{N}{N(x)}\,\ket{\psi(x)}, \qq{where} \ket{\psi(x)} = \hat{\pi}(x)\ket{\psi}.
\end{equation}
The classical state of $\ket{\psi(x)}$ is $x$. As far as expectation values of classical observables are concerned, in a general state the classical degrees of freedom may be viewed as being distributed with respect to a set of classical probabilities. Indeed, using~\eqref{Equation: approximate delta}, we have
\begin{equation}
    \frac{\mel{\psi}{\hat{A}}{\psi}}{\braket{\psi}{\psi}} = \int_{\mathcal{N}}\dd{\mu(x)}\,\frac{N}{N(x)}\, p(x)\, A(x),\qq{where} p(x) = \frac{\braket{\psi(x)}{\psi(x)}}{\braket{\psi}{\psi}}.
\end{equation}
Thus, $\dd{\mu(x)} \frac{N}{N(x)}\, p(x)$ is the probability measure for the state of the classical degrees of freedom.

The above is all that is required to approximately embed classical physics within quantum physics. However, in the kind of classical limit we have described, although some degrees of freedom behave classically, not all of them are guaranteed to do so. More commonly, we want \emph{all} of the degrees of freedom to behave classically, in a classical limit. We will call this a `complete classical limit'.

A more precise way of characterising a complete classical limit is as follows. Suppose the quantum state $\ket{\psi}$ of the system obeys~\eqref{Equation: definite classical state} for some $x$. Thus, the classical degrees of freedom are in some definite classical state. If we are able to determine $x$, then in a \emph{complete} classical limit this knowledge should suffice to completely determine $\ket{\psi}$ (up to phase factors and normalisation) --- since the system contains no degrees of freedom besides those determined by $x$. In other words, for each $x\in\mathcal{N}$ there is some normalised quantum state $\ket{x}$ such that
\begin{equation}
    x\text{ is the classical state of }\ket{\psi} \implies \ket{\psi}\approx C \ket{x}, \text{ some } C\in\mathbb{C}.
    \label{Equation: complete classical limit}
\end{equation}

In a complete classical limit, we can choose for the projection operators $\hat\pi(x)$ to be rank 1, and they may be written $\hat\pi(x)=\ket{x}\bra{x}$. Then~\eqref{Equation: approximate delta} implies
\begin{equation}
    N \braket{x}{y} \approx \delta_\mu(x,y),
\end{equation}
and the classical observable $A(x)$ is related to its corresponding quantum operator $\hat{A}$ via
\begin{equation}
    A(x) \approx \mel{x}{\hat{A}}{x}.
    \label{Equation: classical observable expectation}
\end{equation}
Also, \eqref{Equation: approximate resolution} simplifies to
\begin{equation}
    \mathds{1} \approx \int_{\mathcal{N}}\dd{\mu(x)} \, N \, \ket{x}\bra{x},
\end{equation}
and~\eqref{Equation: approximate classical operator} simplifies to
\begin{equation}
    \hat{A} \approx \int_{\mathcal{N}}\dd{\mu(x)} \, N \, \ket{x}\bra{x} \, A(x).
    \label{Equation: classical limit operator}
\end{equation}

\section{Local structures in the classical limit}
\label{Section: Local structures in the classical limit}

Suppose a quantum system has a Hilbert space $\mathcal{H}$, and a local structure $\mathscr{S}$. Let us take a complete classical limit of this system, to obtain a space $\mathcal{N}$ of classical states $x$ corresponding to quantum states $\ket{x}$. In general, $\mathscr{S}$ will \emph{not} be valid as a local structure for the system obtained in the classical limit. For example, we could consider a continuum QFT which is weakly coupled above a certain lengthscale $\Lambda$, but strongly coupled below that lengthscale. Then quantum fluctuations will prohibit any subsystems smaller than $\Lambda$ from being part of a local structure for the classical theory -- but there is nothing preventing them from being part of the local structure $\mathscr{S}$ for the quantum theory.

In this section, we will study what happens to $\mathscr{S}$ in the classical limit. In particular, we will define precisely what it means for $\mathscr{S}$ to remain valid in this limit, in terms of the `classical resolvability' of its subsystems, and we will find that the classical validity of $\mathscr{S}$ leads to strong constraints on the reduced states of its subsystems.

\subsection{Classical degrees of freedom in quantum subsystems}

Suppose $s\in\mathscr{S}$ is a subsystem of the quantum system, so that $\mathcal{H}$ factorises as in~\eqref{Equation: Hilbert space factorise}:
\begin{equation}
    \mathcal{H}=\mathcal{H}_s\otimes\mathcal{H}_{\bar{s}}.
\end{equation}
Consider a quantum operator $\hat{A}$ associated with subsystem $s$. This operator must be an element of $\mathcal{A}_s=\mathcal{B}(\mathcal{H}_s)\otimes\mathds{1}_{\bar{s}}$, and so may be written as $\hat{A}=\hat{A}_s\otimes\mathds{1}_{\bar{s}}$, for some $\hat{A}_s\in\mathcal{B}(\mathcal{H}_s)$. On the other hand, if $\hat{A}$ is a classical operator of the full system, then it may also be written in the form~\eqref{Equation: classical limit operator}. Thus, any classical operator $\hat{A}$ for the full system that acts only on quantum subsystem $s$ should obey
\begin{equation}
    \hat{A}=\hat{A}_s\otimes\mathds{1}_{\bar{s}} \approx \int_{\mathcal{N}}\dd{\mu(x)}\,N\,\ket{x}\bra{x}\,A(x).
    \label{Equation: classical operators and subsystem operators}
\end{equation}

Any operator $\hat{A}_s\in\mathcal{B}(\mathcal{H}_s)$ that obeys the approximate equality in~\eqref{Equation: classical operators and subsystem operators} for some $A(x)$ is a classical operator for subsystem $s$. Let us use $\mathcal{C}_s$ to denote the set of all such $\hat{A}_s$. In the classical limit $\mathcal{C}_s$ may be treated as a commutative unital $\mathrm{C}^*$-algebra, so by the Gelfand-Naimark theorem it is isomorphic to a space of continuous functions on some Hausdorff space $\mathcal{N}_s$. Thus, the classical observables of subsystem $s$ are in one-to-one correspondence with these functions --- so we should think of $\mathcal{N}_s$ as the space of states for the classical degrees of freedom in $s$.

According to the Gelfand-Naimark theorem, $\mathcal{N}_s$ may be constructed as the space of non-zero `characters' of $\mathcal{C}_s$. These are linear functionals $\phi_s:\mathcal{C}_s\to\mathbb{C}$ obeying
\begin{equation}
    \phi_s(\hat{A}_s\hat{B}_s) = \phi_s(\hat{A}_s)\,\phi_s(\hat{B}_s) \qq{and} \phi_s(\hat{A}_s^\dagger) = \phi_s(\hat{A}_s)^*.
\end{equation}
In other words they are unital $*$-homomorphisms. The function on $\mathcal{N}_s$ corresponding to $\hat{A}_s$ is then defined as
\begin{equation}
    A_s(\phi_s) = \phi_s(\hat{A}_s).
\end{equation}
Suppose the quantum subsystem $s$ is in a state described by a density matrix $\rho_s$. If for all $\hat{A}_s\in\mathcal{C}_s$ this density matrix obeys
\begin{equation}
    \hat{A}_s\,\rho_s \approx A_s(\phi_s) \,\rho_s \approx \rho_s\,\hat{A}_s
    \label{Equation: subsystem classical degrees of freedom state}
\end{equation}
for some $\phi_s\in\mathcal{N}_s$, then this means that the classical degrees of freedom in $s$ are in the state $\phi_s$.

Note that each classical state $x\in\mathcal{N}$ of the full system gives a linear functional $x_s:\mathcal{C}_s\to\mathbb{C}$ defined by
\begin{equation}
x_s(\hat{A}_s) \approx \mel{x}{(\hat{A}_s\otimes\mathds{1}_{\bar{s}})}{x} =  \tr_s(\hat{A}_s\rho_s(x)),
    \label{Equation: subsystem character}
\end{equation}
where
\begin{equation}
    \rho_s(x) = \tr_{\bar{s}} \qty(\ket{x}\bra{x}),
\end{equation}
is the reduced density matrix of subsystem $s$ when the full system is in the state $\ket{x}$ ($\tr_{\bar{s}}$ is a partial trace over $\mathcal{H}_{\bar{s}}$). If there is entanglement between $i$ and the other subsystems, then $\rho_s(x)$ will be mixed.
By~\eqref{Equation: classical observable expectation}, we may note that $A(x) \approx x_s(\hat{A}_s) = A_s(x_s)$. This is a character since
\begin{equation}
    x_s(\hat{A}_s\hat{B}_s) \approx \mel{x}{(\hat{A}_s\otimes\mathds{1}_{\bar{s}})(\hat{B}_s\otimes\mathds{1}_{\bar{s}})}{x} \approx \mel{x}{(\hat{A}_s\otimes\mathds{1}_{\bar{s}})}{x}\mel{x}{(\hat{B}_s\otimes\mathds{1}_{\bar{s}})}{x}\approx x_s(\hat{A}_s)x_s(\hat{B}_s),
\end{equation}
where we used the fact that $\hat{A}_s\otimes\mathds{1}_{\bar{s}}$ and $\hat{B}_s\otimes\mathds{1}_{\bar{s}}$ are classical operators for the full system. For any $\hat{A}_s\in\mathcal{C}_s$, we have
\begin{equation}
    \hat{A}_s\,\rho_s(x) = \tr_{\bar{s}}\qty((\hat{A}_s\otimes\mathds{1}_{\bar{s}})\ket{x}\bra{x}) \approx \tr_{\bar{s}}(A(x)\ket{x}\bra{x}) \approx A_s(x_s)\, \rho_s(x).
\end{equation}
This matches~\eqref{Equation: subsystem classical degrees of freedom state} (we similarly have $\rho_s(x)\hat{A}_s\approx A_s(x_s)\rho_s(x)$). So if $\rho_s(x)$ is the state of quantum subsystem $s$, then the classical degrees of freedom in $s$ are in the state $x_s$. Thus, \eqref{Equation: subsystem character} defines a map 
\begin{equation}
    \quad\mathcal{N}\to\mathcal{N}_s, \quad x\mapsto x_s
    \label{Equation: x to x_s}
\end{equation}
such that if $x$ is the classical state of the full system, then $x_s$ is the classical state of $s$.

\subsection{Classical resolvability}

The construction of $\mathcal{N}_s$ described above works for any quantum subsystem. However, in general the space $\mathcal{N}_s$ will be `too small' to adequately account for the physics in subsystem $s$. In particular, it is not guaranteed that knowledge of the state $\phi_s\in\mathcal{N}_s$ of the \emph{classical} degrees of freedom in $s$ suffices to determine the complete state of subsystem $s$. This indicates that there are still some \emph{quantum} degrees of freedom in $s$. Thus, a complete classical limit for the full system does not necessarily imply that its subsystems behave in a completely classical way. 

To determine which subsystem degrees of freedom are classical, and which are quantum, we must study the properties of the set $\mathcal{C}_s$ of classical subsystem operators. The defining property~\eqref{Equation: classical operators and subsystem operators} of this set is quite non-trivial. In an extreme case, it could be that the only operators which satisfy it are those proportional to the identity $\mathds{1}_s$. This would then imply that $\mathcal{N}_s$ only contains a single element, so there would be only one possible state for the classical degrees of freedom in $s$, which is clearly not enough to describe the physics in a non-trivial subsystem. Thus, such a subsystem would have to be described in a completely quantum way.

In this paper, we are interested in the opposite case: subsystems which can be described using classical degrees of freedom alone. In particular, suppose we know that the full system is in some classical state, although we do not necessarily know which one. Then we call subsystem $s$ `classically resolvable' if, in the classical limit, we can determine its reduced density matrix $\rho_s$ using measurements only of its classical degrees of freedom. If this were not the case, then there would be more than one reduced density matrix $\rho_s$ (i.e.\ quantum state of $s$) yielding the same exact classical observations, which is equivalent to there being unaccounted-for quantum degrees of freedom.

More precisely, suppose $x,y\in\mathcal{N}$ are any two classical states of the full system, and let their corresponding subsystem states be $x_s,y_s\in\mathcal{N}_s$ respectively. Since $x_s,y_s$ are defined by~\eqref{Equation: subsystem character}, we already know that
\begin{equation}
    \rho_s(x)\approx\rho_s(y)\implies x_s=y_s.
\end{equation}
Subsystem $s$ is `classically resolvable' if the implication also goes the other direction:
\begin{equation}
    x_s=y_s\implies \rho_s(x)\approx\rho_s(y).
    \label{Equation: classically resolvable definition}
\end{equation}
We will say that a local structure is `classically resolvable' if all of its subsystems are classically resolvable. 

Unsurprisingly, classical resolvability has non-trivial implications for subsystem physics. The rest of the paper is devoted to determining some of these implications. 

\subsection{States and observables of classically resolvable subsystems}

Suppose $s$ is a classically resolvable subsystem, and consider its classical operators $\hat{A}_s\in\mathcal{C}_s$. We can extract $\hat{A}_s$ from~\eqref{Equation: classical operators and subsystem operators} by taking a partial trace over $\mathcal{H}_{\bar{s}}$ and then dividing by $N_{\bar{s}} = \dim(\mathcal{H}_{\bar{s}})$. This yields
\begin{equation}
    \hat{A}_s \approx \int_{\mathcal{N}}\dd{\mu(x)}\,N_{s}\,\rho_s(x)\,A_s(x_s),
    \label{Equation: classical subsystem operators}
\end{equation}
where $N_{s} = N/N_{\bar{s}}=\dim(\mathcal{H}_i)$, $x_s\in\mathcal{N}_s$ is the subsystem state corresponding to $x\in\mathcal{N}$, and we have used the fact that $A_s(x_s) \approx A(x)$. By the Gelfand-Naimark isomorphism, any function $A_s(x_s)$ will give a valid classical operator. Let us fix a $y\in\mathcal{N}$, and set
\begin{equation}
    A_s(x_s) \propto
    \begin{cases}
        1 & \text{if }\rho_s(x)\approx\rho_s(y),\\
        0 & \text{otherwise},
    \end{cases}
    \label{Equation: A_s y_s}
\end{equation}
which is possible only because the subsystem is classically resolvable.
By~\eqref{Equation: classically resolvable definition} we then approximately have
\begin{equation}
    \rho_s(x)\, A_s(x_s)  \propto
    \begin{cases}
        \rho_s(y) & \text{if }\rho_s(x)=\rho_s(y),\\
        0 & \text{otherwise}.
    \end{cases}
\end{equation}
Thus, the integrand in~\eqref{Equation: classical subsystem operators} is either zero or approximately proportional to $\rho_s(y)$. Performing the integral, we find that
\begin{equation}
    \hat{A}_s\approx \alpha\,\rho_s(y)
\end{equation}
for some constant $\alpha$. On the other hand,~\eqref{Equation: classical operators and subsystem operators} and~\eqref{Equation: A_s y_s} imply that
\begin{equation}
    \alpha\,\rho_s(y)\otimes\mathds{1}_{\bar{s}} \approx \hat{A}_s\otimes\mathds{1}_{\bar{s}} \approx \int_{\mathcal{N}}\dd{\mu(x)}\,N\,\ket{x}\bra{x}\,A_s(x_s)
\end{equation}
is approximately proportional to a sum over approximately mutually orthogonal projection operators, and thus is approximately proportional to a projection operator. This finally implies that $\rho_s(y)$ itself is also approximately proportional to a projection operator. 

So, in a classically resolvable subsystem $s$, for each $y\in\mathcal{N}$ we may write
\begin{equation}
    \rho_s(y) \approx \frac{\hat{\pi}_s(y_s)}{N_s(y_s)}
    \label{Equation: rho approximately projection}
\end{equation}
where $\hat{\pi}_s(y_s)$ is some projection operator with rank $N_s(y_s)$, and we are using the fact that $\rho_s(y)$ only depends on the classical subsystem state $y_s\in\mathcal{N}_s$. This is the first key result of this paper.

Note that~\eqref{Equation: A_s y_s} is the classical subsystem observable that tells us if the subsystem state is $y_s$. Thus, we have the remarkable result that $\rho_s(y)$ itself is the classical subsystem operator that we should use to determine if $\rho_s(y)$ is the state of the subsystem. This is related to the following key property of the projection operators just defined:
\begin{equation}
    \hat{\pi}_s(x_s)\hat{\pi}_s(y_s) \approx \delta_{x_sy_s}\,\hat\pi_s(x_s),
    \label{Equation: approximately orthogonal subsystem}
\end{equation}
where $x_s,y_s\in\mathcal{N}_s$. This property holds because the subsystem cannot be in more than one classical state at a time.

Using~\eqref{Equation: rho approximately projection}, we can write a general classical subsystem operator~\eqref{Equation: classical subsystem operators} as
\begin{equation}
    \hat{A}_s\approx\int_{\mathcal{N}}\dd{\mu(x)} \frac{N_s}{N_s(x_s)}\,\hat\pi_s(x_s) A_s(x_s).
\end{equation}
Actually, because the integrand only depends on $x_s$, this simplifies to
\begin{equation}
    \hat{A}_s\approx\int_{\mathcal{N}_s}\dd{\mu_s(x_s)} \frac{N_s}{N_s(x_s)}\,\hat\pi_s(x_s) A_s(x_s),
    \label{Equation: approximate classical operator subsystem}
\end{equation}
where $\mu_s$ is the measure on $\mathcal{N}_s$ obtained by taking the pushforward of $\mu$ through the map~\eqref{Equation: x to x_s}. Setting $A_s(x_s)=1$, we get an approximate resolution of the identity
\begin{equation}
    \frac{\mathds{1}_s}{N_s} \approx \int_{\mathcal{N}_s}\dd{\mu_s(x_s)}\frac{\hat\pi_s(x_s)}{N_s(x_s)}.
\end{equation}
Acting on both sides with $\hat\pi_s(y_s)$ and using~\eqref{Equation: approximately orthogonal subsystem}, one finds
\begin{equation}
    \frac{N_s}{N_s(x)}\,\hat\pi_s(x_s)\hat\pi_s(y_s) \approx \delta_{\mu_s}(x_s,y_s)\hat\pi(x_s),
    \label{Equation: approximate delta subsystem}
\end{equation}
where $\delta_{\mu_s}(x_s,y_s)$ is a delta function for the measure $\mu_s$.

At this stage, it is clear that subsystem $s$ may be treated using a self-contained classical limit of the kind defined in Section~\ref{Subsection: classical limit}. Indeed, the above equations also appear in that section, just without the ${}_s$ subscript on everything. However, here there is no requirement for the projection operators $\hat\pi_s(x_s)$ to be rank 1, so this is not a \emph{complete} classical limit in the sense defined in that section. This may be puzzling --- the point of a complete classical limit was that it was required for there to be no quantum degrees of freedom remaining in a classical limit. But this is exactly what we wanted to be true of a classically resolvable subsystem. 

The reason it is self-consistent for the rank of $\hat\pi_s(x_s)$ to be greater than 1, even though there are no quantum degrees of freedom remaining in the classical limit, is that in Section~\ref{Subsection: classical limit} we were only considering the classical limit of an isolated system, with no assumptions about its relation to anything else. In contrast, here we are making the assumption that subsystem $s$ is part of a larger total system in some classical state. This yields extra information about the state of the subsystem, which, it turns out, is enough to eliminate any possible leftover quantum degrees of freedom, in the way we have described. This apparent dependence on a non-local relationship between subsystem $s$ and the rest of the total system is the first hint of gauge symmetry.

\section{Gauge symmetry from entanglement}
\label{Section: Gauge symmetry from entanglement}

Let us take a completely classical limit of a quantum system, with classical states $x\in\mathcal{N}$ corresponding to quantum states $\ket{x}\in\mathcal{H}$, and let us assume from now on that the system has a classically resolvable local structure $\mathscr{S}$. 

For any subdivision $\{s_i\in\mathscr{S}\mid i=1,\dots,n\}$, we can decompose Hilbert space as in~\eqref{Equation: quantum tensor factorisation}:
\begin{equation}
    \mathcal{H} = \mathcal{H}_1\otimes\dots\otimes\mathcal{H}_i\otimes\dots\otimes\mathcal{H}_n,
    \label{Equation: quantum tensor factorisation 2}
\end{equation}
where $\mathcal{H}_i$ is the Hilbert space of subsystem $s_i$. Let us summarise the results of the previous section, which apply for each subsystem $s_i$:
\begin{itemize}
    \item There is a classical space of subsystem states $\mathcal{N}_i$, and a map $\mathcal{N}\to\mathcal{N}_i$ taking each classical state $x$ of the full system to the corresponding classical state $x_i$ of subsystem $s_i$.
    \item For any classical state $x$ of the full system, the reduced state of subsystem $s_i$ is approximately proportional to a projection operator $\hat\pi_i(x_i)$ that depends only on $x_i$:
        \begin{equation}
            \rho_i(x) = \tr_{\bar{i}}\qty(\ket{x}\bra{x}) \approx \frac{\hat{\pi}_i(x_i)}{N_i(x_i)}.
            \label{Equation: subdivision density matrices}
        \end{equation}
        Here, $N_i(x_i)$ is the rank of $\hat\pi_i(x_i)$, and $\tr_{\bar{i}}$ is a partial trace over $\mathcal{H}_1,\dots,\mathcal{H}_{i-1},\mathcal{H}_{i+1},\dots,\mathcal{H}_n$.
    \item The classical operators of subsystem $s_i$ take the form
        \begin{equation}
            \hat{A}_i \approx \int_{\mathcal{N}_i}\dd{\mu_i(x_i)} \frac{N_i}{N_i(x_i)}\,\hat\pi_i(x_i) \, A_i(x_i),
            \label{Equation: local classical operators}
        \end{equation}
        where $N_i=\dim(\mathcal{H}_i)$, $\mu_i$ is some measure on $\mathcal{N}_i$, and $A_i(x_i)$ is any function on $\mathcal{N}_i$.
    \item The projection operators obey
        \begin{equation}
            \frac{N_i}{N_i(x_i)}\hat\pi_i(x_i)\hat\pi_i(y_i) \approx \delta_{\mu_i}(x_i,y_i) \hat\pi_i(x_i),
            \label{Equation: subdivision orthogonality}
        \end{equation}
        where $x_i,y_i\in\mathcal{N}_i$, and $\delta_{\mu_i}(x_i,y_i)$ is a delta function for the measure $\mu_i$.
\end{itemize}

The reduced density matrix $\rho_i(x)$ of each subsystem $s_i$ tell us about the way in which the subsystem is entangled with its complement. Here,~\eqref{Equation: subdivision density matrices} means that this can be entirely understood as being due to some part of the subsystem being maximally entangled with some part of its complement. Roughly speaking, the subsystem $s_i$ and its complement may be thought of as sharing approximately $\log_2(N_i(x_i))$ maximally entangled qubits, or Bell pairs.

Usually we think of entanglement as being an altogether quantum phenomenon. It might therefore be surprising that it seems to play a role in classically resolvable subsystems. After all, in such subsystems quantum degrees of freedom are completely eliminated. 

In this section, we will explain how entanglement does indeed play a role at the classical level, but in a different guise than the usual one: as an emergent classical gauge symmetry.

\subsection{The unentangled case}

As a sanity check, let us first consider the case where the subsystems are separable, i.e.\ they are not entangled with each other. There should be no gauge symmetry in this case, and it is not difficult to confirm that this is true --- we will do so now.

When the subsystems are unentangled, the reduced density matrices are rank 1, so we can write
\begin{equation}
    \rho_i(x) \approx \hat\pi_i(x_i) = \ket{x_i}\bra{x_i}
\end{equation}
for some normalised state $\ket{x_i}\in\mathcal{H}_i$ that depends only on $x_i$. The approximate orthogonality of the projection operators~\eqref{Equation: subdivision orthogonality} implies that these states obey
\begin{equation}
    \braket{x_i}{y_i} \approx \delta_{x_iy_i}.
\end{equation}
Also, the classical operators of each subsystem may be written
\begin{equation}
    \hat{A}_i \approx \int_{\mathcal{N}_i}\dd{\mu_i(x_i)} N_i\, \ket{x_i}\bra{x_i} \, A_i(x_i).
\end{equation}
Thus, each classically resolvable separable subsystem undergoes a \emph{complete} classical limit, in contrast to the entangled case.

For each $i$, $x_i\in\mathcal{N}_i$ determines the subsystem quantum state $\ket{x_i}$, and we have
\begin{equation}
    \ket{x}\bra{x}=\ket{x_1}\bra{x_1}\otimes\dots\otimes\ket{x_n}\bra{x_n}.
\end{equation}
We can measure this operator to ascertain whether the full state of the system is $x$, so the collection $x_1,\dots,x_n$ determines $x$. Since $x$ also determines $x_1,\dots, x_n$, it is clear that the space of classical states $\mathcal{N}$ is in bijection with
\begin{equation}
    \mathcal{N}_1\times\dots\times\mathcal{N}_n.
\end{equation}
Thus, the classical space of states for the full system decomposes into a product of the classical space of states for each subsystem. This mirrors the structure of the Hilbert space~\eqref{Equation: quantum tensor factorisation 2}, and indicates that there is indeed no gauge symmetry at the classical level.

\subsection{Non-local degrees of freedom}

Let us now allow the subsystems to be entangled again. Note that in general it is possible for subsystems to be entangled in some classical states $x\in\mathcal{N}$, but not in others. As explained in the introduction, gauge symmetry is used to account for non-local degrees of freedom. We will show that there are non-local classical degrees of freedom whenever subsystems are entangled.

We need to have a precise way of characterising the existence of such non-local degrees of freedom. Recall that the classical operators of subsystem $s_i$ may be written in the form~\eqref{Equation: local classical operators}. Let $\mathcal{C}_i$ be the set of such operators and consider the set 
\begin{equation}
    \mathcal{C}_{\text{local}} = \mathcal{C}_1\otimes\dots\otimes \mathcal{C}_n
\end{equation}
consisting of operators which can be obtained by taking linear combinations of products of the classical operators for each subsystem. We call these `local' operators, because they measure only the classical states $x_1,\dots,x_n$ local to the subsystems $s_1,\dots,s_n$. The most general local operator $\hat{A}_{\text{local}}\in\mathcal{C}_{\text{local}}$ takes the form
\begin{equation}
    \hat{A}_{\text{local}} \approx \int_{\mathcal{N}_1}\dd{\mu_1(x_1)}\dots\int_{\mathcal{N}_n}\dd{\mu_n(x_n)}\, N \, \hat\pi_{\text{local}}(x_1,\dots,x_n)\, A_{\text{local}}(x_1,\dots,x_n),
    \label{Equation: local operator}
\end{equation}
where
\begin{equation}
    \hat\pi_{\text{local}}(x_1,\dots,x_n) = \hat\pi_1(x_1)\otimes\dots\otimes\hat\pi_n(x_n)
\end{equation}
is a projection operator acting on $\mathcal{H}$. Since $\mathcal{C}_i$ consists of classical operators, so too does $\mathcal{C}_{\text{local}}$. Indeed, the classical observable that $\hat{A}_{\text{local}}$ corresponds to is
\begin{equation}
    A(x) = A_{\text{local}}(x_1,\dots,x_n),
\end{equation}
where on the right-hand side $x_i$ is the state of subsystem $s_i$ corresponding to the state $x\in\mathcal{N}$ of the full system. Thus, we have $\mathcal{C}_{\text{local}}\subset\mathcal{C}$, where $\mathcal{C}$ is the set of classical operators for the total system, i.e.\ those which may be written in the form~\eqref{Equation: classical limit operator}. 

Operators in $\mathcal{C}$ which are not in $\mathcal{C}_{\text{local}}$ correspond to classical observables which \emph{cannot} be written in terms of the local classical states $x_1,\dots,x_n$. Thus, they must depend on \emph{non-local} classical degrees of freedom. Let
\begin{equation}
    \mathcal{C}_{\text{non-local}} = \mathcal{C}\setminus\mathcal{C}_{\text{local}}
\end{equation}
be the set of these non-local observables.

There are non-local observables whenever there are entangled classical states. To show this, note that
\begin{equation}
    \hat\pi_{\text{local}}(x_1,\dots,x_n) \hat\pi_{\text{local}}(y_1,\dots,y_n) \approx \delta_{x_1y_1}\dots\delta_{x_ny_n}\, \hat\pi_{\text{local}}(x_1,\dots,x_n),
\end{equation}
so these projection operators are approximately orthogonal. This implies that the rank of the general local operator~\eqref{Equation: local operator} obeys
\begin{equation}
    \operatorname{rank}(\hat{A}_{\text{local}}) \ge \max_{(x_1,\dots,x_n)\in\operatorname{supp}(A_{\text{local}})} \operatorname{rank}(\hat{\pi}_{\text{local}}(x_1,\dots,x_n))
\end{equation}
in the classical limit. Now suppose $y$ is a state in which the subsystems are entangled, and suppose $\hat{A}_\text{local}\ket{y} \approx \ket{y}$. Then $(y_1,\dots, y_n)\in\operatorname{supp}(A_{\text{local}})$, which gives
\begin{equation}
    \operatorname{rank}(\hat{A}_{\text{local}}) \ge \operatorname{rank}(\hat{\pi}_{\text{local}}(y_1,\dots,y_n)) = \operatorname{rank}(\hat\pi_1(y_1))\dots\operatorname{rank}(\hat\pi_n(y_n)) > 1.
\end{equation}
On the other hand, $\hat\pi(y) = \ket{y}\bra{y}$ is a rank 1 classical operator obeying $\hat\pi(y)\ket{y}\approx \ket{y}$. Thus, we must have $\hat\pi(y)\in\mathcal{C}$ and $\hat\pi(y)\not\in\mathcal{C}_{\text{local}}$, i.e.\ $\hat\pi(y)\in\mathcal{C}_{\text{non-local}}$ is non-local.

Thus, for each state $y$ in which the subsystems are entangled, the operator $\ket{y}\bra{y}$ must be a non-local classical operator. This operator measures whether the classical state of the full system is $y$, so to know with complete precision whether the state is $y$, we must measure some non-local degrees of freedom.

It is interesting that the presence of these non-local degrees of freedom depends upon whether or not, and how, the subsystems are entangled. Indeed, as shown in the last subsection, when the subsystems are separable there are no non-local degrees of freedom. It is possible for the subsystems to be separable for some states $x\in\mathcal{N}$, but entangled in others $y\in\mathcal{N}$. This would mean that there are non-local degrees of freedom when the state is $y$, but not when the state is $x$. Moreover, for different entangled states, it may be that the subsystems are entangled with each other in different ways, and thus that the set of non-local degrees of freedom varies. As we will discuss in Section~\ref{Section: Gravitational properties}, this has a natural gravitational interpretation in terms of a variable `bulk' topology.

Let us now dial down a bit more on the structure of the non-local degrees of freedom. In particular, let us ask the following question: when do two subsystems $s_i,s_j\in\mathscr{S}$ share non-local degrees of freedom? This is the case if there are classical degrees of freedom in $s_{ij}$ which cannot be measured in terms of the local degrees of freedom $x_i,x_j$ in $s_i,s_j$ respectively, where $s_{ij}$ is the union of $s_i$ and $s_j$. 

By the analysis of the previous section, there is a space $\mathcal{N}_{ij}$ of classical states in $s_{ij}$, and classical operators acting on $s_{ij}$ may be written in the form
\begin{equation}
    \hat{A}_{ij} = \int_{\mathcal{N}_{ij}}\dd{\mu_{ij}(x_{ij})} \frac{N_{ij}}{N_{ij}(x_{ij})}\,\hat\pi_{ij}(x_{ij}) \, A_{ij}(x_{ij}),
\end{equation}
where $N_{ij} = \dim(\mathcal{H}_{i}\otimes\mathcal{H}_j)=N_iN_j$, and $\hat\pi_{ij}(x_{ij})$ is a projection operator of rank $N_{ij}(x_{ij})$ to which the reduced density matrix of $s_{ij}$ is approximately proportional:
\begin{equation}
    \rho_{ij}(x) = \tr_{\overline{ij}}\qty(\ket{x}\bra{x}) \approx \frac{\hat\pi_{ij}(x_{ij})}{N_{ij}(x_{ij})}
    \label{Equation: subdivision density matrices 2}
\end{equation}
Note that the density matrices of $s_i$ and $s_j$ can be obtained by taking partial traces of the density matrix for $s_{ij}$:
\begin{equation}
    \rho_i(x) = \tr_j\qty(\rho_{ij}(x)), \quad \rho_j(x)=\tr_i\qty(\rho_{ij}(x)).
    \label{Equation: density matrix hierarchy}
\end{equation}
Since classical resolvability means there are 1-to-1 correspondences between the density matrices $\rho_i(x),\rho_j(x),\rho_{ij}(x)$ and the classical states $x_i,x_j,x_{ij}$, we have maps $\mathcal{N}_{ij}\mapsto \mathcal{N}_i$, $\mathcal{N}_{ij}\mapsto \mathcal{N}_j$ which take the classical state $x_{ij}$ of subsystem $s_{ij}$ to the classical states $x_i,x_j$ of subsystems $s_i,s_j$ respectively. 

The classical operator that measures whether the state in $s_{ij}$ is $x_{ij}$ is the projection operator $\hat\pi_{ij}(x_{ij})$. If there are no non-local degrees of freedom shared between $s_i$ and $s_j$, then we should be able to determine $x_{ij}$ using $x_i,x_j$. If this is the case, then we should also be able to write the classical operator that measures whether the state in $s_{ij}$ is $x_{ij}$ as $\hat\pi_i(x_i)\otimes\hat\pi_j(x_j)$, since this operator measures whether the states in $s_i,s_j$ are $x_i,x_j$ respectively. Thus, in the absence of non-local degrees of freedom shared between $s_i$ and $s_j$, we must have
\begin{equation}
    \hat\pi_{ij}(x_{ij}) \approx \hat\pi_i(x_i)\otimes\hat\pi_j(x_j).
    \label{Equation: projection factorisation}
\end{equation}
In terms of the ranks of these operators, this implies
\begin{equation}
    \log(N_i(x_i)) + \log(N_j(x_j)) - \log(N_{ij}(x_{ij}))  \approx 0,
\end{equation}
or in terms of the density matrices~\eqref{Equation: subdivision density matrices} and~\eqref{Equation: subdivision density matrices 2}, we may write
\begin{equation}
    I_{i:j}(x) = S(\rho_i(x)) + S(\rho_j(x)) - S(\rho_{ij}(x)) \approx 0,
\end{equation}
where $S(\rho) = -\tr(\rho\log\rho)$ is the von Neumann entropy of $\rho$.

The quantity $I_{i:j}(x)$ has a name in quantum information theory --- it is the `mutual information' of subsystems $s_i$ and $s_j$. Thus, we find that subsystems $s_i$ and $s_j$ share emergent non-local degrees of freedom in the state $x$ only if their mutual information $I_{i:j}(x)$ does not vanish in the classical limit. Actually, the vanishing of the mutual information suffices to guarantee the existence of the factorisation~\eqref{Equation: projection factorisation}, so the implication goes both ways.

The mutual information $I_{i:j}(x)$ tells us when pairs of subsystems share degrees of freedom. In other words, it tells us about `bilocal' degrees of freedom. However, more generally there could be emergent `multilocal' degrees of freedom, i.e.\ degrees of freedom which are shared between more than two systems. It would be interesting to understand the extent to which these multilocal degrees of freedom can be characterised by generalisations of the mutual information (e.g.\ those described in~\cite{multiparty}), but we will not comment further on this here.

\subsection{Purifications as entanglement edge modes}

Let us now describe how to account for the non-local degrees of freedom using gauge symmetry and edge modes. Because the non-local degrees of freedom come from entanglement, the edge modes must be capable of adequately describing the way in which each subsystem is entangled with the others. 

There are different ways one could do this --- right now we will only describe one. Of course, at the physical level, the type of gauge symmetry one introduces does not matter so much, so long as it allows one to adequately account for the right non-local degrees of freedom. What we give here does so in a completely general setting. Later, in Sections~\ref{Section: Example} and~\ref{Section: group}, we will see different examples of more specialised kinds of gauge symmetry which can be used to account for the non-local degrees of freedom emerging from entanglement.

The edge modes we will add to subsystem $s_i$ are `purifications' of $\rho_i(x)$. When the classical state of the subsystem is $x_i\in\mathcal{N}_i$, these are quantum states
\begin{equation}
    \ket{\psi_i} \in \mathcal{H}_i(x_i)\otimes\widetilde{\mathcal{H}}_i(x_i)
\end{equation}
satisfying $\widetilde{\tr}_{i}\qty(\ket{\psi_i}\bra{\psi_i}) \approx \rho_i(x)$, where $\mathcal{H}_i(x_i)$ is the image of $\hat\pi_i(x_i)$, $\widetilde{\mathcal{H}}_i(x_i)$ is an auxiliary Hilbert space, and $\widetilde{\tr}_{i}$ denotes a partial trace over $\widetilde{\mathcal{H}}_i(x_i)$.

It will be more notationally convenient for us to think of the purification as a map $\Psi_i:\mathcal{H}_i(x_i)\to \widetilde{\mathcal{H}}_i(x_i)^*$, related to $\ket{\psi_i}$ by a partial dualisation:
\begin{equation}
    \Big(\Psi_i \ket{\phi}\Big)\ket*{\tilde\phi} = \bra{\psi_i}(\ket{\phi}\otimes\ket*{\tilde\phi}), \qquad \ket{\phi}\in\mathcal{H}_i(x_i),\quad \ket*{\tilde\phi}\in\widetilde{\mathcal{H}}_i(x_i),
\end{equation}
The condition $\widetilde{\tr}_{i}\qty(\ket{\psi_i}\bra{\psi_i}) \approx \rho_i(x)$ may then be written $\Psi_i^\dagger\Psi_i\approx\rho_i(x)$. Since $\rho_i(x)\approx\hat\pi_i(x_i)/N_i(x_i)$, this means that $\sqrt{N_i(x_i)}\Psi_i$ is an approximate isometry. 

The dimension of the auxiliary space must be greater than or equal to the rank of $\hat\pi_i(x_i)$ for purifications to exist. We will take the simplest possible choice, which is $\widetilde{\mathcal{H}}_i(x_i)^*=\mathbb{C}^{N_i(x_i)}$. Then the auxiliary space has the same dimension as $\mathcal{H}_i(x_i)$, which implies that $\sqrt{N_i(x_i)}\Psi_i$ is approximately unitary. Let $\mathcal{P}_i(x_i)$ be the space of all such purifications, i.e.
\begin{equation}
    \mathcal{P}_i(x_i)=\qty{\Psi_i:\mathcal{H}_i(x_i)\to\mathbb{C}^{N_i(x_i)}\mid N_i(x_i)\Psi_i\Psi_i^\dagger \approx \mathds{1}_{N_i(x_i)}},
\end{equation}
where $\mathds{1}_{N_i(x_i)}$ is the identity acting in $\mathbb{C}^{N_i(x_i)}$.

When the subsystem state is $x_i$, we take the possible states of the edge modes to be elements of $\mathcal{P}_i(x_i)$, which means that the space of kinematical states for subsystem $s_i$ is
\begin{equation}
    \mathcal{N}^{\text{kin.}}_i = \qty{\qty(x_i,\Psi_i)\mid x_i\in\mathcal{N}_i, \, \Psi_i\in\mathcal{P}_i(x_i)},
\end{equation}
with the map $R_i:\mathcal{N}^{\text{kin.}}_i\to \mathcal{N}_i$ from each kinematical subsystem state to the corresponding physical subsystem state given by
\begin{equation}
    R_i: (x_i,\Psi_i) \mapsto x_i.
\end{equation}

The Hilbert space $\mathbb{C}^{N_i(x_i)}$ may be thought of as (roughly speaking) the space of states of approximately $\log_2 N_i(x_i)$ qubits. Each purification $\Psi_i$ describes the way in which these qubits are entangled with the subsystem. In the context of entanglement distillation, these qubits are sometimes called `ebits', which is an abbreviation of `entanglement bits'. It is amusing to note that, in the present context, `ebits' could also be an abbreviation of `edge bits'.

\subsection{Constraints, gluing, and gauge reduction}

\label{Subsection: entanglement gauge symmetry}

Given the kinematical state spaces $\mathcal{N}_i^\text{kin.}$ for the subsystems $s_i$, the kinematical state space for the full system is defined as
\begin{equation}
    \mathcal{N}^{\text{kin.}} = \mathcal{N}_1^{\text{kin.}}\times\dots\times\mathcal{N}_n^{\text{kin.}}.
\end{equation}
The next step in the gauge symmetry procedure is to identify the constraint surface $\overline{\mathcal{N}^{\text{kin.}}}\subset\mathcal{N}^{\text{kin.}}$. So we need to know: when can a given kinematical state
\begin{equation}
    \big((x_1,\Psi_1),\dots,(x_n,\Psi_n)\big)\in\mathcal{N}^{\text{kin.}}
    \label{Equation: kinematical state}
\end{equation}
be consistently `glued together' to form a physical state?

Let 
\begin{equation}
    X: \quad \mathcal{N}\to\mathcal{N}_1\times\dots\times\mathcal{N}_n,
\end{equation}
be the map which takes each physical state $x$ for the full system to the corresponding physical states $(x_1,\dots,x_n)$ for the subsystems, and let
\begin{equation}
    \overline{\mathcal{N}_1\times\dots\times\mathcal{N}_n} = \operatorname{image}(X)
\end{equation}
be the image of this map. Clearly, in order for the kinematical state~\eqref{Equation: kinematical state} to yield a sensible physical state, the $x_1,\dots,x_n$ appearing in it should come from some $x\in\mathcal{N}$, i.e.\ they should obey 
\begin{equation}
    (x_1,\dots,x_n)\in \overline{\mathcal{N}_1\times\dots\times\mathcal{N}_n}.
    \label{Equation: first constraint}
\end{equation}
This is the first constraint that will define $\overline{\mathcal{N}^{\text{kin.}}}$.

To describe the rest of the constraints, and the gluing procedure, it is useful to introduce some additional structures. The first is a section of $X$, i.e.\ a map $Y:\overline{\mathcal{N}_1\times\dots\times\mathcal{N}_n}\to\mathcal{N}$ such that $X\circ Y$ is the identity acting on $\overline{\mathcal{N}_1\times\dots\times\mathcal{N}_n}$. In other words, to every collection of physical subsystem states $(x_1,\dots,x_n)$ which \emph{can} come from a physical state of the total system, we pick one such state $Y(x_1,\dots,x_n)\in\mathcal{N}$ obeying $X\big(Y(x_1,\dots,x_n)\big) = (x_1,\dots,x_n)$. The other structure that we will use is a choice of purification $\Phi_i(x_i)\in\mathcal{P}_i(x_i)$ for each $x_i\in\mathcal{N}_i$, for all subsystems $s_i$. The choices of section $Y$ and purifications $\Phi_i(x_i)$ do not affect the final physical content of the theory.

Given $Y$ and $\Phi_i(x_i)$, for each $(x_1,\dots,x_n)\in\overline{\mathcal{N}_1\times\dots\times\mathcal{N}_n}$ we can construct
\begin{equation}
    \ket{\sigma(x_1,\dots,x_n)} = N_1(x_1)\dots N_n(x_n)\,\big(\Phi_1(x_1)\otimes\dots\otimes\Phi_n(x_n)\big) \ket{Y(x_1,\dots,x_n)}.
\end{equation}
This is an (entangled) state in $\mathbb{C}^{N_1(x_1)}\otimes\dots\otimes\mathbb{C}^{N_n(x_n)}$, which is the tensor product of the auxiliary Hilbert spaces of the subsystems. We can directly use this state to glue together the kinematical subsystem states. More precisely, suppose a kinematical state~\eqref{Equation: kinematical state} of the full system obeys~\eqref{Equation: first constraint}. Then we define a quantum state $\ket{x_i,\Psi_i}\in\mathcal{H}$ of the full system via
\begin{equation}
    \ket{x_i,\Psi_i} = \big(\Psi_1^\dagger\otimes\dots\otimes\Psi_n^\dagger\big)\ket{\sigma(x_1,\dots,x_n)}.
    \label{Equation: purification gluing}
\end{equation}
Essentially, we obtain $\ket{x_i,\Psi_i}$ by taking the tensor product of the purifications $\Psi_i$ for each subsystem, and then projecting the state of the auxiliary degrees of freedom onto $\ket{\sigma(x_1,\dots,x_n)}$. Some diagrams representing this procedure are given in Figure~\ref{Figure: tensor diagrams}.

\begin{figure}
    \centering
    \begin{subfigure}{\linewidth}
        \centering
        \begin{tikzpicture}
            \draw[thick,green!40!black] (0,0.35) -- (0,0.6);
            \draw[thick,green!40!black] (0,-0.35) -- (0,-0.6);
            \draw[very thick, fill=green!8] (-0.55,-0.35) rectangle (0.55,0.35);
            \node at (0,0) {$\rho_i(x)$};
            \node at (1.08,0) {$=$};
            \begin{scope}[shift={(2,0.5)}]
                \draw[very thick, dotted, red] (0,-0.35) -- (0,-0.65);
                \draw[thick,green!40!black] (0,0.35) -- (0,0.6);
                \draw[very thick, fill=blue!10] (-0.35,-0.35) rectangle (0.35,0.35);
                \node at (0,0) {$\Psi_i^\dagger$};
            \end{scope}
            \begin{scope}[shift={(2,-0.5)}]
                \draw[thick,green!40!black] (0,-0.35) -- (0,-0.6);
                \draw[very thick, fill=blue!10] (-0.35,-0.35) rectangle (0.35,0.35);
                \node at (0,0) {$\Psi_i$};
            \end{scope}
        \end{tikzpicture}
        \caption{}
    \end{subfigure}
    \vspace{\baselineskip}

    \begin{subfigure}{\linewidth}
        \centering
        \begin{tikzpicture}
            \begin{scope}[shift={(4,-0.5)}]
                \draw[thick,green!40!black] (-1,0.35) -- (-1,0.6);
                \draw[thick,green!40!black] (-0.6,0.35) -- (-0.6,0.5);
                \draw[thick,green!40!black] (-0.2,0.35) -- (-0.2,0.5);
                \draw[thick,green!40!black] (0.2,0.35) -- (0.2,0.5);
                \draw[thick,green!40!black] (0.6,0.35) -- (0.6,0.5);
                \draw[thick,green!40!black] (1,0.35) -- (1,0.6);
                \draw[very thick, fill=green!8] (-1.4,-0.35) rectangle (1.4,0.35);
                \node at (0,0) {$\ket{Y(x_1,\dots,x_n)}$};
            \end{scope}
            \begin{scope}[shift={(5,0.5)}]
                \draw[very thick,dotted,red] (0,0.35) -- (0,0.7);
                \draw[thick,green!40!black] (0,-0.35) -- (0,-0.6);
                \draw[very thick, fill=blue!10] (-0.57,-0.35) rectangle (0.57,0.35);
                \node at (0,0) {\footnotesize$\Phi_n(x_n)$};
            \end{scope}
            \node at (4,0.5) {$\dots$};
            \begin{scope}[shift={(3,0.5)}]
                \draw[very thick,dotted,red] (0,0.35) -- (0,0.7);
                \draw[thick,green!40!black] (0,-0.35) -- (0,-0.6);
                \draw[very thick, fill=blue!10] (-0.57,-0.35) rectangle (0.57,0.35);
                \node at (0,0) {\footnotesize$\Phi_1(x_1)$};
            \end{scope}
            \begin{scope}[shift={(-3.2,0)}]
                \draw[very thick,red,dotted] (-1,0.35) -- (-1,0.6);
                \draw[very thick,red,dotted] (-0.6,0.35) -- (-0.6,0.6);
                \draw[very thick,red,dotted] (-0.2,0.35) -- (-0.2,0.6);
                \draw[very thick,red,dotted] (0.2,0.35) -- (0.2,0.6);
                \draw[very thick,red,dotted] (0.6,0.35) -- (0.6,0.6);
                \draw[very thick,red,dotted] (1,0.35) -- (1,0.6);
                \draw[very thick, fill=red!8] (-1.4,-0.35) rectangle (1.4,0.35);
                \node at (0,0) {$\ket{\sigma(x_1,\dots,x_n)}$};
            \end{scope}
            \node at (.3,0) {$=\,N_1(x_1)\dots N_n(x_n)$};
        \end{tikzpicture}
        \caption{}
    \end{subfigure}
    \vspace{\baselineskip}

    \begin{subfigure}{\linewidth}
        \centering
        \begin{tikzpicture}
            \begin{scope}[shift={(5,0.5)}]
                \draw[thick,green!40!black] (0,0.35) -- (0,0.7);
                \draw[very thick,dotted,red] (0,-0.35) -- (0,-0.65);
                \draw[very thick, fill=blue!10] (-0.35,-0.35) rectangle (0.35,0.35);
                \node at (0,0) {$\Psi_n^\dagger$};
            \end{scope}
            \node at (4,0.5) {$\dots$};
            \begin{scope}[shift={(3,0.5)}]
                \draw[thick,green!40!black] (0,0.35) -- (0,0.7);
                \draw[very thick,dotted,red] (0,-0.35) -- (0,-0.65);
                \draw[very thick, fill=blue!10] (-0.35,-0.35) rectangle (0.35,0.35);
                \node at (0,0) {$\Psi_1^\dagger$};
            \end{scope}
            \begin{scope}[shift={(4,-0.5)}]
                \draw[very thick,dotted,red] (-0.6,0.35) -- (-0.6,0.5);
                \draw[very thick,dotted,red] (-0.2,0.35) -- (-0.2,0.5);
                \draw[very thick,dotted,red] (0.2,0.35) -- (0.2,0.5);
                \draw[very thick,dotted,red] (0.6,0.35) -- (0.6,0.5);
                \draw[very thick, fill=red!8] (-1.4,-0.35) rectangle (1.4,0.35);
                \node at (0,0) {$\ket{\sigma(x_1,\dots,x_n)}$};
            \end{scope}
            \draw[thick,green!40!black] (-1,0.35) -- (-1,0.6);
            \draw[thick,green!40!black] (-0.6,0.35) -- (-0.6,0.6);
            \draw[thick,green!40!black] (-0.2,0.35) -- (-0.2,0.6);
            \draw[thick,green!40!black] (0.2,0.35) -- (0.2,0.6);
            \draw[thick,green!40!black] (0.6,0.35) -- (0.6,0.6);
            \draw[thick,green!40!black] (1,0.35) -- (1,0.6);
            \draw[very thick, fill=green!8] (-1.4,-0.35) rectangle (1.4,0.35);
            \node at (0,0) {$\ket{x_i;\Psi_i}$};
            \node at (1.9,0) {$=$};
        \end{tikzpicture}
        \caption{}
    \end{subfigure}
    \caption{Tensor diagrams for entanglement edge modes and the gluing procedure. Red dotted lines {\protect\tikz[baseline=-0.5ex,very thick,red,dotted]{\protect\draw (0,0) -- (0.5,0);}} represent indices in the auxiliary Hilbert spaces $\widetilde{\mathcal{H}}_i(x_i)$, while green solid lines {\protect\tikz[baseline=-0.5ex, thick,green!40!black]{\protect\draw (0,0) -- (0.5,0);}} represent indices in the original Hilbert spaces $\mathcal{H}_i$. \mbox{\textbf{(a)}}\, The edge modes we introduce are purifications of the subsystem states, i.e.\ maps $\Psi_i:\mathcal{H}_i\to\widetilde{\mathcal{H}}_i(x_i)$ obeying $\rho_i(x)=\Psi_i^\dagger\Psi_i$. \mbox{\textbf{(b)}}\, We construct a special state $\ket{\sigma(x_1,\dots,x_n)}$ for the auxiliary degrees of freedom by picking some reference purifications $\Phi_i(x_i)$ and a function $Y:\overline{\mathcal{N}_1\times\dots\times\mathcal{N}_n}\to\mathcal{N}$. \mbox{\textbf{(c)}}\, We can use the state $\ket{\sigma(x_1,\dots,x_n)}$ to `glue together' a collection of purifications into a state for the full system. }
    \label{Figure: tensor diagrams}
\end{figure}

Such a gluing procedure is only physically meaningful if it can be used to obtain all classical states, i.e.\ if for all $y\in\mathcal{N}$ there is a kinematical state~\eqref{Equation: kinematical state} which yields $\ket{x_i,\Psi_i}=\ket{y}$. Let us now show that this does in fact hold.

Suppose $X(x)=X(y)$, i.e.\ $x,y\in\mathcal{N}$ have the same physical subsystem states $x_i=y_i$. It will be useful to define the operator
\begin{equation}
    \hat{V}_i(x,y) = N_i(x_i) \tr_{\bar{i}}\qty(\ket{y}\bra{x}),
\end{equation}
where $\tr_{\bar{i}}$ is a partial trace over all $\mathcal{H}_j$ with $j\ne i$. This operator has certain useful properties. It can be used to map $\ket{x}$ to $\ket{y}$ via
\begin{multline}
    \qquad\qty(\hat{\pi}_1(x_1)\otimes\dots\otimes\hat{V}_i(x,y)\otimes\dots\otimes\hat\pi_n(x_n))\ket{x} \\
    = \qty(\hat{\pi}_1(x_1)\otimes\dots\otimes\hat{\pi}_i(x_i)\otimes\dots\otimes\hat\pi_n(x_n))\ket{y} \approx \ket{y},\qquad
\end{multline}
where the first equality is a simple rearrangement of the positions of $\ket{x}$ and $\ket{y}$. Also, if $\Psi_i\in\mathcal{P}_i(x_i)$, then we have
\begin{nalign}
    \qty(\Psi_i \hat{V}_i(x,y)^\dagger)^\dagger\qty(\Psi_i \hat{V}_i(x,y)^\dagger) 
    &= \hat{V}_i(x,y) \underbrace{\Psi_i^\dagger\Psi_i}_{\mathclap{\approx \hat\pi_i(x_i)/N_i(x_i)}} \qty(\hat{V}_i(x,y))^\dagger \\
    &\approx \hat{V}_i(x,y)\hat\pi_i(x_i)\tr_{\bar{i}}\qty(\ket{x}\bra{y}) \\
    &\approx \tr_{\bar{i}}\big(\underbrace{\big(\hat\pi_1(x_1)\otimes\dots\otimes\hat{V}_i(x,y)\hat\pi_i(x_i)\otimes\dots\otimes\hat\pi_n(x_n)\big)\ket{x}}_{\approx \ket{y}}\bra{y}\big) \\
    &\approx \frac{\hat\pi_i(y_i)}{N_i(y_i)}= \frac{\hat\pi_i(x_i)}{N_i(x_i)}.
\end{nalign}
Thus, $\Psi_i \hat{V}_i(x,y)^\dagger\in\mathcal{P}_i(x_i)$.

Let $y\in\mathcal{N}$ be any physical state of the full system with corresponding physical subsystem states $y_1,\dots,y_n$, and let us define a kinematical state~\eqref{Equation: kinematical state} for the full system via $x_i=y_i$ and
\begin{equation}
    \Psi_i = 
    \begin{cases}
        \Phi_j(y_j)\, \hat{V}_j\qty(Y(y_1,\dots,y_n),y)^\dagger & \text{ if $i=j$},\\
        \Phi_i(y_i) & \text{ otherwise},
    \end{cases}
    \label{Equation: kinematical to be glued j}
\end{equation}
for some fixed $j$. Then we have
\begin{nalign}
    \ket{x_i,\Psi_i} &= \qty(\Phi_1(y_1)^\dagger\otimes\dots\otimes\hat{V}_j(Y(y_1,\dots,y_n),y) \Phi_j(y_j)^\dagger\otimes\dots\otimes\Phi_n(y_n)^\dagger)\ket{\sigma(y_1,\dots,y_n)} \\
                     & 
                     \begin{multlined}
                         =N_1(y_1)\dots N_n(y_n)
                         \Big( \Phi_1(y_1)^\dagger\Phi_1(y_1)\otimes\dots
                             \otimes\hat{V}_j(Y(y_1,\dots,y_n),y)\Phi_j(y_j)^\dagger\Phi_j(y_j)\otimes\dots\\\dots\otimes\Phi_n(y_n)^\dagger\Phi_n(y_n)\Big)
                         \ket{Y(y_1,\dots,y_n)}
                     \end{multlined}\\
                     &= \qty( \hat\pi_1(y_1)\otimes\dots\otimes\hat{V}_j(Y(y_1,\dots,y_n),y)\otimes\dots\otimes\hat\pi_n(y_n))\ket{Y(y_1,\dots,y_n)} 
                     \approx \ket{y}
\end{nalign}
Thus, after gluing, this kinematical state yields the classical state $\ket{y}$. Since this works for all $y$, the gluing procedure can produce all classical states, and so is physically meaningful.

Not all kinematical states in $\overline{\mathcal{N}_1\times\dots\times\mathcal{N}_n}$ will yield sensible classical states when glued together. Thus, we should restrict to kinematical states satisfying
\begin{equation}
    \ket{x_i,\Psi_i} = \ket{y} \text{ for some } y\in \mathcal{N}.
\end{equation}
This is the only remaining constraint we need. Therefore, the constraint surface is defined to be
\begin{multline}
    \overline{\mathcal{N}^\text{kin.}} = \Big\{\big((x_1,\Psi_1),\dots,(x_n,\Psi_n)\big)\in \mathcal{N}^{\text{kin.}} \bigm\vert (x_1,\dots,x_n)\in \overline{\mathcal{N}_1\times\dots\times\mathcal{N}_n}\\
    \text{ and }\ket{x_i,\Psi_i}=\ket{y}\text{ for some }y\in\mathcal{N}\Big\}.
\end{multline}
The gauge reduction map is now also clear:
\begin{equation}
    R: \big((x_1,\Psi_1),\dots,(x_n,\Psi_n)\big) \mapsto y \text{ such that } \ket{x_i,\Psi_i} = \ket{y}.
\end{equation}

Note also that there are many kinematical states which can be glued together to form a given physical state. For example, different choices of $j\in\{1,\dots,n\}$ in~\eqref{Equation: kinematical to be glued j} yield different kinematical states -- but once glued together they give the same physical state $y$.

Before moving on, let us comment on the role of the structures $Y,\Phi_i(x_i)$ that we introduced to carry out this construction. The set of physical states $\mathcal{N}$ for the full system clearly does not depend on these structures. However, the constraint surface $\overline{\mathcal{N}^{\text{kin.}}}$ and gauge reduction map $R$ do. This is entirely analogous to what happens in other gauge theories. For example, the analysis of subregions in gravity is greatly aided by the introduction of systems of spacetime coordinates. One discusses the physics in each subregion relative to the coordinates, and one uses the coordinates when one wishes to understand how different subregions are related (i.e.\ glued) to each other. Similarly, in Yang-Mills theory one often picks a local trivialisation of a principal bundle over spacetime, and works with the gauge connection relative to that trivialisation. In general, a coordinate system or local trivialisation, or whatever structure one similarly introduces, is an imaginary, non-physical \emph{reference frame}, relative to which one describes subsystem physics. From the point of view of the total system, this reference frame is irrelevant. In this sense, the objects $Y,\Phi_i(x_i)$ constitute a reference frame for the entanglement in classical states. The purifications $\Phi_i(x_i)$ are like `coordinate systems' for each subsystem, while the map $Y$ describes how these coordinate systems are related to each other. It would be interesting to understand how this fits into the bigger picture of physical reference frames, which has attracted recent renewed interest in both the classical and quantum contexts~\cite{Giacomini:2017zju,AhmadAli:2021ajb,delaHamette:2021oex,Castro-Ruiz:2021vnq,Carrozza:2021gju,Goeller:2022rsx}.

\subsection{Gauge transformations, observables, and Wilson lines}

A gauge transformation of a classical system with gauge symmetry is a transformation of its edge modes which does not affect the physical state. In this case, the edge modes are the purifications $\Psi_i$, and the physical state is determined by~\eqref{Equation: purification gluing}, so a gauge transformation is a change $\Psi_i\to \tilde\Psi_i$ that preserves~\eqref{Equation: purification gluing}. Since 
\begin{equation}
    \sqrt{N_i(x_i)}\Psi_i,\quad \sqrt{N_i(x_i)}\tilde\Psi_i\quad:\quad\mathcal{H}_i(x_i)\to \mathbb{C}^{N_i(x_i)}
\end{equation}
are both unitary, we can write $\tilde\Psi_i= U_i^\dagger\Psi_i$ for some unitary map $U_i:\mathbb{C}^{N_i(x_i)}\to\mathbb{C}^{N_i(x_i)}$. To preserve~\eqref{Equation: purification gluing}, these maps must obey
\begin{equation}
    \ket{\sigma(x_1,\dots,x_n)} = \big(U_1\otimes\dots\otimes U_n\big)\ket{\sigma(x_1,\dots,x_n)}.
\end{equation}
The gauge group is therefore
\begin{equation}
    G(x_1,\dots,x_n) = \operatorname{Stab}_{U(N_1(x_1))\times\dots\times U(N_n(x_n))}\Big(\ket{\sigma(x_1,\dots,x_n)}\Big),
\end{equation}
i.e.\ the stabiliser of $\ket{\sigma(x_1,\dots,x_n)}$ in $U(N_1(x_1))\times\dots\times U(N_n(x_n))$.\footnote{Elements of $U(N_1(x_1))\times\dots\times U(N_n(x_n))$ but not $G(x_1,\dots,x_n)$ may be thought of as `large' gauge transformations, because locally they look like gauge transformations, but globally they are not. This is analogous with the large gauge transformations and large diffeomorphisms of field theories and gravity.} The gauge group depends on the subsystem states $x_i$. This may seem strange, but it also happens, for example, in gravity, where the gauge group is the diffeomorphism group of a state-dependent manifold. In this case, it simply reflects the fact that there are different amounts of entanglement in different states, and so different kinds of non-local degrees of freedom.

Let us now see what kinds of observables we can construct from kinematical states, and how they transform under gauge transformations.

First, from the kinematical state $(x_i,\Psi_i)$ of a single subsystem $s_i$, we can construct observables from $x_i$ and $\Psi_i$. Clearly, $x_i$ is gauge-invariant, while $\Psi_i$ transforms under the right action $\Psi_i\to U_i^\dagger \Psi_i$ of the gauge group (by definition). A gauge transformation can map any $\Psi_i\in\mathcal{P}(x_i)$ to any other $\Psi_i\in\mathcal{P}(x_i)$, so no gauge-invariant observables may be formed from $\Psi_i$ alone.

Next, suppose $s_i$, $s_j$ are two subsystems with respective kinematical states $(x_i,\Psi_i)$, $(x_j,\Psi_j)$, which we will assume are consistent with the constraints. Again $x_i,x_j$ are gauge-invariant, while $\Psi_i,\Psi_j$ transform under the right action of the gauge group, and so individually carry no gauge-invariant information. Consider the object
\begin{equation}
    w_{ij} = \big(\Psi_i\otimes\Psi_j\big)\tr_{\overline{ij}}\big(\ket{\sigma(x_1,\dots,x_n)}\bra{\sigma(x_1,\dots,x_n)}\big) \big(\Psi_i^\dagger \otimes \Psi_j^\dagger\big).
\end{equation}
Note that although $\ket{\sigma(x_1,\dots,x_n)}$ depends on all of $x_1,\dots,x_n$, the object above only depends on $x_i,x_j$ and $\Psi_i,\Psi_j$. Under a gauge transformation we have
\begin{equation}
    w_{ij} \to \big(U_i\otimes U_j\big)^\dagger w_{ij}\big(U_i\otimes U_j\big).
\end{equation}
Note that $w_{ij}$ is a map $\mathbb{C}^{N_i(x_i)}\otimes\mathbb{C}^{N_j(x_j)}\to\mathbb{C}^{N_i(x_i)}\otimes\mathbb{C}^{N_j(x_j)}$. However, it is useful to recast it as a map 
\begin{equation}
    \mathcal{W}_{ij}:\quad \mathrm{GL}(N_j(x_j),\mathbb{C}) \to \mathrm{GL}(N_i(x_i),\mathbb{C})
\end{equation}
defined by
\begin{equation}
    \mathcal{W}_{ij}(f_j) = \tr_j\big(w_{ij}(\mathds{1}_i\otimes f_j)\big).
\end{equation}
Under a gauge transformation we have
\begin{equation}
    \mathcal{W}_{ij} \mapsto \operatorname{Ad}[U_i^\dagger] \circ \mathcal{W}_{ij} \circ \operatorname{Ad}[U_j] ,
\end{equation}
where $\operatorname{Ad}$ is the adjoint action, defined by $\operatorname{Ad}[U]: f\mapsto U f U^\dagger$. Thus $\mathcal{W}_{ij}$ transforms under the adjoint actions of the local gauge groups of subsystem $s_i$ and $s_j$. We may think of it as a `Wilson line' in the adjoint representation, connecting $s_i$ to $s_j$. Clearly, the Wilson lines connecting the subsystems are generally not gauge-invariant. However, we may form gauge-invariant observables from them. In particular, $w_{ij}$ is related to the reduced density matrix $\rho_{ij}(x)$ of the union $s_{ij}$ of the two subsystems via
\begin{equation}
    \rho_{ij}(x) = N_i(x_i) N_j(x_j) w_{ij}.
\end{equation}
Thus, the Wilson line $\mathcal{W}_{ij}$ may be used to construct any physical observable non-locally shared between $s_i$ and $s_j$.

Similarly, for any collection of three or more subsystems $s_i,s_j,\dots$, we can construct
\begin{equation}
    w_{ij\dots} = \big(\Psi_i\otimes\Psi_j\otimes\dots\big)\tr_{\overline{ij\dots}}\big(\ket{\sigma(x_1,\dots,x_n)}\bra{\sigma(x_1,\dots,x_n)}\big) \big( \Psi_i^\dagger \otimes \Psi_j^\dagger\otimes\dots\big)
\end{equation}
from the kinematical states $(x_i,\Psi_i),(x_j,\Psi_j),\dots$, and we may view $w_{ij\dots}$ as a kind of `multivalent Wilson line'. Any non-local observable shared between the subsystems may be constructed from $w_{ij\dots}$, since it is proportional to the reduced state $\rho_{ij\dots}(x)$ of the union of the subsystems.

Usually, Wilson lines in gauge theory measure parallel transport with respect to some gauge connection. The current setup is analogous to that situation. It would be interesting to understand the properties of this `emergent gauge connection', and to compare them with~\cite{Czech:2017zfq,Czech:2018kvg,Czech:2019vih,Jafferis:2020ora}. In Section~\ref{Section: Example}, we will see an example where the connection is flat.

\section{Toy model: classical limit of three entangled spins}
\label{Section: Example}

We shall now give an explicit example of the phenomenon described in this paper: three spins at high angular momentum, entangled in a particular way. This example provides a good toy model for the mechanism of emergent gauge symmetry. It may also be viewed as being based upon the simplest possible spin network, and so may be a good prototype for gaining insight into the classical limits of more general spin network states in loop quantum gravity.

\subsection{Ordinary spin coherent states}

Let $\mathcal{H}_i$ be the Hilbert space of a spin with total angular momentum $j_i$, and let $\hat{J}_a^i$, $a=1,2,3$ be the angular momentum operators acting on $\mathcal{H}_i$. Thus, these operators obey $\comm*{\hat{J}_a^i}{\hat{J}_b^i} = i\epsilon_{abc}\hat{J}_c^i$, and a basis of $\mathcal{H}_i$ is provided by the eigenvectors $\ket{j_i,m_i}$ of $\hat{J}_3^i$ with eigenvalues $m_i=-j_i,\dots,j_i$. We can define a set of spin coherent states $\ket{\mathbf{n}_i}\in\mathcal{H}_i$ via
\begin{equation}
    \ket{\mathbf{n}_i} = \hat{D}_i(\mathbf{n}_i)\ket{j_i,-j_i}, \qquad \hat{D}_i(\mathbf{n}_i) = \exp(i\frac{\theta_i}{\sin(\theta_i)}(\mathbf{n}_i\times\mathbf{e}^3)\cdot \hat{\mathbf{J}^i}),
\end{equation}
where $\mathbf{n}_i$ is some unit 3-vector, and $\theta_i$ is the angle between $\mathbf{n}_i$ and $\mathbf{e}^3=(0,0,1)$. These states have the following well-known properties (see e.g.~\cite{Coherent3j}). First, they satisfy
\begin{equation}
    \mathds{1}_i = \int_{S^2} \frac{\dd[2]{\mathbf{n}_i}}{4\pi}\, N_i\,\ket{\mathbf{n}_i}\bra{\mathbf{n}_i}
    \label{Equation: spin coherent resolution}
\end{equation}
where $\dd[2]{\mathbf{n}_i}$ is the standard measure on the unit sphere, $\mathds{1}_i$ is the identity on $\mathcal{H}_i$, and $N_i=\dim(\mathcal{H}_i)=2j_i+1$. Second, their inner product is given by
\begin{equation}
    \braket{\mathbf{n}_i}{\mathbf{n}'_i} = e^{ij_if(\mathbf{e}^3,\mathbf{n}_i,\mathbf{n}'_i)} \qty(\frac{1+\mathbf{n}_i\cdot\mathbf{n}'_i}2)^{j_i},
    \label{Equation: spin coherent overlap}
\end{equation}
where the function $f(\mathbf{e}^3,\mathbf{n}_i,\mathbf{n}'_i)$ in the phase is the oriented area of the spherical triangle whose vertices are $\mathbf{e}^3,\mathbf{n}_i,\mathbf{n}'_i$. For large enough $j_i$, it is clear that~\eqref{Equation: spin coherent overlap} can be made arbitrarily small, unless $\mathbf{n}_i=\mathbf{n}'_i$. Thus, $j_i\to \infty$ is a classical limit, with classical state space equal to the unit sphere $\mathcal{N}_i=S^2$, with each classical state $\mathbf{n}_i\in\mathcal{N}_i$ corresponding to the quantum state $\ket{\mathbf{n}_i}$. The classical picture of a quantum spin is thus just a classical spin, as expected.

The full quantum system we consider has the Hilbert space $\mathcal{H}=\mathcal{H}_1\otimes\mathcal{H}_2\otimes\mathcal{H}_3$, with each tensor factor $\mathcal{H}_i$ being the Hilbert space of a spin with total angular momentum $j_i$, as in the previous paragraph. One very simple way to construct a classical limit for this system is to consider the set of states $\ket{\mathbf{n}_1}\otimes\ket{\mathbf{n}_2}\otimes\ket{\mathbf{n}_3}$, where $\mathbf{n}_i$, $i=1,2,3$ are unit vectors. By~\eqref{Equation: spin coherent resolution}, these states give a resolution of the identity for $\mathcal{H}$:
\begin{equation}
    \mathds{1}=\int_{S^2\times S^2 \times S^2} \frac{\dd[2]{\mathbf{n}_1}}{4\pi}\frac{\dd[2]{\mathbf{n}_2}}{4\pi}\frac{\dd[2]{\mathbf{n}_3}}{4\pi} \, N \,\ket{\mathbf{n}_1}\otimes\ket{\mathbf{n}_2}\otimes\ket{\mathbf{n}_3}\bra{\mathbf{n}_1}\otimes\bra{\mathbf{n}_2}\otimes\bra{\mathbf{n}_3},
    \label{Equation: three spin coherent resolution}
\end{equation}
where $N=\operatorname{dim}(\mathcal{H})=N_1N_2N_3=(2j_1+1)(2j_2+1)(2j_3+1)$. By~\eqref{Equation: spin coherent overlap}, their inner product is given by 
\begin{multline}
    \qty(\bra{\mathbf{n}_1}\otimes\bra{\mathbf{n}_2}\otimes\bra{\mathbf{n}_3}) \qty(\ket{\mathbf{n}'_1}\otimes\ket{\mathbf{n}'_2}\otimes\ket{\mathbf{n}'_3}) \\
    = e^{i\qty(j_1f(\mathbf{e}^3,\mathbf{n}_1,\mathbf{n}'_1)+j_2f(\mathbf{e}^3,\mathbf{n}_2,\mathbf{n}'_2)+j_3f(\mathbf{e}^3,\mathbf{n}_3,\mathbf{n}'_3))} \qty(\frac{1+\mathbf{n}_1\cdot\mathbf{n}'_1}2)^{j_1}\qty(\frac{1+\mathbf{n}_2\cdot\mathbf{n}'_2}2)^{j_2}\qty(\frac{1+\mathbf{n}_3\cdot\mathbf{n}'_3}2)^{j_3}.
\end{multline}
If we set $j_i=K_i/\chi$ for some fixed $K_i>0$, $i=1,2,3$, and consider a limit in which $\chi\to 0$, this inner product vanishes for $(\mathbf{n}_1,\mathbf{n}_2,\mathbf{n}_3)\ne(\mathbf{n}'_1,\mathbf{n}'_2,\mathbf{n}'_3)$. Thus, $\chi\to0$ is a classical limit, in which the classical space of states is the product of three unit spheres $\mathcal{N}=\mathcal{N}_1\times\mathcal{N}_2\times\mathcal{N}_3$, where $\mathcal{N}_i=S^2$. This classical limit describes three independent classical spins. There is no gauge symmetry, as is to be expected from the lack of entanglement.

This classical limit is perfectly valid, but there may be reasons it is not useful in a given physical scenario. For example, the dynamics may be such that significant amounts of entanglement can be generated between the three spins, in which case the classical description given by the set of states $\ket{\mathbf{n}_1}\otimes\ket{\mathbf{n}_2}\otimes\ket{\mathbf{n}_3}$ cannot remain deterministic at all times. For this reason, it is worth trying to understand different kinds of classical limits --- in particular, ones which involve entanglement between the spins. 

\subsection{Entangled spin coherent states}

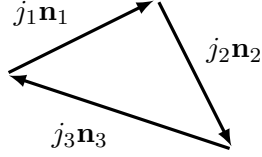
\begin{figure}
    \centering
    \begin{tikzpicture}[very thick,-latex,shorten <=0.05cm,shorten >=0.05cm]
        \draw (0,0) -- (2,1) node[midway, above left] {$j_1\mathbf{n}_1$};
        \draw (2,1) -- (3,-1) node[midway, above right] {$j_2\mathbf{n}_2$};
        \draw (3,-1) -- (0,0) node[midway, below left] {$j_3\mathbf{n}_3$};
    \end{tikzpicture}
    \caption{In a separable coherent state basis $\ket{\mathbf{n}_1}\otimes\ket{\mathbf{n}_2}\otimes\ket{\mathbf{n}_3}$, the zero total angular momentum state $\ket{0,0}$ of three spins is sharply peaked when the closure condition~\eqref{Equation: closure condition} holds, in a classical limit $j_i=\order{1/\chi}$, $\chi\to 0$. Thus, the vectors $j_1\mathbf{n}_1,j_2\mathbf{n}_2,j_3\mathbf{n}_3$ give three sides of a triangle.}
    \label{Figure: closure condition}
\end{figure}

Let us now exhibit one such different classical limit. Our construction starts with the unique state $\ket{0,0}\in\mathcal{H}$ of total angular momentum zero. This state may be written in terms of the $\hat{J}_3^i$ eigenstates as
\newcommand\Wigner[6]{\begin{pmatrix}#1&#2&#3\\#4&#5&#6\end{pmatrix}}
\newcommand\wigner[6]{\begin{psmallmatrix}#1&#2&#3\\#4&#5&#6\end{psmallmatrix}}
\begin{equation}
    \ket{0,0} = \sum_{m_1=-j_1}^{j_1}\sum_{m_1=-j_2}^{j_2}\sum_{m_3=-j_3}^{j_3}\Wigner{j_1}{j_2}{j_3}{m_1}{m_2}{m_3} \ket{j_1,m_1}\otimes\ket{j_2,m_2}\otimes\ket{j_3,m_3},
\end{equation}
where $\wigner{j_1}{j_2}{j_3}{m_1}{m_2}{m_3}$ is the Wigner $3j$-symbol. As shown in~\cite{Livine:2007vk,Coherent3j}, the inner product between $\ket{0,0}$ and the separable coherent states $\ket{\mathbf{n}_1}\otimes\ket{\mathbf{n}_2}\otimes\ket{\mathbf{n}_3}$ is given by
\begin{multline}
    \bra{0,0}\qty(\ket{\mathbf{n}_1}\otimes\ket{\mathbf{n}_2}\otimes\ket{\mathbf{n}_3}) \\
    = 
    N_{j_1 j_2 j_3} e^{ig(\mathbf{n}_1,\mathbf{n}_2,\mathbf{n}_3)} \qty(\frac{1-\mathbf{n}_1\cdot\mathbf{n}_2}2)^{\frac{j_1+j_2-j_3}{2}} \qty(\frac{1-\mathbf{n}_2\cdot\mathbf{n}_3}2)^{\frac{j_2+j_3-j_1}{2}} \qty(\frac{1-\mathbf{n}_3\cdot\mathbf{n}_1}2)^{\frac{j_3+j_1-j_2}{2}},
    \label{Equation: coherent 3j}
\end{multline}
where $N_{j_1j_2j_3}$ is a normalisation constant parametrised in some combinatorial way by the three spins $j_1,j_2,j_3$, and $g(\mathbf{n}_1,\mathbf{n}_2,\mathbf{n}_3)\in\RR$. In the classical limit $j_i=K_i/\chi$, $\chi\to 0$, the norm of~\eqref{Equation: coherent 3j} is sharply peaked when the so-called `closure condition'
\begin{equation}
    j_1 \mathbf{n}_1 + j_2 \mathbf{n}_2 + j_3 \mathbf{n}_3 = 0
    \label{Equation: closure condition}
\end{equation}
is satisfied (here we are assuming that $j_1,j_2,j_3$ satisfy the triangle condition, so that~\eqref{Equation: closure condition} can be satisfied), depicted in Figure~\ref{Figure: closure condition}. This is a reflection of the fact that the total angular momentum of $\ket{0,0}$ is zero.

We wish to generate a set of coherent states from $\ket{0,0}$. To that end, let us parametrise elements of $\operatorname{SU}(2)$ as
\begin{equation}
    g(\psi,\mathbf{m}) = \mathds{1} \cos(\psi) + i(\mathbf{m}\cdot\sigma) \sin(\psi) = \exp(i\psi (\mathbf{m}\cdot\sigma)),
\end{equation}
where $\psi\in[0,\pi)$, $\mathbf{m}$ is a unit 3-vector, and $\sigma_a$, $a=1,2,3$ are the Pauli matrices. In terms of the variables $\psi,\mathbf{m}$, the invariant measure on $\mathrm{SU}(2)$ (normalised so that $\mathrm{SU}(2)$ has unit volume) is given by $\frac1{2\pi^2}\sin^2(\psi)\dd{\psi}\dd[2]{\mathbf{m}}$, where $\dd[2]{\mathbf{m}}$ is the standard measure on the unit 2-sphere. The action of $g(\psi_i,\mathbf{m}_i)$ on $\mathcal{H}_i$ is given by the unitary operator 
\begin{equation}
    U_i(\psi_i,\mathbf{m}_i) = \exp(2i\psi_i(\mathbf{m}_i\cdot \hat{\mathbf{J}}^i))
\end{equation}
Since this is an irreducible representation, Schur's lemma implies that
\begin{multline}
    \mathds{1}= \int_{\mathrm{SU}(2)}\frac{\dd{\psi_1}\dd[2]{\mathbf{m}_1} \sin^2(\psi_1)}{2\pi^2}\int_{\mathrm{SU}(2)}\frac{\dd{\psi_2}\dd[2]{\mathbf{m}_2}\sin^2(\psi_2)}{2\pi^2}\int_{\mathrm{SU}(2)}\frac{\dd{\psi_3}\dd[2]{\mathbf{m}_3}\sin^2(\psi_3)}{2\pi^2}\, N \\
    \qty\big(U_1(\psi_1,\mathbf{m}_1)\otimes U_2(\psi_2,\mathbf{m}_2)\otimes U_3(\psi_3,\mathbf{m}_3))
        \ket{0,0}\\\bra{0,0}
    \qty\big(U_1(\psi_1,\mathbf{m}_1)\otimes U_2(\psi_2,\mathbf{m}_2)\otimes U_3(\psi_3,\mathbf{m}_3))^\dagger
\end{multline}
holds. Since $\ket{0,0}$ has zero angular momentum, we have
\begin{equation}
    \qty\big(\mathds{1}_1\otimes\mathds{1}_2\otimes U_3(\psi_3,\mathbf{m}_3))\ket{0,0} = \qty\big(U_1(\psi_3,\mathbf{m}_3)^\dagger\otimes U_2(\psi_3,\mathbf{m}_3)^\dagger\otimes\mathds{1}_3)\ket{0,0},
\end{equation}
and we can use this, along with the fact that the measures are invariant under \begin{align}
    U_1(\psi_1,\mathbf{m}_1)&\to U_1(\psi_1,\mathbf{m}_1)U_1(\psi_3,\mathbf{m}_3),\\
    U_1(\psi_1,\mathbf{m}_1)&\to U_1(\psi_1,\mathbf{m}_1)U_1(\psi_3,\mathbf{m}_3),
\end{align}
to perform the third $\mathrm{SU}(2)$ integral, writing
\begin{multline}
    \mathds{1}= \int_{\mathrm{SU}(2)}\frac{\dd{\psi_1}\dd[2]{\mathbf{m}_1}\sin^2(\psi_1)}{2\pi^2} \int_{\mathrm{SU}(2)}\frac{\dd{\psi_2}\dd[2]{\mathbf{m}_2}\sin^2(\psi_2)}{2\pi^2}\, N\\
        \qty\big(U_1(\psi_1,\mathbf{m}_1)\otimes U_2(\psi_2,\mathbf{m}_2)\otimes \mathds{1}_3)
        \ket{0,0}\bra{0,0}
        \qty\big(U_1(\psi_1,\mathbf{m}_1)\otimes U_2(\psi_2,\mathbf{m}_2)\otimes \mathds{1}_3)^\dagger
\end{multline}
Let us therefore define 
\begin{equation}
    \ket{\psi_1,\psi_2;\mathbf{m}_1,\mathbf{m}_2} = \qty\big(U_1(\psi_1,\mathbf{m}_1)\otimes U_2(\psi_2,\mathbf{m}_2)\otimes \mathds{1}_3)\ket{0,0}.
\end{equation}
These will be our coherent states. By the above, they give a resolution of the identity
\begin{equation}
    \mathds{1} = \int_{S^3\times S^3} \frac{\dd{\psi_1}\dd[2]{\mathbf{m}_1}\sin^2(\psi_1)}{2\pi^2}\frac{\dd{\psi_2}\dd[2]{\mathbf{m}_2}\sin^2(\psi_2)}{2\pi^2}\, N \, \ket{\psi_1,\psi_2;\mathbf{m}_1,\mathbf{m}_2} \bra{\psi_1,\psi_2;\mathbf{m}_1,\mathbf{m}_2},
\end{equation}
where now we are writing $\mathrm{SU}(2)$ in terms of its group manifold $S^3$. The space $S^3\times S^3$ will be the space of classical states for the classical limit we are studying.

\subsection{Classical orthogonality of states}

For the classical limit described above to be meaningful, when $(\psi_1,\psi_2;\mathbf{m}_1,\mathbf{m}_2)\ne(\psi_1,\psi_2;\mathbf{m}_1,\mathbf{m}_2)$ we should have that $\ket{\psi_1,\psi_2;\mathbf{m}_1,\mathbf{m}_2}$ and $\ket{\psi_1',\psi_2';\mathbf{m}_1',\mathbf{m}_2'}$ are approximately orthogonal in the large angular momentum limit $\chi\to 0$. Let us now show that this is true. 

We can use~\eqref{Equation: three spin coherent resolution} to write
\begin{multline}
    \braket{\psi_1,\psi_2;\mathbf{m}_1,\mathbf{m}_2}{\psi'_1,\psi'_2;\mathbf{m}'_1,\mathbf{m}'_2} 
    = \int_{S^2\times S^2\times S^2}\frac{\dd[2]{\mathbf{n}_1}}{4\pi}\frac{\dd[2]{\mathbf{n}_2}}{4\pi}\frac{\dd[2]{\mathbf{n}_3}}{4\pi}\\
    \bra{0,0}\qty(U_1(\psi_1,\mathbf{m}_1)^\dagger U_1(\psi'_1,\mathbf{m}'_1)\ket{\mathbf{n}_1}\otimes U_2(\psi_2,\mathbf{m}_2)^\dagger U_2(\psi'_2,\mathbf{m}'_2)\ket{\mathbf{n}_2}\otimes\ket{\mathbf{n}_3})\\
    \qty(\bra{\mathbf{n}_1}\otimes\bra{\mathbf{n}_2}\otimes\bra{\mathbf{n}_3})\ket{0,0}.
\end{multline}
Each operator of the form $U_i(\psi_i,\mathbf{m}_i)$ acts on states of the form $\ket{\mathbf{n}_i}$ by rotating the vector $\mathbf{n}_i$ around $\mathbf{m}_i$ by an angle of $2\psi_i$. It also multiplies the state by a phase that we do not need to determine here. Thus, we can write
\begin{equation}
    e^{i\gamma_i}\ket{\mathbf{n}'_i} = U_i(\psi_i,\mathbf{m}_i)^\dagger U_i(\psi'_i,\mathbf{m}'_i)\ket{\mathbf{n}_i},
\end{equation}
where $\gamma_i\in\RR$, and $\mathbf{n}'_i$ is obtained by rotating $\mathbf{n}_i$ around $\mathbf{m}'_i$ by an angle $2\psi'_i$, and then around $\mathbf{m}_i$ by an angle $-2\psi_i$. We then have
\begin{multline}
    \abs{\braket{\psi_1,\psi_2;\mathbf{m}_1,\mathbf{m}_2}{\psi'_1,\psi'_2;\mathbf{m}'_1,\mathbf{m}'_2}} \\
    \le \int_{S^2\times S^2\times S^2} \frac{\dd[2]{\mathbf{n}_1}}{4\pi}\frac{\dd[2]{\mathbf{n}_2}}{4\pi}\frac{\dd[2]{\mathbf{n}_3}}{4\pi}
    \abs{\bra{0,0}(\ket{\mathbf{n}'_1}\otimes \ket{\mathbf{n}'_2}\otimes\ket{\mathbf{n}_3})}\abs{
\qty(\bra{\mathbf{n}_1}\otimes\bra{\mathbf{n}_2}\otimes\bra{\mathbf{n}_3})\ket{0,0}}.
\end{multline}
The integration here is dominated by contributions where the closure condition~\eqref{Equation: closure condition} holds for $\mathbf{n}_1,\mathbf{n}_2,\mathbf{n}_3$. Moreover, when the closure condition holds, the rotational invariance of $\bra{0,0}$ implies that 
\begin{equation}
    C=\abs{\bra{0,0}(\ket{\mathbf{n}_1}\otimes \ket{\mathbf{n}_2}\otimes\ket{\mathbf{n}_3})}
\end{equation}
is a constant. Thus, we may write
\begin{equation}
    \abs{\braket{\psi_1,\psi_2;\mathbf{m}_1,\mathbf{m}_2}{\psi'_1,\psi'_2;\mathbf{m}'_1,\mathbf{m}'_2}} \lessapprox \frac1 C\int_{\mathcal{T}} \dd{\tau(\mathbf{n}_1,\mathbf{n}_2,\mathbf{n}_3)} \abs{\bra{0,0}\ket{\mathbf{n}'_1}\otimes \ket{\mathbf{n}'_2}\otimes\ket{\mathbf{n}_3})},
\end{equation}
where $\mathcal{T}$ is the space of vectors $(\mathbf{n}_1,\mathbf{n}_2,\mathbf{n}_3)$, with a measure $\tau$ that is invariant under simultaneous rotations of those vectors, and $\lessapprox$ denotes an inequality that holds in the classical limit. The normalisation $\frac1 C$ has been chosen so that $\mathcal{T}$ has unit volume according to $\tau$.\footnote{This can be verified by setting $\psi_i=\psi'_i$ and $\mathbf{n}_i=\mathbf{n}'_i$, in which case $\braket{\psi_1,\psi_2;\mathbf{m}_1,\mathbf{m}_2}{\psi'_1,\psi'_2;\mathbf{m}'_1,\mathbf{m}'_2}=1$ and $\mathbf{n}_i=\mathbf{n}'_i$.} This integral is now dominated by contributions where $\mathbf{n}'_1,\mathbf{n}'_2,\mathbf{n}_3$ obey the closure condition; let $\widetilde{\mathcal{T}}\subset\mathcal{T}$ be the space where this is true. Within $\widetilde{\mathcal{T}}$, we have 
\begin{equation}
    C=\abs{\bra{0,0}\ket{\mathbf{n}'_1}\otimes \ket{\mathbf{n}'_2}\otimes\ket{\mathbf{n}_3})},
\end{equation}
so we may write
\begin{equation}
    \abs{\braket{\psi_1,\psi_2;\mathbf{m}_1,\mathbf{m}_2}{\psi'_1,\psi'_2;\mathbf{m}'_1,\mathbf{m}'_2}} \lessapprox \tau(\widetilde{\mathcal{T}}),
    \label{Equation: inner product measure bound spin}
\end{equation}
i.e.\ in the classical limit the absolute value of the inner product is bounded above by the measure of $\widetilde{\mathcal{T}}$ according to $\tau$. 

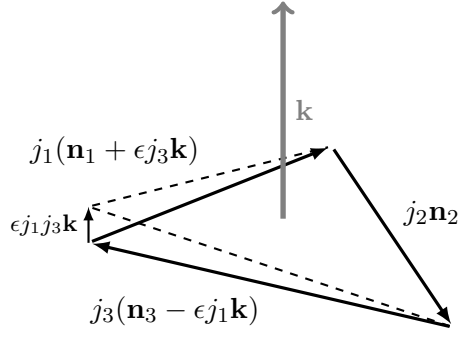
\begin{figure}
    \centering
    \begin{tikzpicture}[scale=1.6]
        \begin{scope}[dashed,thick,,shorten <=0.05cm,shorten >=0.05cm]
            \draw (0,0) -- (2,0.5) node[midway, above left] {$j_1(\mathbf{n}_1+\epsilon j_3\mathbf{k})$};
            \draw (3,-1) -- (0,0);
        \end{scope}
        \begin{scope}[very thick,-latex,shorten <=0.05cm,shorten >=0.05cm]
            \draw (0,-0.3) -- (2,0.5);
            \draw (2,0.5) -- (3,-1) node[midway, above right] {$j_2\mathbf{n}_2$};
            \draw (3,-1) -- (0,-0.3) node[midway, below left] {$j_3(\mathbf{n}_3-\epsilon j_1\mathbf{k})$};
        \end{scope}
        \draw[line width=2pt,->,black!50!white] (1.6,-0.1) -- (1.6, 1.7) node[midway,right] {$\mathbf{k}$};
        \draw[thick,-latex] (0,-0.3) -- (0,0) node[midway,left] {\footnotesize$\epsilon j_1j_3\mathbf{k}$};
    \end{tikzpicture}
    \caption{We consider a perturbation to the closure condition of the form~\eqref{Equation: triangle perturbation 2}. This amounts to a small rotation of the triangle around its second side.}
    \label{Figure: closure condition perturbation}
\end{figure}

We claim that $\widetilde{\mathcal{T}}$ is measure zero, unless the rotations given by $U_i(\psi_i,\mathbf{m}_i)^\dagger U_i(\psi'_i,\mathbf{m}'_i)$ are trivial. To see this, suppose that it is measure non-zero, and let $(\mathbf{n}_1,\mathbf{n}_2,\mathbf{n}_3)\in\widetilde{\mathcal{T}}$. Then we must have
\begin{align}
    j_1\mathbf{n}_1+j_2\mathbf{n}_2+j_3\mathbf{n}_3 &= 0, \label{Equation: T definition}\\
    j_1R_1\cdot\mathbf{n}_1+j_2R_2\cdot\mathbf{n}_2 + j_3\mathbf{n}_3 &= 0, \label{Equation: tilde T definition}
\end{align}
where $R_1,R_2$ are the orthogonal 3-matrices defining the transformations $\mathbf{n}_1\to\mathbf{n}'_1$, $\mathbf{n}_2\to\mathbf{n}'_2$ respectively. These conditions define $\mathcal{T}$ and $\widetilde{\mathcal{T}}$ as closed submanifolds of $S^2\times S^2\times S^2$, and the measure $\tau$ is just proportional to the induced volume form from $S^2\times S^2\times S^2$. Since we are supposing $\widetilde{\mathcal{T}}$ is measure non-zero, it must have zero codimension as a submanifold of $\mathcal{T}$, so it must be open when considered as a subset of $\mathcal{T}$. Thus, any small perturbation
\begin{equation}
    (\mathbf{n}_1,\mathbf{n}_2,\mathbf{n}_3) \to (\mathbf{n}_1+\delta\mathbf{n}_1,\mathbf{n}_2+\delta\mathbf{n}_2,\mathbf{n}_3+\delta\mathbf{n}_3)
    \label{Equation: triangle perturbation}
\end{equation}
which preserves~\eqref{Equation: T definition} must also preserve~\eqref{Equation: tilde T definition}. Consider in particular the perturbation with 
\begin{equation}
    \delta\mathbf{n}_1 = \epsilon\,j_3\mathbf{k}, \quad \delta\mathbf{n}_2 = 0, \quad \delta\mathbf{n}_3 = -\epsilon\,j_1\mathbf{k},
    \label{Equation: triangle perturbation 2}
\end{equation}
where $\mathbf{k}$ is normal to the plane spanned by $\mathbf{n}_1,\mathbf{n}_2,\mathbf{n}_3$, and $\epsilon\ll 1$. This is illustrated in Figure~\ref{Figure: closure condition perturbation}. Such a perturbation clearly satisfies~\eqref{Equation: T definition} and preserves the normalisation of the vectors $\mathbf{n}_1,\mathbf{n}_2,\mathbf{n}_3$, so it stays within $\mathcal{T}$. For it to also stay within $\widetilde{\mathcal{T}}$, it must also satisfy~\eqref{Equation: tilde T definition}, which implies
\begin{equation}
    R_1\cdot \mathbf{k} = \mathbf{k}.
\end{equation}
Thus, $R_1$ fixes $\mathbf{k}$. Actually, by going to second order in the perturbation~\eqref{Equation: triangle perturbation}, we can immediately see that $R_1$ must fix every unit vector within a small neighbourhood of $\mathbf{k}$. This is because doing~\eqref{Equation: triangle perturbation} changes the normal unit vector $\mathbf{k}\to\mathbf{k}+\delta\mathbf{k}$, and by picking $\delta\mathbf{n}_1,\delta\mathbf{n}_2,\delta\mathbf{n}_3$ appropriately we can put $\mathbf{k}+\delta\mathbf{k}$ anywhere within a neighbourhood of $\mathbf{k}$. Thus, $R_1$ fixes $\mathbf{k}+\delta\mathbf{k}$ for arbitrary small $\delta\mathbf{k}$. But since $R_1$ is a linear map, it must be equal to the identity. A similar argument applies to $R_2$. 

Thus, $\widetilde{\mathcal{T}}$ has zero measure, unless the rotations corresponding to $U_i(\psi_i,\mathbf{m}_i)^\dagger U_i(\psi'_i,\mathbf{m}'_i)$ are trivial. The only way for this to happen is $\psi_i=\psi'_i$ and $\mathbf{m}_i=\mathbf{m}'_i$. Therefore,~\eqref{Equation: inner product measure bound spin} implies
\begin{equation}
    \braket{\psi_1,\psi_2;\mathbf{m}_1,\mathbf{m}_2}{\psi'_1,\psi'_2;\mathbf{m}'_1,\mathbf{m}'_2} \approx
    \begin{cases}
        1 & \text{if }\psi_i=\psi'_i \text{ and }\mathbf{m}_i=\mathbf{m}'_i,\\
        0 & \text{otherwise},
    \end{cases}
    \label{Equation: inner product spin}
\end{equation}
which is the classical orthogonality condition we seek.

\subsection{States of individual spins}

If the state of the full three spin system is $\ket{\psi_1,\psi_2;\mathbf{m}_1,\mathbf{m}_2}$, then the reduced states of each individual spin are maximally mixed. To see this, note that they are given by
\begin{align}
    \rho_1(\psi_1,\psi_2;\mathbf{m}_1,\mathbf{m}_2) &= \tr_{23} \qty\big(\ket{\psi_1,\psi_2;\mathbf{m}_1,\mathbf{m}_2}\bra{\psi_1,\psi_2;\mathbf{m}_1,\mathbf{m}_2}) \nonumber\\
                                                    &= U_1(\psi_1,\mathbf{m}_1)\tr_{23}\qty(\ket{0,0}\bra{0,0})U_1(\psi_1,\mathbf{m}_1)^\dagger\\
    \rho_2(\psi_1,\psi_2;\mathbf{m}_1,\mathbf{m}_2) &= \tr_{31} \qty\big(\ket{\psi_1,\psi_2;\mathbf{m}_1,\mathbf{m}_2}\bra{\psi_1,\psi_2;\mathbf{m}_1,\mathbf{m}_2}) \nonumber\\
                                                    &= U_2(\psi_2,\mathbf{m}_2)\tr_{31}\qty(\ket{0,0}\bra{0,0})U_2(\psi_2,\mathbf{m}_2)^\dagger\\
    \rho_3(\psi_1,\psi_2;\mathbf{m}_1,\mathbf{m}_2) &= \tr_{12} \qty\big(\ket{\psi_1,\psi_2;\mathbf{m}_1,\mathbf{m}_2}\bra{\psi_1,\psi_2;\mathbf{m}_1,\mathbf{m}_2})\nonumber\\
                                                    &= \tr_{12}\qty(\ket{0,0}\bra{0,0})
\end{align}
Then a well-known property of the Wigner $3j$-symbol implies (for example)
\begin{nalign}
    \bra{j_1,m_1}\tr_{23}(\ket{0,0}\bra{0,0})\ket{j_1,m'_1} &= \sum_{m_2=-j_2}^{j_2}\sum_{m_3=-j_3}^{j_3}\Wigner{j_1}{j_2}{j_3}{m_1}{m_2}{m_3}\Wigner{j_1}{j_2}{j_3}{m'_1}{m_2}{m_3} \\
                                                            &= \frac1{2j_1+1}\delta_{m_1m'_1}.
\end{nalign}
Similar results hold for the second and third spins. Therefore
\begin{equation}
    \rho_i(\psi_1,\psi_2;\mathbf{m}_1,\mathbf{m}_2) = \frac{\mathds{1}_i}{2j_i+1}.
    \label{Equation: individual spin maximally mixed}
\end{equation}

This trivially implies that each spin $j_i$ is a classically resolvable subsystem $s_i$. However, there is only one possible classical state for each spin. In other words, there are no degrees of freedom associated with any \emph{individual} spin, in the classical limit we are considering.

\subsection{States of pairs of spins}

Let us now consider the subsystems consisting of two out of the three spins. These have reduced states
\begin{align}
    \rho_{23}(\psi_1,\psi_2;\mathbf{m}_1,\mathbf{m}_2) &= \tr_{1} \qty\big(\ket{\psi_1,\psi_2;\mathbf{m}_1,\mathbf{m}_2}\bra{\psi_1,\psi_2;\mathbf{m}_1,\mathbf{m}_2}) \nonumber\\
                                                       &= \big(U_2(\psi_2,\mathbf{m}_2)\otimes\mathds{1}_3\big)\tr_{1}\qty(\ket{0,0}\bra{0,0})\big(U_2(\psi_2,\mathbf{m}_2)^\dagger\otimes\mathds{1}_3\big)\label{Equation: spin rho 23}\\
    \rho_{31}(\psi_1,\psi_2;\mathbf{m}_1,\mathbf{m}_2) &= \tr_{2} \qty\big(\ket{\psi_1,\psi_2;\mathbf{m}_1,\mathbf{m}_2}\bra{\psi_1,\psi_2;\mathbf{m}_1,\mathbf{m}_2}) \nonumber\\
                                                       &= \big(U_1(\psi_1,\mathbf{m}_1)\otimes \mathds{1}_3\big)\tr_{2}\qty(\ket{0,0}\bra{0,0})\big(U_1(\psi_1,\mathbf{m}_1)^\dagger\otimes \mathds{1}_3\big)\\
    \rho_{12}(\psi_1,\psi_2;\mathbf{m}_1,\mathbf{m}_2) &= \tr_{3} \qty\big(\ket{\psi_1,\psi_2;\mathbf{m}_1,\mathbf{m}_2}\bra{\psi_1,\psi_2;\mathbf{m}_1,\mathbf{m}_2})\nonumber\\
                                                       &= \big(U_1(\psi_1,\mathbf{m}_1)\otimes U_2(\psi_2,\mathbf{m}_2)\big)\tr_{3}\qty(\ket{0,0}\bra{0,0})\big(U_1(\psi_1,\mathbf{m}_1)^\dagger\otimes U_2(\psi_2,\mathbf{m}_2)^\dagger\big)\nonumber\\
                                                       &= \big(\mathds{1}_1\otimes U_2(\psi_3,\mathbf{m}_3)\big)\tr_3(\ket{0,0}\bra{0,0}) \big(\mathds{1}_1\otimes U_2(\psi_3,\mathbf{m}_3)^\dagger\big),
\end{align}
where to get the last line we are using the fact that $\ket{0,0}$ has zero angular momentum, and defining $\psi_3,\mathbf{m}_3$ via the group composition
\begin{equation}
    U_2(\psi_3,\mathbf{m}_3) = U_2(\psi_2,\mathbf{m}_2)U_2(\psi_1,\mathbf{m}_1)^\dagger.
    \label{Equation: spin pair constraint}
\end{equation}
Explicitly, $\psi_3$ and $\mathbf{m}_3$ are given by
\begin{nalign}
    \cos(\psi_3) &= \cos(\psi_1)\cos(\psi_2) + \mathbf{m}_1\cdot\mathbf{m}_2 \sin(\psi_1)\sin(\psi_2), \\
    \mathbf{m}_3\sin(\psi_3) &= -\mathbf{m}_1\sin(\psi_1)\cos(\psi_2) +\mathbf{m}_2\sin(\psi_2)\cos(\psi_2) - \mathbf{m}_1\times\mathbf{m}_2 \sin(\psi_1)\sin(\psi_2).
    \label{Equation: spin pair constraint explicit}
\end{nalign}

By~\eqref{Equation: individual spin maximally mixed}, each of these density matrices are exactly proportional to projection operators, of ranks $2j_1+1$, $2j_2+1$ and $2j_3+1$ respectively. Thus, we are already halfway to showing that the subsystems corresponding to pairs of spins are classically resolvable.

Let us consider the composition of density matrices for different classical states. It is useful to note that Schur's lemma implies
\begin{equation}
    \mathds{1}_1\otimes \rho_{23}(\psi_1,\psi_2;\mathbf{m}_1,\mathbf{m}_2) = \int_{S^3}\frac{\dd{\tilde\psi_1}\dd[2]{\tilde{\mathbf{m}}_1}}{2\pi^2}(2j_1+1)\sin^2(\tilde\psi_1)\ket*{\tilde\psi_1,\psi_2;\tilde{\mathbf{m}}_1,\mathbf{m}_2}\bra*{\tilde\psi_1,\psi_2;\tilde{\mathbf{m}}_1,\mathbf{m}_2}.
\end{equation}
If $(\psi_2,\mathbf{m}_2)\ne (\psi'_2,\mathbf{m}'_2)$, then~\eqref{Equation: inner product spin} implies
\begin{multline}
    \Big(\mathds{1}_1\otimes\rho_{23}(\psi_1,\psi_2;\mathbf{m}_1,\mathbf{m}_2)\Big)
    \Big(\mathds{1}_1\otimes\rho_{23}(\psi'_1,\psi'_2;\mathbf{m}'_1,\mathbf{m}'_2)\Big) 
    \\
    = \int_{S^3}\frac{\dd{\tilde\psi_1}\dd[2]{\tilde{\mathbf{m}_1}}}{2\pi^2}(2j_1+1) \sin^2(\tilde\psi_1)\int_{S^3}\frac{\dd{\tilde\psi'_1}\dd[2]{\tilde{\mathbf{m}'_1}}}{2\pi^2}(2j_1+1) \sin^2(\tilde\psi_2)\\
    \ket*{\tilde\psi_1,\psi_2;\tilde{\mathbf{m}}_1,\mathbf{m}_2}\underbrace{\braket*{\tilde\psi_1,\psi_2;\tilde{\mathbf{m}}_1,\mathbf{m}_2}{\tilde\psi'_1,\psi'_2;\tilde{\mathbf{m}}'_1,\mathbf{m}'_2}}_{\approx 0}\bra*{\tilde\psi'_1,\psi'_2;\tilde{\mathbf{m}}'_1,\mathbf{m}'_2} \approx 0.
\end{multline}
On the other hand, if $(\psi_2,\mathbf{m}_2) = (\psi'_2,\mathbf{m}'_2)$, then~\eqref{Equation: spin rho 23} and~\eqref{Equation: individual spin maximally mixed} imply that 
\begin{equation}
    \rho_{23}(\psi_1,\psi_2;\mathbf{m}_1,\mathbf{m}_2)\rho_{23}(\psi'_1,\psi'_2;\mathbf{m}'_1,\mathbf{m}'_2) = \frac1{2j_1+1} \rho_{23}(\psi_1,\psi_2;\mathbf{m}_1,\mathbf{m}_2).
\end{equation}
Similar arguments apply for the other pairs of spins. Writing
\begin{align}
    \rho_{23}(\psi_1,\psi_2;\mathbf{m}_1,\mathbf{m}_2) &= \frac{\hat{\pi}_{23}(\psi_2,\mathbf{m}_2)}{2j_1+1}, \\
    \rho_{31}(\psi_1,\psi_2;\mathbf{m}_1,\mathbf{m}_2) &= \frac{\hat{\pi}_{31}(\psi_1,\mathbf{m}_1)}{2j_2+1}, \\
    \rho_{12}(\psi_1,\psi_2;\mathbf{m}_1,\mathbf{m}_2) &= \frac{\hat{\pi}_{12}(\psi_3,\mathbf{m}_3)}{2j_3+1},
\end{align}
where
\begin{align}
    \hat\pi_{23}(\psi,\mathbf{m}) &= (2j_1+1)U_2(\psi,\mathbf{m})\otimes\mathds{1}_3\tr_{1}\qty(\ket{0,0}\bra{0,0})U_2(\psi,\mathbf{m})^\dagger\otimes\mathds{1}_3 \\
    \hat\pi_{31}(\psi,\mathbf{m}) &= (2j_2+1)U_1(\psi,\mathbf{m})\otimes\mathds{1}_3\tr_{2}\qty(\ket{0,0}\bra{0,0})U_1(\psi,\mathbf{m})^\dagger\otimes\mathds{1}_3 \\
    \hat\pi_{12}(\psi,\mathbf{m}) &= (2j_3+1)\mathds{1}_1\otimes U_2(\psi,\mathbf{m})\tr_{3}\qty(\ket{0,0}\bra{0,0})\mathds{1}_1\otimes U_2(\psi,\mathbf{m})^\dagger
\end{align}
are the projection operators to which these density matrices are proportional, we may conclude that
\begin{equation}
    \hat\pi_{ij}(\psi,\mathbf{m})\hat\pi_{ij}(\psi',\mathbf{m}') \approx
    \begin{cases}
        \hat\pi_{ij}(\psi,\mathbf{m}) & \text{if } \psi=\psi' \text{ and } \mathbf{m}=\mathbf{m}', \\
        0 & \text{otherwise},
    \end{cases}
\end{equation}
where $ij\in  \{12,23,31\}$. Therefore, each pair of spins is a classically resolvable subsystem. Moreover, we may identify the space of classical states for each pair of spins: it is a 3-sphere, parametrised by $\psi,\mathbf{m}$.

\subsection{Non-local degrees of freedom and constraints}

Let us summarise the structure of the classical theory we have produced. We considered the classical limit of three quantum spins with angular momentum $j_i = K_i/\chi$, prepared in the highly entangled state $\ket{0,0}$ with zero total angular momentum. From $\ket{0,0}$ we obtained the set of coherent states $\ket{\psi_1,\psi_2;\mathbf{m}_1,\mathbf{m}_2}$, with $\psi_i,\mathbf{m}_i$ being the parameters of $\mathrm{SU}(2)$ elements, and thus labelling points on 3-spheres. These coherent states are approximately mutually orthogonal, and so they yield a classical limit with the classical space of states 
\begin{equation}
    \mathcal{N} = S^3\times S^3.
    \label{EEquation: total spin space}
\end{equation}
If the state of the full system is $\ket{\psi_1,\psi_2;\mathbf{m}_1,\mathbf{m}_2}$, then the reduced state of any individual spin is always maximally mixed. Thus, each individual spin is classically resolvable, and its classical space of states is a singleton
\begin{equation}
    \mathcal{N}_i = \{e\},
    \label{Equation: individual spin space}
\end{equation}
where $e$ represents the maximally mixed state. The reduced state of any pair of spins is proportional to a projection operator $\hat\pi_{ij}(\psi,\mathbf{m})$, with $\psi,\mathbf{m}$ the parameters of an $\mathrm{SU}(2)$ element, and so labelling a point on a 3-sphere. Moreover, for different points on the 3-sphere, these projection operators are approximately orthogonal. Thus, each pair of spins is a classically resolvable subsystem, with the classical space of states
\begin{equation}
    \mathcal{N}_{ij} = S^3.
    \label{Equation: pair of spins space}
\end{equation}

There are clearly a wealth of non-local degrees of freedom in this classical system. Actually, by~\eqref{Equation: individual spin space}, no individual spin has any degrees of freedom -- so there are in fact \emph{no} local degrees of freedom.

Each point in the state space~\eqref{Equation: pair of spins space} of a pair of spins labels the state of the non-local degrees of freedom shared between the two spins. Thus, the states of bilocal degrees of freedom are parametrised by points in 3-spheres. 

There are no additional degrees of freedom beyond these bilocal ones. In particular, there are no degrees of freedom that are fundamentally trilocal, i.e.\ shared between all three spins. To see this, it suffices to note that knowledge of the classical states of any two pairs of spins determines the classical state of all three spins. Indeed, suppose the classical states of the pairs of spins $23$ and $31$ are $(\psi_{23},\mathbf{m}_{23})$ and $(\psi_{31},\mathbf{m}_{31})$ respectively, so that their density matrices are given by
\begin{equation}
    \rho_{23} = \frac{\hat\pi_{23}(\psi_{23},\mathbf{m}_{23})}{2j_1+1}, \qquad
    \rho_{31} = \frac{\hat\pi_{31}(\psi_{31},\mathbf{m}_{31})}{2j_2+1}.
\end{equation}
There is only one coherent state for the total system that is consistent with these reduced states. This is $\ket{\psi_{31},\psi_{23};\mathbf{m}_{31},\mathbf{m}_{23}}$ --- so the classical state of the full system must be $(\psi_{31},\psi_{23};\mathbf{m}_{31},\mathbf{m}_{23})$. Similarly, the states of any other two pairs of spins suffices to determine the full system state. 

Since we only need the states of two out of the three available pairs of spins $12,23,31$ to determine the full system state, there must be some constraint relating each of these states. Indeed, for consistency each of the pairs of pairs of spins $(12,23)$, $(31,12)$ and $(23,31)$ must imply the same full system state. We have already seen what this constraint is; it is given by~\eqref{Equation: spin pair constraint}, or~\eqref{Equation: spin pair constraint explicit}.

This redundancy is reminiscent of the 3-qutrit code in quantum error correction (QEC). In that case, knowledge of two out of three qutrits is sufficient to recover the single logical qutrit, just like here knowledge of the states of two out of three pairs of spins is sufficient to determine the state of the whole system. We will have a little bit more to say on this in Section~\ref{Section: Gravitational properties}.

A possibly instructive way to understand the constraint is as follows. The state $(\psi,\mathbf{m})$ of each pair of spins yields an $\mathrm{SU}(2)$ element $U(\psi,\mathbf{m})$. We can think of this $\mathrm{SU}(2)$ element as a `Wilson line' connecting the two spins. The constraint~\eqref{Equation: spin pair constraint} then says that the composition of all three Wilson lines is equal to the identity (up to some specific orientation of the lines). Thus, intuitively speaking, the `curvature' measured by the Wilson loop around the entire system is vanishing.

\subsection{Gauge symmetry}

Let us now introduce some edge modes and gauge symmetry into the classical picture, in order to enable a description of the physics in terms of local kinematical degrees of freedom. The form of the gauge symmetry we are about to introduce is not exactly the same as that which was described in Section~\ref{Section: Gauge symmetry from entanglement}. If we wanted, we could instead use that more general method, but what we describe now is just a different, more geometrically motivated option that works well for the case at hand.

We will take the kinematical space of states for each spin to be a 3-sphere, i.e.\ a copy of $\mathrm{SU}(2)$:
\begin{equation}
    \mathcal{N}_i^{\text{kin.}} = \mathrm{SU}(2).
\end{equation}
The global gauge group will be $G=\mathrm{SU}(2)$, and we define a gauge transformation to be the simultaneous right action of $G$ on each spin's kinematical state. 

If we take the quotient of $\mathcal{N}_i^{\text{kin.}}$ by the gauge group, we obtain $\mathrm{SU}(2)/\mathrm{SU}(2)$, which is clearly a singleton. Thus, 
\begin{equation}
    \mathcal{N}_i^{\text{kin.}}/G = \mathcal{N}_i.
\end{equation}
In other words, the local kinematical degrees of freedom, modulo gauge symmetry, are equal to the local physical degrees of freedom.

The kinematical space of states for a pair of spins is
\begin{equation}
    \mathcal{N}_{ij}^{\text{kin.}} = \mathcal{N}_i^{\text{kin.}}\times\mathcal{N}_j^{\text{kin.}} = \mathrm{SU}(2)\times\mathrm{SU}(2),
\end{equation}
where $ij\in\{12,23,31\}$. Clearly, $\mathrm{SU}(2)\times\mathrm{SU}(2)/\mathrm{SU}(2) = \mathrm{SU}(2)$, so quotienting by the gauge group yields the correct physical space of states for each pair of spins. Explicitly, we can implement this via the maps
\begin{nalign}
    \mathcal{N}_{23}^{\text{kin.}}&\to \mathcal{N}_{23}, \quad (V_2,V_3) \mapsto V_2V_3^\dagger = U(\psi_{23},\mathbf{m}_{23}), \\
    \mathcal{N}_{31}^{\text{kin.}}&\to \mathcal{N}_{31}, \quad (V_3,V_1) \mapsto V_1V_3^\dagger = U(\psi_{31},\mathbf{m}_{31}), \\
    \mathcal{N}_{12}^{\text{kin.}}&\to \mathcal{N}_{12}, \quad (V_1,V_2) \mapsto V_2V_1^\dagger = U(\psi_{12},\mathbf{m}_{12}).
    \label{Equation: spin pair reduction}
\end{nalign}

The kinematical space of states for the full system is
\begin{equation}
    \mathcal{N}^{\text{kin.}} = \mathcal{N}_1^{\text{kin.}}\times\mathcal{N}_2^{\text{kin.}}\times\mathcal{N}_3^{\text{kin.}} = \mathrm{SU}(2)\times\mathrm{SU}(2)\times\mathrm{SU}(2).
\end{equation}
Note that, for a given $(V_1,V_2,V_3)\in\mathcal{N}^{\text{kin.}}$, the states of each pair of spins as given by~\eqref{Equation: spin pair reduction} automatically satisfy the constraint~\eqref{Equation: spin pair constraint}. We clearly have $\mathrm{SU}(2)\times \mathrm{SU}(2)\times\mathrm{SU}(2)/\mathrm{SU}(2) = \mathrm{SU}(2)\times\mathrm{SU}(2)$, so quotienting by the gauge group yields the correct physical space of states, and this is explicitly implemented by the map
\begin{equation}
    \mathcal{N}^{\text{kin.}} \to \mathcal{N}, \quad (V_1,V_2,V_3) \mapsto (\psi_1,\psi_2;\mathbf{m}_1,\mathbf{m}_2)
\end{equation}
where
\begin{equation}
    V_1V_3^\dagger = U(\psi_1,\mathbf{m}_1), \quad V_2V_3^\dagger = U(\psi_2,\mathbf{m}_2).
\end{equation}
Thus, we have succeeded in accounting for the non-local degrees of freedom by introducing an $\mathrm{SU}(2)$ gauge symmetry.

Let us summarise what has been shown in this section. We started with the Hilbert space of three spins, with no gauge symmetry. We took a particular classical limit involving entangled states, and obtained a classical system with emergent fundamental non-local degrees of freedom, shared between the spins. Finally, we accounted for this with the gauge symmetry described above. Thus, we have provided an explicit example of classical gauge symmetry emerging from quantum entanglement.

\section{Entangled group coherent states}
\label{Section: group}

In this section, we will focus on a particular type of classical limit with convenient group-theoretic properties. In many ways, what follows is a vast generalisation of the toy model described in the previous section.

\subsection{Classical limits and coadjoint orbits}

Suppose $G$ is a compact Lie group with a unitary representation $U$ on a finite-dimensional Hilbert space $\mathcal{H}$. Let $\chi$ be a parameter which can be taken to be arbitrarily small. We will assume $G$ is independent of $\chi$, but $\mathcal{H}$ and the representation of $G$ can depend on $\chi$. 

If the representation $U$ is irreducible, then~\cite{LargeNLimits} described a large class of classical limits for this system, where in each case the classical space of states $\mathcal{N}$ is a `coadjoint orbit' of $G$.\footnote{For any coadjoint orbit $\mathcal{N}$ of a Lie group $G$, the program of geometric quantisation and the `orbit method'~\cite{orbits} does the reverse of this: it provides one with a unitary irreducible representation of $G$ whose classical limit has state space $\mathcal{N}$.} Let us now very briefly describe what happens.

Let $\mathfrak{g}$ be the Lie algebra of $G$, and let $\mathfrak{g}^*$ be its dual. Let us pick a normalised `base state' $\ket{0}\in\mathcal{H}$, and define $X^0\in\mathfrak{g}^*$ by
\begin{equation}
    X^0(\lambda) = \mel{0}{u(\lambda)}{0},
\end{equation}
where $u$ is the representation of $\mathfrak{g}$ on $\mathcal{H}$ induced by $U$. The `coadjoint orbit' $\mathcal{N}$ of $X^0$ is defined as the orbit of $X^0$ under the coadjoint action of $G$ on $\mathfrak{g}^*$:
\begin{equation}
    \mathcal{N} = \qty{X(g) \mid g\in G}, \qq{where} X(g) = \operatorname{Ad}^*[g](X^0),
\end{equation}
where
\begin{equation}
    \operatorname{Ad}^*[g](X^0) = X^0\circ \operatorname{Ad}[g^{-1}],
\end{equation}
and $\operatorname{Ad}$ is the adjoint action of $G$ on $\mathfrak{g}$, so
\begin{equation}
    X(g)(\lambda) = X^0(g^{-1}\lambda g) = \mel{0}{u(g^{-1}\lambda g)}{0} = \mel{0}{U(g)^\dagger\,u(\lambda)\,U(g)}{0} = \mel{g}{u(\lambda)}{g},
\end{equation}
where for each $g\in G$, we are defining the state
\begin{equation}
    \ket{g}= U(g)\ket{0}.
\end{equation}
If these states obey
\begin{equation}
    \lim_{\chi\to 0}\abs{\braket{g}{g'}}^2 =
        \begin{cases}
            1 & \text{if }X(g)= X(g'), \\
            0 & \text{otherwise},
        \end{cases}
        \label{Equation: coadjoint classical assumption}
\end{equation}
then $\chi\to 0$ is a complete classical limit with $\mathcal{N}$ being the classical space of states. To see this, note that by Schur's lemma, the states $\ket{g}$ form an overcomplete basis for $\mathcal{H}$, with a resolution of the identity given by
\begin{equation}
    \mathds{1} = \int_G \dd{\sigma(g)} N\, \ket{g}\bra{g},
    \label{Equation: Schur group identity}
\end{equation}
where $\sigma$ is the invariant measure on $G$, normalised so that $G$ has unit volume, and $N=\dim(\mathcal{H})$. This holds at arbitrary $\chi$. However, in the classical limit,~\eqref{Equation: coadjoint classical assumption} implies that if $X(g)=X(g')$, then $\ket{g}\approx\ket{g'}$ (up to a phase). Thus, if we pick any function $v:\mathcal{N}\to G$ obeying $v\big(X(g)) = g$, and define $\ket{x}=\ket{v(x)}$ for each $x\in \mathcal{N}$, we can approximate~\eqref{Equation: Schur group identity} via
\begin{equation}
    \mathds{1} \approx \int_{\mathcal{N}} \dd{\mu(x)} N\, \ket{x}\bra{x},
    \label{Equation: coadjoint classical resolution}
\end{equation}
where $\mu$ is proportional to the pushforward of $\sigma_L$ through the map $g\mapsto X(g)$, scaled so that $\mathcal{N}$ has unit volume.\footnote{$\mu$ is proportional to the Liouville measure of the Kostant-Souriau form on $\mathcal{N}$.} Furthermore, by~\eqref{Equation: coadjoint classical assumption} we have $\braket{x}{y}\approx 0$ if $x\ne y$. Thus, $\chi\to 0$ is a complete classical limit with classical states $x\in\mathcal{N}$ corresponding to quantum states $\ket{x}$.

\subsection{Composite systems: an unentangled case}

Suppose now that the system we are considering has some local structure $\mathscr{S}$, with a subdivision $\{s_i\in\mathscr{S}\mid i=1,\dots,n\}$, with respect to which Hilbert space $\mathcal{H}$ and the group $G$ decompose as
\begin{align}
    \mathcal{H} &= \mathcal{H}_1\otimes\dots\otimes\mathcal{H}_i\otimes\dots\otimes\mathcal{H}_n,\\
    G &= G_1\times \dots \times G_i \times \dots\times G_n.
\end{align}
Furthermore, let us take the unitary representation $U$ to be of the form
\begin{equation}
    U(g_1,\dots, g_i,\dots, g_n) = U_1(g_1)\otimes \dots\otimes U_i(g_i)\otimes\dots\otimes U_n(g_n),
\end{equation}
where for each $i$, $U_i$ is a unitary irreducible representation of $G_i$ acting on $\mathcal{H}_i$.

From here on, we will not assume that $U$ itself is irreducible, so the arguments of~\cite{LargeNLimits} will not necessarily apply to the total system. On the other hand, we will assume that the subsystems are classically resolvable, and that $U_i$ are irreducible, which means we will (to a certain extent) be able to apply the machinery of~\cite{LargeNLimits} to the subsystems.

Let us first consider a complete classical limit in which the classical states $\ket{x}=\ket{x_1}\otimes\dots\otimes\ket{x_n}$ are separable. Each of the subsystems then also undergoes a complete classical limit, since we are assuming they are classically resolvable. We will now show that, even though $U$ is not assumed to be irreducible, the space of states $\mathcal{N}$ for the total system can still be a coadjoint orbit of $G$.

Since, for each $i$, $U_i$ is irreducible, we can take the space of classical states for subsystem $s_i$ to be a coadjoint orbit $\mathcal{N}_i$ of $G_i$. Let $\ket{0_i}$, $i=1,\dots,n$ be the base states which generate these coadjoint orbits, and let
\begin{equation}
    \ket{0}=\ket{0_1}\otimes\dots\otimes\ket{0_n}.
    \label{Equation: separable base state}
\end{equation}
Let $\mathcal{N}$ be the coadjoint orbit of $G$ generated by this base state. Thus, elements of $\mathcal{N}$ may be written as
\begin{equation}
    X(g) = \operatorname{Ad}^*[g](X^0), \qq{where} X^0(\lambda) = \mel{0}{\lambda}{0},
\end{equation}
where $g\in G$ and $\lambda\in\mathfrak{g}$. In terms of the states
\begin{equation}
    \ket{(g_1,\dots,g_n)} = \ket{g_1}\otimes\dots\otimes\ket{g_n},\qq{where} \ket{g_i}=U_i(g_i)\ket{0_i},
\end{equation}
we have
\begin{equation}
    X(g)(\lambda) = \mel{(g_1,\dots,g_n)}{(\lambda_1\oplus\dots\oplus\lambda_n)}{(g_1,\dots,g_n)} = X_1(g_1)(\lambda_1) + \dots + X_n(g_n)(\lambda_n),
\end{equation}
where we are using the decomposition $\mathfrak{g}=\mathfrak{g}_1\oplus\dots\oplus\mathfrak{g}_n$ to write $\lambda=\lambda_1\oplus\dots\oplus\lambda_n$, where $\mathfrak{g}_i$ is the Lie algebra of $G_i$, and
\begin{equation}
    X_i(g_i)(\lambda_i) = \mel{g_i}{\lambda_i}{g_i}.
\end{equation}
More concisely, we have
\begin{equation}
    X(g) = \bigoplus_{i=1}^n X_i(g_i).
    \label{Equation: subsystem coadjoint orbit decomposition}
\end{equation}
Since $\mathcal{N}_i$ consists of objects of the form $X_i(g_i)$,~\eqref{Equation: subsystem coadjoint orbit decomposition} furnishes a bijection between $\mathcal{N}$ and $\mathcal{N}_1\times\dots\times \mathcal{N}_n$.

Defining functions $v_i:\mathcal{N}_i\to G_i$ obeying $v_i(X_i(g_i))=g_i$ for each $i$, we get resolutions of the identity for each subsystem as in~\eqref{Equation: coadjoint classical resolution}:
\begin{equation}
    \mathds{1}_i \approx \int_{\mathcal{N}_i} \dd{\mu_i(x_i)} N_i\, \ket{x_i}\bra{x_i},
\end{equation}
in terms of states
\begin{equation}
    \ket{x_i}=\ket{v_i(x_i)} ,\quad x_i\in\mathcal{N}_i,
\end{equation}
which obey
\begin{equation}
    \lim_{\chi\to 0} \braket{x_i}{x_i'} = 
    \begin{cases}
        1 & \text{if }x_i=x_i', \\
        0 & \text{otherwise}.
    \end{cases}
    \label{Equation: subsystem coadjoint orthogonality}
\end{equation}
Thus, defining 
\begin{equation}
    \ket{x}=\ket{x_1}\otimes\dots\otimes\ket{x_n}=\ket{(v_1(x_1),\dots,v_n(x_n))},
\end{equation}
where $x$ is related to $(x_1,\dots,x_n)$ via the bijection $\mathcal{N}\to\mathcal{N}_1\times\dots\times\mathcal{N}_n$, we have
\begin{nalign}
    \mathds{1} 
    &= \mathds{1}_1\otimes\dots\otimes\mathds{1}_n \\
    &\approx \int_{\mathcal{N}_1} \dd{\mu_1(x_1)} N_1\, \ket{x_1}\bra{x_1} \,\otimes\dots\otimes\,\int_{\mathcal{N}_n} \dd{\mu_n(x_n)} N_n\, \ket{x_n}\bra{x_n} \\
    &= \int_{\mathcal{N}}\dd{\mu}(x)\,N\, \ket{x}\bra{x},
\end{nalign}
where $\mu$ is the pushforward of the product measure $\mu_1\times\dots\times\mu_n$ through the bijection $\mathcal{N}_1\times\dots\times\mathcal{N}_n\to\mathcal{N}$. Also,~\eqref{Equation: subsystem coadjoint orthogonality} implies
\begin{equation}
    \lim_{\chi\to 0} \braket{x}{x'} = 
    \begin{cases}
        1 & \text{if }x=x', \\
        0 & \text{otherwise}.
    \end{cases}
\end{equation}
Thus, the classical set of states for the system is the coadjoint orbit $\mathcal{N}$, as claimed, with each classical state $x\in\mathcal{N}$ corresponding to the quantum state $\ket{x}$.

Moreover, the bijection $\mathcal{N}\to\mathcal{N}_1\times\dots\times\mathcal{N}_n$ implies that there is no emergent gauge symmetry, as is to be expected from the lack of entanglement in the states $\ket{x}$.

\subsection{Including entanglement}

We now wish to consider a more interesting case, where there is some entanglement.

Suppose a separable base state of the form $\ket{0}=\ket{0_1}\otimes\dots\otimes\ket{0_n}$ yields a classical limit as described in the previous subsection. Thus, for each $i$ we have a coadjoint orbit $\mathcal{N}_i$ of $G_i$, and a map $X_i: G_i\to\mathcal{N}_i$ defined by $X_i(g_i)=\operatorname{Ad}^*[g_i](X_i^0)$ for some $X_i^0\in\mathfrak{g}_i^*$, such that the states $\ket{g_i}=U_i(g_i)\ket{0_i}$ obey
\begin{equation}
    \lim_{\chi\to 0} \abs{\braket{g_i}{g_i'}}^2 = 
    \begin{cases}
        1 & \text{if }X_i(g_i)=X_i(g_i'), \\
        0 & \text{otherwise}.
    \end{cases}
    \label{Equation: coadjoint classical assumption start}
\end{equation}
From this starting point, we can obtain a different classical limit (one with emergent gauge symmetry) by picking a subgroup $H\subset G$ with certain properties. We will take $H$ to be a connected Lie group, but it is possible that the following analysis can be generalised in a natural way if this assumption is weakened.

It will be useful to define a homomorphism $\phi_i: G \to G_i$ by
\begin{equation}
    \phi_i(g_1,\dots,g_i,\dots,g_n)= g_i.
\end{equation}
The image $H_i=\phi_i(H)$ of $H$ is a subgroup of $G_i$. We will assume that $H_i$ is a normal subgroup. We will also assume that $X_i^0$ has trivial stabiliser under the coadjoint action of $H_i$, i.e.
\begin{align}
    h_i\in H_i, \, X_i(h_i) = X_i^0\, \implies\, h_i = \operatorname{Id}.
    \label{Equation: trivial stabiliser H}
\end{align}
A consequence of~\eqref{Equation: trivial stabiliser H} and~\eqref{Equation: coadjoint classical assumption start} is
\begin{equation}
    h_i,h_i' \in H_i \,\implies\, \abs{\braket{h_i}{h_i'}}^2 \approx
    \begin{cases}
        1 & \text{if }h_i=h_i', \\
        0 & \text{otherwise}.
    \end{cases}
    \label{Equation: H orthogonality}
\end{equation}

We define a new base state in $\mathcal{H}_1\otimes\dots\otimes\mathcal{H}_n$ by averaging $\ket{0_1}\otimes\dots\otimes\ket{0_n}$ over $H$:
\begin{equation}
    \ket{0_H} = \frac{\ket{\tilde 0_H}}{\sqrt{\alpha}}, \qq{where} 
    \ket{\tilde 0_H} = \hat\Pi_H\ket{0_1}\otimes\dots\otimes\ket{0_n}, \quad \alpha = \braket{\tilde 0_H}{\tilde 0_H},
    \label{Equation: 0 H and Pi H}
\end{equation}
and
\begin{equation}
    \hat\Pi_H = \int_{H}\dd{\tau(h)} U_1\qty\big(\phi_1(h))\otimes\dots\otimes U_n\qty\big(\phi_n(h)),
\end{equation}
where $\tau$ is the invariant measure on $H$, normalised so that $H$ has unit volume. Note that $\hat\Pi_H$ is the projection operator onto the subspace of $\mathcal{H}$ that is invariant under
\begin{equation}
    U_1\qty\big(\phi_1(h))\otimes\dots\otimes U_n\qty\big(\phi_n(h))
    \label{Equation: H action on 0}
\end{equation}
for all $h\in H$. Thus, $\ket{0_H}$ is also invariant under the action of $H$ defined by this operator. Here we are assuming that $\ket{0_1}\otimes\dots\otimes\ket{0_2}$ is not orthogonal to this subspace, so that $\alpha\ne 0$ and $\ket{0_H}$ is well-defined.

By acting on the base state $\ket{0_H}$, with $G_1\times \dots\times G_n$ via its unitary representation $U_1\otimes \dots\otimes U_n$, we get a set of states
\begin{equation}
    \ket{(g_1,\dots,g_n)_H} = U_1(g_1)\otimes\dots\otimes U_n(g_n) \ket{0_H}.
\end{equation}
Like $\ket{0_H}$, each state $\ket{(g_1,\dots,g_n)_H}$ is also invariant under an action of $H$ --- but it is not the same action as the one defined by~\eqref{Equation: H action on 0}. Instead, it is defined by
\begin{equation}
    U_1\qty\big(g_1\phi_1(h)g_1^{-1})\otimes \dots\otimes U_n\qty\big(g_n\phi_n(h)g_n^{-1}),
    \label{Equation: H more general action}
\end{equation}
which depends on $g_i$.

The states $\ket{(g_1,\dots,g_n)_H}$ form an overcomplete basis for $\mathcal{H}$, with a resolution of the identity provided by Schur's lemma applied to each of the $n$ irreducible representations separately:\footnote{Note the distinction between these states $
\ket{(g_1,\dots,g_n)_H} = (U_1(g_1)\otimes\dots\otimes  U_n(g_n))\hat\Pi_H(\ket{0_1}\otimes\dots\otimes\ket{0_n})$ and the alternate set of states $\ket{(g_1,\dots,g_n)'_H)}=\hat\Pi_H(U_1(g_1)\otimes\dots\otimes U_n(g_n))(\ket{0_1}\otimes\dots\otimes\ket{0_2})$. The latter only provide an overcomplete basis for the image of $\hat\Pi_H$.}
\begin{equation}
    \mathds{1} = \int_{G_1}\dd{\mu_1(g_1)}\dots\int_{G_n}\dd{\mu_n(g_n)} N\, \ket{(g_1,\dots,g_n)_H}\bra{(g_1,\dots,g_n)_H}.
    \label{Equation: coadjoint H identity}
\end{equation}

These states generically contain entanglement between the subsystems. We claim that they lead to a classical limit with an emergent gauge symmetry, in a way that is consistent with the mechanism described in this paper. The rest of this section is devoted to justifying this claim. We shall start in Sections~\ref{Section: group / emergent gauge} and~\ref{Section: group / states} by simply describing the system that is obtained in the classical limit, deferring a proof that this description is accurate until Sections~\ref{Section: group / orthogonal} and~\ref{Section: group / subsystems}

\subsection{Emergent gauge group}
\label{Section: group / emergent gauge}

One might expect that $H$ will be the gauge group of the classical system, since this was the group that we averaged over when constructing the states $\ket{(g_1,\dots, g_n)_H}$. This intuition is almost correct. Indeed, the elements of $H$ do correspond to gauge symmetries, as we will show. However, it turns out that there are some gauge transformations which are \emph{not} contained within $H$ --- so the true gauge group is actually larger than $H$. Let us now describe it.

The gauge group $K$ is defined by
\begin{equation}
    K = \{k\in G \mid k H k^{-1} = H,\,\exists\, h\in H \text{ such that }X_i(\phi_i(k))=X_i(\phi_i(h)) \text{ for all $i$}\}.
    \label{Equation: K definition}
\end{equation}
Let us confirm that this is a group. It clearly contains the identity, and the property $kHk^{-1}=H$ is clearly preserved under inverses and composition. If $k\in K$, then acting with $\operatorname{Ad}^*[\phi_i(k^{-1}h^{-1})]$ on both sides of $X_i(\phi_i(k))=X_i(\phi_i(h))$ yields
\begin{equation}
    X_i(\phi_i(k^{-1}h^{-1}k)) = X_i(\phi_i(k^{-1}))
    \label{Equation: K inverse}
\end{equation}
for all $i$. Note that $k^{-1}h^{-1}k\in k^{-1}H k = H$, so $k^{-1}\in K$. If $k,k'\in K$, then 
\begin{nalign}
    X_i(\phi_i(k k'))
    &= \operatorname{Ad}^*[\phi_i(k)](X_i(\phi_i(k'))) \\
    &= \operatorname{Ad}^*[\phi_i(k)](X_i(\phi_i(h'))) & \text{(some } h'\in H\text{)} \\
    &= \operatorname{Ad}^*[\phi_i(kh'k^{-1})](X_i(\phi_i(k))) \\
    &= \operatorname{Ad}^*[\phi_i(kh'k^{-1})](X_i(\phi_i(h))) & \text{(some }h\in H\text{)} \\
    &= X_i(\phi_i(kh'k^{-1}h)).
    \label{Equation: K composition}
\end{nalign}
holds for all $i$. Since $kh'k^{-1}h\in kHk^{-1}H = H$, we have $kk'\in K$. Thus, $K$ is a group, as claimed.

It is simple to check that $H$ is a subgroup of $K$. Thus, each element of $H$ is a gauge transformation, but not all gauge transformations are necessarily elements of $H$.

Note that~\eqref{Equation: trivial stabiliser H} implies that there is a unique $h_i\in H_i$ satisfying $X_i(\phi_i(k))=X_i(h_i)$ for each $k\in K$, which we can use to define a function
\begin{equation}
    f_i:\quad K \to H_i, \quad k \mapsto h_i \text{ such that } X_i(\phi_i(k))=X_i(h_i).
\end{equation}

\subsection{Kinematical and physical states}
\label{Section: group / states}

Having defined the classical gauge groups, we will now describe the classical kinematical states in each subsystem. We can then obtain physical states by taking quotients with respect to the gauge groups.

For each $i$ let us define
\begin{equation}
    \mathcal{Y}_i = \{Y_i^{g_i}\mid g_i\in G_i\} \subset \operatorname{Aut}(H_i),
\end{equation}
where $Y_i^{g_i}:H_i\to H_i$ is defined by $Y_i^{g_i}:h_i\mapsto g_ih_ig_i^{-1}$ (this is an automorphism of $H_i$ because $H_i$ is a normal subgroup of $G_i$). The kinematical state space for subsystem $s_i$ is then given by
\begin{equation}
    \mathcal{N}^{H,\text{kin.}}_i = \qty\big{X_i^{H,\text{kin.}}(g_i)\,\big\vert\, g_i\in G_i} \subset \mathcal{N}_i\times\mathcal{Y}_i,\qq{where} X_i^{H,\text{kin.}}(g_i) = \big(X_i(g_i),Y_i^{g_i}\big).
    \label{Equation: N H kin definition}
\end{equation}
The gauge group $K$ has a right action on $\mathcal{N}^{H,\text{kin.}}_i$ defined by
\begin{equation}
    (x_i, y_i) \mapsto (x_i,y_i)\triangleleft k = \big(\!\operatorname{Ad}^*[y_i(f_i(k))](x_i),\, y_i\circ Y_i^{\phi_i(k)}\big).
\end{equation}
Note that 
\begin{nalign}
    X_i^{H,\text{kin.}}(g_i)\triangleleft k 
    &= (X_i(g_i),Y_i^{g_i})\triangleleft k \\
    &= \big(\!\operatorname{Ad}^*[Y_i^{g_i}(f_i(k))](X_i(g_i)),Y_i^{g_i}\circ Y_i^{\phi_i(k)}\big) \\
    &= \big(\!\operatorname{Ad}^*[g_if_i(k)g_i^{-1}](X_i(g_i)),Y_i^{g_i\phi_i(k)}\big) \\
    &= \big(X_i(g_if_i(k)),Y_i^{g_i\phi_i(k)}\big) \\
    &= \big(X_i(g_i\phi_i(k)),Y_i^{g_i\phi_i(k)}\big) 
    = X_i^{H,\text{kin.}}(g_i \phi_i(k))
\end{nalign}
Thus, the right action of $K$ on $\mathcal{N}_i^{H,\text{kin.}}$ may be thought of representing the right multiplication action of $\phi_i(K)$ on $G_i$. This right action is a \emph{local gauge transformation} of the kinematical state in subsystem $s_i$, and so we may think of $\phi_i(K)$ as the local gauge group of $s_i$.

If $s_{ij\dots}$ is the union of some collection of the subsystems $s_i,s_j,\dots$, then the space of kinematical states in $s_{ij\dots}$ is given by 
\begin{equation}
    \mathcal{N}_{ij\dots}^{H,\text{kin.}} = \mathcal{N}_i^{H,\text{kin.}}\times\mathcal{N}_j^{H,\text{kin.}}\times\dots.
\end{equation}
The gauge group $K$ acts from the right on $\mathcal{N}_{ij\dots}^{H,\text{kin.}}$ via
\begin{equation}
    \big((x_i,\zeta_i),(x_j,\zeta_j),\dots\big) \mapsto \big((x_i,\zeta_i),(x_j,\zeta_j),\dots\big)\triangleleft k
    = \big((x_i,\zeta_i)\triangleleft k, (x_j,\zeta_j) \triangleleft k,\dots\big).
    \label{Equation: group union subsystem gauge group}
\end{equation}
Defining
\begin{equation}
    X^{H,\text{kin.}}_{ij\dots}(g_i,g_j,\dots) = (X^{H,\text{kin.}}_i(g_i),X^H_j(g_j),\dots),
\end{equation}
we note that $X^{H,\text{kin.}}_{ij\dots}(g_i,g_j,\dots)\triangleleft k=X^{H,\text{kin.}}_{ij\dots}(g_i\phi_i(k),g_j\phi_j(k),\dots)$. This is a local gauge transformation of the kinematical state in $s_{ij\dots}$.

The entire system is obtained by taking the union of all the subsystems $s_1,\dots, s_n$. In this case we denote the space of kinematical states as
\begin{equation}
    \mathcal{N}^{H,\text{kin.}} = \mathcal{N}^{H,\text{kin.}}_{1\dots n} = \mathcal{H}^{H,\text{kin.}}_1\times\dots\times\mathcal{N}^{H,\text{kin.}}_n.
\end{equation}
The gauge group $K$ acts on this space as in~\eqref{Equation: group union subsystem gauge group}. Defining 
\begin{equation}
    X^{H,\text{kin.}}(g)=X^{H,\text{kin.}}_{1\dots n}(\phi_1(g),\dots,\phi_n(g)),
\end{equation}
we have $X^{H,\text{kin.}}(g)\triangleleft k = X^{H,\text{kin.}}(gk)$. This is a \emph{global gauge transformation}.

We obtain the physical space of states in any subsystem (or union of subsystems, or the full system) by taking the quotient of its kinematical state space by the right action of the gauge group. Thus, the physical spaces of states in subsystem $s_i$, subsystem $s_{ij\dots}$, and the full system, are given respectively by
\begin{align}
    \mathcal{N}_i^H &= \mathcal{N}_i^{H,\text{kin.}}/K, \\
    \mathcal{N}_{ij\dots}^H &= \mathcal{N}_{ij\dots}^{H,\text{kin.}}/K,\\
    \mathcal{N}^H &= \mathcal{N}^{H,\text{kin.}}/K.
\end{align}

It is useful to define functions $X^H_i:G_i\to \mathcal{N}^H_i$, $X^H_{ij\dots}:G_i\times G_j\times\dots \to \mathcal{N}^H_{ij\dots}$, $X^H: G\to \mathcal{N}^H$ such that
\begin{equation}
    X^H_i(g_i), \quad X^H_{ij\dots}(g_i,g_j,\dots), \quad X^H(g_1,\dots,g_n)
\end{equation}
are the equivalence classes of
\begin{equation}
    X^{H,\text{kin.}}_i(g_i), \quad X^{H,\text{kin.}}_{ij\dots}(g_i,g_j,\dots), \quad X^{H,\text{kin.}}(g_1,\dots,g_n)
\end{equation}
within $\mathcal{N}_i^H$, $\mathcal{N}_{ij\dots}^H$, $\mathcal{N}^H$ respectively. These functions are surjective.

Given a physical state of the full system $x^H\in \mathcal{N}^H$, we can determine the corresponding physical state $x^H_{ij\dots}\in \mathcal{N}^H_{ij\dots}$ of a subsystem $s_{ij\dots}$ in the following way. First we let $x^{H,\text{kin.}} \in \mathcal{N}^{H,\text{kin.}}$ be a member of the equivalence class $x^H\subset\mathcal{N}^{H,\text{kin.}}$. Then, writing
\begin{equation}
    x^{H,\text{kin.}} = (x_1^{H,\text{kin.}},\dots,x_n^{H,\text{kin.}}),
\end{equation}
we set
\begin{equation}
    x^{H,\text{kin.}}_{ij\dots} = (x_i^{H,\text{kin.}},x_j^{H,\text{kin.}},\dots).
\end{equation}
Finally, $x^H_{ij\dots}$ is defined as the equivalence class of $x^{H,\text{kin.}}_{ij\dots}$. This procedure yields a map
\begin{equation}
    b_{ij\dots}:\quad \mathcal{N}^H \to \mathcal{N}^H_{ij\dots},\quad x^H\mapsto x^H_{ij\dots}
\end{equation}
which is well-defined by the properties of the right actions of $K$. Note that
\begin{equation}
    b_{ij\dots}(X^H(g)) = X^H_{ij\dots}(g_i,g_j,\dots),
\end{equation}
where $g=(g_1,\dots,g_n)$.

\subsection{Classical orthogonality of states}
\label{Section: group / orthogonal}

Let us now explain exactly how the above structure emerges when we take the classical limit.

Our first course of action is to demonstrate that the states $\ket{(g_1,\dots,g_n)_H}$ yield a good classical limit. To that end, let us compute the inner product of two of them:
\begin{equation}
    \braket{(g_1,\dots,g_n)_H}{(g_1',\dots,g_n')_H} = \frac1\alpha\int_H\dd{\tau(h)}\int_H\dd{\tau(h')}\prod_{i=1}^n \braket{g_i\phi_i(h)}{g'_i\phi_i(h')}.
    \label{Equation: group entangled inner product}
\end{equation}
It will be useful to change variables from $h'\in H$ to 
\begin{equation}
    p = g^{-1} g' h' g'^{-1} g\in H' = g^{-1} g' H g'^{-1} g,
\end{equation}
where $g=(g_1,\dots,g_n)$ and $g'=(g'_1,\dots,g'_n)$. Since $H_i$ is a normal subgroup of $G$, we have
\begin{equation}
    \phi_i(p) = g_i^{-1} g'_i \phi_i(h') g_i^{-1} g_i \in H_i.
\end{equation}
Also,
\begin{nalign}
    \braket*{g_i\phi_i(h)}{g'_i\phi_i(h')} &= \braket*{\phi_i(h)}{g_i^{-1}g'_i\phi_i(h)} \\
                                           &= \braket*{\phi_i(h)}{\phi_i(p)g_i^{-1} g_i'} \\
                                           &= \braket*{\phi_i(p)^{-1}\phi_i(h)}{g_i^{-1}g_i'} \\
                                           &= \braket*{\phi_i(p^{-1}h)}{g_i^{-1}g_i'}.
\end{nalign}
Thus, we may write
\begin{equation}
    \braket{(g_1,\dots,g_n)_H}{(g_1',\dots,g_n')_H} = \frac1\alpha\int_H\dd{\tau(h)}\int_{H'}\dd{\tau'(p)}\prod_{i=1}^n \braket*{\phi_i(p^{-1}h)}{g_i^{-1}g'_i},
    \label{Equation: group entangled inner product change of variables}
\end{equation}
where $\tau'$ is the pushforward of $\tau$ through the map $h'\mapsto p$. It can be checked that $\tau'$ is the invariant measure on $H'$, normalised so that $H'$ has unit volume.

From~\eqref{Equation: coadjoint classical assumption start}, the integrand is dominated by pairs $h\in H$, $p\in H'$ obeying
\begin{equation}
    X_i(\phi_i(p^{-1}h)) = X_i(g_i^{-1}g_i').
    \label{Equation: classical equivalence entangled coadjoint 1}
\end{equation}
If no such pairs exist, then the integrand in~\eqref{Equation: group entangled inner product change of variables} is approximately vanishing, and one can conclude that the states are approximately orthogonal. Let us consider the other case, where at least one pair $(h,p)=(h_0,p_0)$ satisfies~\eqref{Equation: classical equivalence entangled coadjoint 1}. Then in the classical limit we have 
\begin{equation}
    \ket*{g_i^{-1}g_i'} \approx e^{i\gamma_i}\ket*{\phi_i(p_0^{-1}h_0)},
    \label{Equation: g g' phase}
\end{equation}
where $\gamma_i\in\RR$, and we may write the inner product as
\begin{nalign}
    \braket{(g_1,\dots,g_n)_H}{(g_1',\dots,g_n')_H} &\approx \frac{e^{i\gamma}}\alpha\int_H\dd{\tau(h)}\int_{H'}\dd{\tau'(p)}\prod_{i=1}^n \braket*{\phi_i(p^{-1}h)}{\phi_i(p_0^{-1}h_0)} \\
                                                    &= \frac{e^{i\gamma}}\alpha\int_H\dd{\tau(h)}\int_{H'}\dd{\tau'(p)}\prod_{i=1}^n \braket*{\phi_i(h)}{\phi_i(p p_0^{-1}h_0)},
    \label{Equation: group entangled inner product 2}
\end{nalign}
where $\gamma=\sum_i\gamma_i$. By~\eqref{Equation: H orthogonality}, this integral is dominated by contributions where $h=pp_0^{-1} h_0$. In fact, the integral on the left-hand side of
\begin{equation}
    \int_H\dd{\tau}(h) \prod_{i=1}^n\braket*{\phi_i(h)}{\phi_i(p p_0^{-1} h_0)} = (\bra{0_1}\otimes\dots\otimes\bra{0_n})\hat\Pi_H(\ket*{\phi_1(p p_0^{-1} h_0)}\otimes\dots\otimes\ket*{\phi_n(p p_0^{-1} h_0)})
\end{equation}
is also dominated by $h=p p_0^{-1} h_0^{-1}$, so we can write
\begin{align}
    \prod_{i=1}^n\braket*{\phi_i(h)}{\phi_i(p p_0^{-1} h_0)} 
    &\approx \delta_\tau(h,p p_0^{-1} h_0)\,(\bra{0_1}\otimes\dots\otimes\bra{0_n})\hat\Pi_H(\ket*{\phi_1(p p_0^{-1} h_0)}\otimes\dots\otimes\ket*{\phi_n(p p_0^{-1} h_0)}) \nonumber\\
    &= \delta_\tau(h h_0^{-1},p p_0^{-1})\,(\bra{0_1}\otimes\dots\otimes\bra{0_n})\hat\Pi_H(\ket{\phi_1(h)}\otimes\dots\otimes\ket{\phi_n(h)})\nonumber\\
    &= \delta_\tau(h h_0^{-1},p p_0^{-1})\,(\bra{0_1}\otimes\dots\otimes\bra{0_n})\hat\Pi_H(\ket{0_1}\otimes\dots\otimes\ket{0_n}) \\
    &= \delta_\tau(h h_0^{-1},p p_0^{-1})\,\underbrace{\braket{\tilde 0_H}{\tilde 0_H}}_{=\alpha},\nonumber
\end{align}
where in the third line we used the invariance of the measure $\tau$ in the definition~\eqref{Equation: 0 H and Pi H} of $\hat\Pi_H$. Substituting this into~\eqref{Equation: group entangled inner product 2}, and changing variables $h\to h h_0$, $p \to p p_0$, yields
\begin{nalign}
    \braket{(g_1,\dots,g_n)_H}{(g_1',\dots,g_n')_H} &\approx e^{i\gamma}\int_H\dd{\tau(h)}\int_{H'}\dd{\tau'(p)} \delta_\tau(h,p) \\
                                                    &= e^{i\gamma} \,\tau(H\cap H').
    \label{Equation: group entangled inner product 3}
\end{nalign}
Thus, up to a phase, the inner product is approximately equal to the volume of $H\cap H'\subset H$ according to the measure $\tau$. Note that $H$ and $H'$ are both Lie subgroups of $G$, so their intersection is also a Lie subgroup of $G$. Thus, $H\cap H'$ is a submanifold of $H$. Since $H$ is connected, we must either have that $H=H\cap H'$, or that $H\cap H'$ has positive codimension in $H$. In the latter case we have $\tau(H\cap H')=0$, since the measure $\tau$ can be written in terms of a volume form on $H$. The former case is equivalent to the map
\begin{equation}
    h' \mapsto p = g^{-1}g' h' g'^{-1} g
    \label{Equation: g-1g' normalises}
\end{equation}
being an automorphism of $H$. Then $\tau(H\cap H')=\tau(H)=1$, so the inner product is approximately a pure phase. In this case we have $p_0\in H$, so $p_0^{-1}h_0\in H$. Since $X_i(g_i^{-1}g_i)=X_i(\phi_i(p_0^{-1}h_0))$, and additionally~\eqref{Equation: g-1g' normalises} holds, we must by definition have $g^{-1}g'\in K$.

We have so far succeeded in showing that the inner product~\eqref{Equation: group entangled inner product} is approximately non-vanishing only if $g^{-1}g'\in K$. Actually, the reverse is true too. If $g^{-1}g' = k\in K$, then we may write the inner product~\eqref{Equation: group entangled inner product} as
\begin{nalign}
    \braket{(g_1,\dots,g_n)_H}{(g_1',\dots,g_n')_H} 
    &= \frac1\alpha\int_H\dd{\tau(h)}\int_H\dd{\tau(h')}\prod_{i=1}^n \braket{g_i\phi_i(h)}{g_i\phi_i(k)\phi_i(h')} \\
    &= \frac1\alpha\int_H\dd{\tau(h)}\int_H\dd{\tau(h')}\prod_{i=1}^n \braket{\phi_i(h)}{\phi_i(kh')} \\
    &= \frac1{\alpha}\int_H\dd{\tau(h')} \qty\Big(\bra{0_1}\otimes\dots\otimes\bra{0_n})\hat\Pi_H U(kh'k^{-1})\qty\big(\ket{\phi_1(k)}\otimes\dots\otimes\ket{\phi_n(k)})\\
    &\approx \frac{e^{i\gamma}}{\alpha}\int_H\dd{\tau(h')} \underbrace{\qty\big(\bra{0_1}\otimes\dots\otimes\bra{0_n})\hat\Pi_H\qty\big(\ket{0_1}\otimes\dots\otimes\ket{0_n})}_{=\alpha} = e^{i\gamma}.
\end{nalign}
In the fourth line we used the fact that $kh'k^{-1}\in H$, and used~\eqref{Equation: g g' phase}, which here implies
\begin{equation}
    \ket{\phi_i(k)} = \ket*{g_i^{-1}g_i'} \approx e^{i\gamma_i} U_i(\phi_i(\tilde{h})) \ket{0_i},
\end{equation}
where $\tilde{h}=p_0^{-1}h_0\in H$. Therefore,
\begin{equation}
    \abs{\braket{(g_1,\dots,g_n)_H}{(g_1',\dots,g_n')_H}}^2 \approx
    \begin{cases}
        1 & \text{if } g^{-1}g'\in K\\
        0 & \text{otherwise.}
    \end{cases}
    \label{Equation: group entangled states orthogonality G}
\end{equation}

We may alternatively write this in terms of the classical physical states $X^H(g)\in\mathcal{N}^H$. To see this, note that $g^{-1}g'\in K$ implies
\begin{equation}
    X^{H,\text{kin.}}(g)\triangleleft k = X^{H,\text{kin.}}(g')
\end{equation}
for some $k\in K$, i.e.\ $X^H(g)=X^H(g')$. The reverse is also true, since $X^H(g)=X^H(g')$ implies there exists a $k\in K$ such that
\begin{equation}
    X_i(\phi_i(g')) = X_i(\phi_i(g k)).
\end{equation}
Acting with $\operatorname{Ad}^*[\phi_i(g^{-1})]$ on both sides yields
\begin{equation}
    X_i(\phi_i(g^{-1}g')) = X_i(\phi_i(k)) = X_i(\phi_i(h))
\end{equation}
for some $h\in H$. This implies $g^{-1}g'\in K$ by definition.
We thus have
\begin{equation}
    \abs{\braket{(g_1,\dots,g_n)_H}{(g_1',\dots,g_n')_H}}^2 \approx
    \begin{cases}
        1 & \text{if } X^H(g)=X^H(g'),\\
        0 & \text{otherwise.}
    \end{cases}
    \label{Equation: group entangled states orthogonality}
\end{equation}

Now let $v:\mathcal{N}^H\to G$ be a function such that $X^H\circ v$ is the identity on $\mathcal{N}^H$; such a function exists because $X_H$ is surjective. For each $x^H\in\mathcal{N}_H$, let us define the state $\ket*{x^H} = \ket*{(v(x^H))_H}$. Then by~\eqref{Equation: group entangled states orthogonality} we have
\begin{equation}
    \braket*{x^H}{x'^H} \approx
    \begin{cases}
        1 & \text{if } x^H=x'^H,\\
        0 & \text{otherwise.}
    \end{cases}
\end{equation}
Also~\eqref{Equation: group entangled states orthogonality} implies that $\ket*{x^H}\bra*{x^H} \approx  \ket{(g_1,\dots,g_n)_H}\bra{(g_1,\dots,g_n)_H}$ if $X^H(g)=x^H$, so by~\eqref{Equation: coadjoint H identity} we have
\begin{equation}
    \mathds{1} \approx \int_{\mathcal{N}^H} \dd{\mu^H(x^H)} N\, \ket*{x^H}\bra*{x^H},
\end{equation}
where $\mu^H$ is the pushforward of the product measure $\mu_1\times\dots\times\mu_n$ through the map $X^H:G\to \mathcal{N}^H$. 

Therefore, as claimed, the states $\ket{(g_1,\dots,g_n)_H}$ lead to a classical limit whose classical space of states is given by $\mathcal{N}^H$.

\subsection{States of subsystems}
\label{Section: group / subsystems}

Our next course of action is to understand what happens to the local structure of this system in the classical limit.

Consider the subsystem $s_{i\dots}$ given by the union of a collection of subsystems $s_i,\dots$. When the state of the full system is $\ket*{x^H}$, the state of $s_{i\dots}$ is described by the reduced density matrix
\begin{align}
    \rho_{i\dots}(x^H) 
    &= \tr_{\overline{i\dots}}\qty\big(\ket*{x^H}\bra*{x^H}) \\
    &\approx \tr_{\overline{i\dots}}\qty\big(\ket{(g_1,\dots,g_n)_H}\bra{(g_1,\dots,g_n)_H}) \\
    & 
    \begin{multlined}
        =\qty\Big(U_1(g_1)\otimes\dots\otimes U_{i-1}(g_{i-1})\otimes U_{i+1}(g_{i+1})\otimes \dots \otimes U_n(g_n))\tr_{\overline{i\dots}}(\ket{0_H}\bra{0_H})\\
            \qty\Big(U_1(g_1)\otimes\dots\otimes U_{i-1}(g_{i-1})\otimes U_{i+1}(g_{i+1})\otimes \dots \otimes U_n(g_n))^\dagger,
    \end{multlined}
\end{align}
where $(g_1,\dots, g_n)=v(x^H)$. Note that this density matrix is determined fully by $x^H_{i\dots}=b_{i\dots}(x^H)$.

We can use Schur's lemma to write (up to a rearrangement of the tensor factors in $\mathcal{H}=\mathcal{H}_1\otimes\dots\otimes\mathcal{H}_n$)
\begin{equation}
    \mathds{1}_{\overline{i\dots}} \otimes \rho_{i\dots}(x^H) \approx \qty{\int_{G_j}\dd{\mu_j(\tilde g_j)}N_j\, \int_{G_k}\dd{\mu_k(\tilde g_k)N_k}\,\dots} \ket{(\tilde g_1,\dots,\tilde g_n)_H}\bra{(\tilde g_1,\dots,\tilde g_n)_H}\Big|_{\tilde g_i=g_i,\dots},
\end{equation}
where the braced integration is done for all subsystems $s_j,s_k,\dots$ that make up the complement of $s_{i\dots}$, and at the end we set $\tilde g_i=g_i$ for all subsystems $s_i$ that make up $s_{i\dots}$. We may thus write
\begin{align}
    \MoveEqLeft\mathds{1}_{\overline{i\dots}} \otimes \qty\Big(\rho_{i\dots}(x^H)\rho_{i\dots}(x'^H))\\
    &\begin{multlined}
        \approx \qty{\int_{G_j}\dd{\mu_j(\tilde g_j)}N_j\, \int_{G_k}\dd{\mu_k(\tilde g_k)N_k}\,\dots} \ket{(\tilde g_1,\dots,\tilde g_n)_H}\bra{(\tilde g_1,\dots,\tilde g_n)_H}\Big|_{\tilde g_i=g_i,\dots}\\
        \qty{\int_{G_j}\dd{\mu_j(\tilde g'_j)}N_j\, \int_{G_k}\dd{\mu_k(\tilde g'_k)N_k}\,\dots} \ket{(\tilde g'_1,\dots,\tilde g'_n)_H}\bra{(\tilde g'_1,\dots,\tilde g'_n)_H}\Big|_{\tilde g'_i=g'_i,\dots}
    \end{multlined}\\
    &\begin{multlined}
        = \qty{\int_{G_j\times G_j}\dd{\mu_j(\tilde g_j)}\dd{\mu_j(\tilde g'_j)}N_j^2\,\int_{G_k\times G_k}\dd{\mu_k(\tilde g_k)}\dd{\mu_k(\tilde g'_k)}N_k^2} \\
        \ket{(\tilde g_1,\dots,\tilde g_n)_H}\braket{(\tilde g_1,\dots,\tilde g_n)_H}{(\tilde g'_1,\dots,\tilde g'_n)_H}\bra{(\tilde g'_1,\dots,\tilde g'_n)_H}\Big|_{\substack{\tilde g_i=g_i,\dots\\\tilde g'_i=g'_i,\dots}}
    \end{multlined}\label{Equation: subsystem group composition}
\end{align}
where $(g'_1,\dots,g'_n)=v(x'^H)$. Taking the partial trace over $\mathcal{H}_{\overline{i\dots}}$ of~\eqref{Equation: subsystem group composition}, and dividing by $N_jN_k\dots$, we then have
\begin{multline}
    \rho_{i\dots}(x^H)\rho_{i\dots}(x'^H) = \qty{\int_{G_j\times G_j}\dd{\mu_j(\tilde g_j)}\dd{\mu_j(\tilde g'_j)}N_j\,\int_{G_k\times G_k}\dd{\mu_k(\tilde g_k)}\dd{\mu_k(\tilde g'_k)}N_k} \\
    \tr_{\overline{i\dots}}\Big(\ket{(\tilde g_1,\dots,\tilde g_n)_H}\braket{(\tilde g_1,\dots,\tilde g_n)_H}{(\tilde g'_1,\dots,\tilde g'_n)_H}\bra{(\tilde g'_1,\dots,\tilde g'_n)_H}\Big)\Big|_{\substack{\tilde g_i=g_i,\dots\\\tilde g'_i=g'_i,\dots}}
    \label{Equation: subsystem group composition 2}
\end{multline}
By~\eqref{Equation: group entangled states orthogonality G}, the integrand is approximately vanishing, unless there exist $\tilde{g}_j,\tilde{g}_k,\dots$ and $\tilde{g}'_j,\tilde{g}'_k,\dots$ such that
\begin{equation}
    (\tilde{g}_1,\dots,\tilde{g}_n)^{-1}(\tilde{g}'_1,\dots,\tilde{g}'_n)\Big|_{\substack{\tilde g_i=g_i,\dots\\\tilde g'_i=g'_i,\dots}} \in K.
    \label{Equation: subsystem K dominate}
\end{equation}
This condition is equivalent to
\begin{equation}
    (g_i,\dots)^{-1}(g_i',\dots) \in \phi_{i\dots}(K),
\end{equation}
which holds if and only if 
\begin{equation}
    X^H_{i\dots}(g_i,\dots) = X^H_{i\dots}(g_i',\dots), \text{ i.e.\ } x^H_{i\dots} = x'^H_{i\dots}.
    \label{Equation: subsystem physical dominate}
\end{equation}
If~\eqref{Equation: subsystem physical dominate} does hold, then the integral~\eqref{Equation: subsystem group composition 2} is dominated by contributions where~\eqref{Equation: subsystem K dominate} is true. But note that for these contributions~\eqref{Equation: group entangled states orthogonality G} implies that
\begin{equation}
    \ket{(\tilde g_1,\dots,\tilde g_n)_H}\bra{(\tilde g_1,\dots,\tilde g_n)_H}\Big|_{\tilde{g}_i=g_i,\dots} \approx \ket{(\tilde g'_1,\dots,\tilde g'_n)_H}\bra{(\tilde g'_1,\dots,\tilde g'_n)_H}\Big|_{\tilde{g}'_i=g_i',\dots},
\end{equation}
in which case the second line in~\eqref{Equation: subsystem group composition 2} is approximately equal to 
\begin{equation}
    \tr_{\overline{i\dots}}\Big(\ket{(\tilde g_1,\dots,\tilde g_n)_H}\bra{(\tilde g_1,\dots,\tilde g_n)_H}\Big)\Big|_{\tilde{g}_i=g_i,\dots} \approx \rho_{i\dots}(x^H).
\end{equation}
This is constant over the range of integration, and so $\rho_{i\dots}(x^H)\rho_{i\dots}(x'^H)$ is approximately proportional to $\rho_{i\dots}(x^H)$. But note also that~\eqref{Equation: subsystem physical dominate} implies $\rho_{i\dots}(x^H)=\rho_{i\dots}(x'^H)$, so we have $\qty\big(\rho_{i\dots}(x^H))^2=\rho_{i\dots}(x^H)$. In other words, $\rho_{i\dots}(x^H)$ is approximately proportional to a projection operator. We use $\hat\pi_{i\dots}(x^H_{i\dots})$ to denote this projection operator, so that
\begin{equation}
    \rho_{i\dots}(x^H) = \frac{\hat\pi_{i\dots}(x^H_{i\dots})}{N_{i\dots}(x_{i\dots})},
\end{equation}
where $N_{i\dots}(x_{i\dots})$ is the rank of $\hat\pi_{i\dots}(x^H_{i\dots})$. If~\eqref{Equation: subsystem physical dominate} does not hold, then the integrand in~\eqref{Equation: subsystem group composition 2} approximately vanishes, and so $\rho_i(x^H)\rho_i(x'^H) \approx 0$, which implies $\hat\pi_{i\dots}(x^H_{i\dots})\hat\pi_{i\dots}(x'^H_{i\dots})\approx 0$.

To summarise, we have shown that the density matrices $\rho_{i\dots}(x^H)$ of subsystem $s_{i\dots}$ are approximately proportional to projection operators $\hat\pi_{i\dots}(x^H_{i\dots})$, and that these projection operators obey
\begin{equation}
    \hat\pi_{i\dots}(x^H_{i\dots}) \hat\pi_{i\dots}(x'^H_{i\dots}) \approx \delta_{x^H_{i\dots}\,x'^H_{i\dots}}\hat\pi_{i\dots}(x^H_{i\dots}).
\end{equation}
Therefore, the subsystem $s_{i\dots}$ is classically resolvable, and its physical state space is $\mathcal{N}^H_{i\dots}$. This result holds for all subsystems $s_{i\dots}$. Thus, we have succeeded in demonstrating that the local structure of this system is classically resolvable.

\subsection{Summary of structures}

In this section, we have encountered a large family of models based on unitary representations of Lie groups. Let us summarise the structure of these models.

We assume that Hilbert space factorises into subsystem Hilbert spaces as
\begin{equation}
    \mathcal{H}=\mathcal{H}_1\otimes\dots\otimes\mathcal{H}_n,
\end{equation}
with a Lie group $G_i$ acting unitarily and irreducibly on each factor $\mathcal{H}_i$. The ingredients of the model are then:
\begin{enumerate}
    \item A classical limit for each subsystem individually, where the classical space of states for the subsystem is a coadjoint orbit $\mathcal{N}_i\subset\mathfrak{g}_i^*$ of $G_i$.
    \item A connected Lie subgroup 
        \begin{equation}
            H\subseteq G = G_1\times\dots\times G_n,
        \end{equation}
        such that $H_i=\phi_i(H)$ is a normal subgroup of $G_i$ with the property that its stabiliser for the corresponding subsystem base state is trivial, i.e.~\eqref{Equation: trivial stabiliser H}.
\end{enumerate}

We have shown how to construct a family of coherent states for the full system by averaging over $H$ in a certain way, and demonstrated that these states give a well-defined classical limit. 

The averaging over $H$ determines the way in which the coherent states are entangled, and we have demonstrated that this entanglement is consistent with the classical resolvability of all the subsystems. Moreover, we have described the emergent gauge symmetry that this yields. The kinematical state space for each subsystem is defined by
\eqref{Equation: N H kin definition}, and the kinematical state space for the full system is defined by the usual
\begin{equation}
    \mathcal{N}^{H,\text{kin.}} = \mathcal{N}_1^{H,\text{kin.}}\times\dots\times\mathcal{N}_n^{H,\text{kin.}}.
\end{equation}
We constructed the emergent gauge group $K$ in~\eqref{Equation: K definition}. It satsifies $H \subseteq K \subseteq G$, and has a natural right action on the kinematical state spaces. We demonstrated that quotienting by this gauge group yields the correct physical space of states for each subsystem.

Thus, we have exhibited a large family of systems where classical gauge symmetry emerges from entanglement. This means that the mechanism we have described is not just a theoretical possibility, or a curiosity of some very specific models. Rather, it is a fairly generic phenomenon.

The toy model involving three spins is a special case of this family, where the Lie groups $G_i$ are each $\mathrm{SU}(2)$, and $H$ is the diagonal subgroup of $G_1\times G_2\times G_3$. In the case of the toy model it turns out that $K=H$.

\section{`Gravitational' properties of the mechanism}
\label{Section: Gravitational properties}

Let us submit the following conjecture:
\begin{quote}
    Diffeomorphism invariance in the classical limit of quantum gravity emerges from entanglement via the mechanism described in this paper (or some close relative of it).
\end{quote}
This short section is devoted to presenting a few pieces of evidence in favour of the conjecture, and to describing some more general properties of the mechanism we are describing that are reminiscent of what happens in gravity.

\begin{figure}
    \centering
    \begin{subfigure}{\linewidth}
        \centering
        \begin{tikzpicture}
            \begin{scope}[shift={(6,0)}]
                \foreach \x in {0,1,2,3,4} {
                    \draw (\x*72:2) -- ({(\x+1)*72}:2);
                    \draw (\x*72:2) .. controls ({(\x+1)*72}:1.4) .. ({(\x+2)*72}:2);
                }
                \foreach \x in {0,1,2,3,4} {
                    \fill[white] (\x*72:2) circle (0.25);
                    \fill (\x*72:2) circle (0.15);
                }
            \end{scope}
            \foreach \x in {0,1,2,3,4,5} {
                \draw (\x*60:2) -- ({(\x+1)*60}:2);
                \draw (\x*60:2) .. controls ({(\x+1)*60}:1.4) .. ({(\x+2)*60}:2);
            }
            \foreach \x in {0,1,2,3,4,5} {
                \fill[white] (\x*60:2) circle (0.25);
                \fill (\x*60:2) circle (0.15);
            }
        \end{tikzpicture}
        \caption{}
    \end{subfigure}
    \vspace{\baselineskip}

    \begin{subfigure}{\linewidth}
        \centering
        \begin{tikzpicture}
            \begin{scope}[shift={(8,0)}]
                \foreach \x in {0,1,2,3,4} {
                    \coordinate (a\x) at (\x*72:2);
                    \draw (\x*72:2) -- ({(\x+1)*72}:2);
                    \draw (\x*72:2) .. controls ({(\x+1)*72}:1.4) .. ({(\x+2)*72}:2);
                }
            \end{scope}
            \foreach \x in {0,1,2,3,4,5} {
                \coordinate (b\x) at (\x*60:2);
                \draw (\x*60:2) -- ({(\x+1)*60}:2);
                \draw (\x*60:2) .. controls ({(\x+1)*60}:1.4) .. ({(\x+2)*60}:2);
            }
            \begin{scope}[very thick]
                \draw (a2) .. controls (4,1) and (3,1) .. (b1);
                \draw (a3) .. controls (4,-1) and (3,-1) .. (b5);
                \draw (a2) .. controls (5.5,0) and (3,0) .. (b5);
                \draw (a3) .. controls (5.5,0) and (3,0) .. (b1);
                \draw (a3) .. controls (3.5,-0.5) and (3,0) .. (b0);
                \draw (a2) .. controls (3.5,0.5) and (3,0) .. (b0);
            \end{scope}
            \begin{scope}[shift={(8,0)}]
                \foreach \x in {0,1,2,3,4} {
                    \fill[white] (\x*72:2) circle (0.25);
                    \fill (\x*72:2) circle (0.15);
                }
            \end{scope}
            \foreach \x in {0,1,2,3,4,5} {
                \fill[white] (\x*60:2) circle (0.25);
                \fill (\x*60:2) circle (0.15);
            }
        \end{tikzpicture}
        \caption{}
    \end{subfigure}
    \caption{An example of two possible different `bulk topologies', as determined by the sets of emergent non-local degrees of freedom arising from entanglement in the classical limit. A dot {\protect\tikz{\protect\fill (0,0) circle (.5ex);}} represents each subsystem, and we draw a line {\protect\tikz[baseline=-0.5ex]{\protect\draw (0,0) -- (0.5,0.1);}} between subsystems when their mutual information is non-vanishing in the classical limit (as this determines when they share non-local degrees of freedom, as descibed in Section~\ref{Section: Gauge symmetry from entanglement}). These lines can roughly be thought of as determining the bulk topology. \mbox{\textbf{(a)}}\, A bulk topology coming from a state where the subsystems split into two unentangled sets, which thus share no non-local degrees of freedom. \mbox{\textbf{(b)}}\, A bulk topology coming from a different state of the same system, in which some of the subsystems in the first set are now entangled with some of those in the second set. Thus, the two sets now share some non-local degrees of freedom, indicated by thicker lines {\protect\tikz[baseline=-0.5ex]{\protect\draw[very thick] (0,0) -- (0.5,0.1);}} connecting their subsystems. This can be interpreted as the bulk topology including a `wormhole' that was not there before.
    }
    \label{Figure: emergent wormhole}
\end{figure}
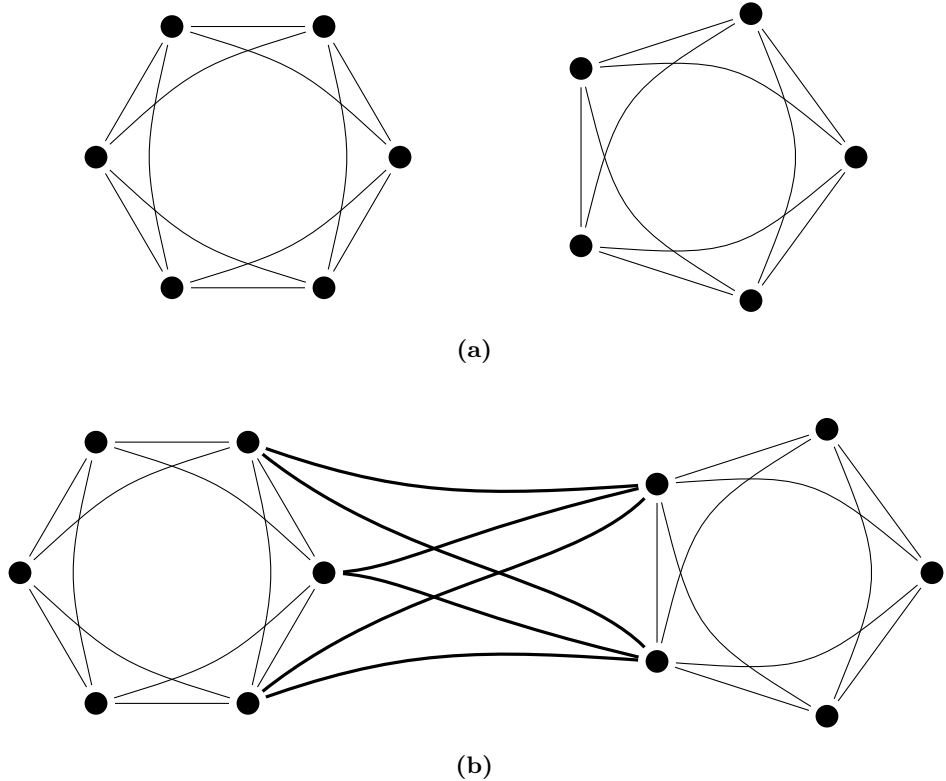

\begin{itemize}
    \item As mentioned in the introduction, there is by now a widely held expectation that the bulk spacetime in holography, and in quantum gravity more generally, emerges from the structure of entanglement in the quantum state. Thus, it is only natural that the diffeomorphism invariance associated with that spacetime should also emerge from entanglement~\cite{Harlow:2015lma,Witten:2017hdv,Maldacena:2001kr,VanRaamsdonk:2010pw,Faulkner:2013ica,Swingle:2014uza,Jacobson:2015hqa,Verlinde:2016toy}. The mechanism that we have described is a very general way in which this can happen, and it is not too much of a stretch to suggest that it is general enough to include the gravitational case.\footnote{Although it probably needs to be extended to account for pre-existing quantum gauge symmetries, and to work with infinite-dimensional Hilbert spaces --- we comment on this further in the conclusion.}
    \item In semiclassical treatments of quantum gravity, the density matrix of a spacetime subregion can usually be written in the form $\rho = \exp\big(-\frac{\hat{A}}{4G} + \dots\big)/\mathcal{Z}$, where $\hat{A}$ is an operator that measures the area of some surface, $\mathcal{Z} = \tr\big(\exp\big(-\frac{\hat{A}}{4G}+\dots\big)\big)$ is a normalisation factor, and the dots $\dots$ contain subleading in $G$ corrections~\cite{Jafferis:2015del}. The classical limit is $G\to 0$. In this limit, $\rho$ becomes approximately proportional to a projection operator whose image contains states in which the expectation value of $\hat{A}$ is arbitrarily close to its minimum (see also~\cite{BeyondToyModels}). Moreover, if $\rho'$ is the density matrix of the same spacetime subregion in a different state, then we have $\rho\rho' \approx 0$ in the classical limit. There are various ways to show this --- for example, one may compute the fidelity $\tr(\sqrt{\sqrt{\rho}\rho'\sqrt{\rho}})$ of the two states, and show that it is $e^{-\order{1/G}}$, and thus that it vanishes in the classical limit~\cite{Kirklin:2019ror}. Thus, spacetime subregions in quantum gravity are classically resolvable, in a way that is consistent with what we have described here.
    \item In holographic theories, gravity usually only emerges in a certain limit. For example, in AdS/CFT, the classical gravitational regime is a strong coupling limit in the boundary theory. However, we can instead consider a weak coupling regime, and then take a direct classical limit of the boundary theory. Thus, there are two different classical limits -- one for the bulk, and one for the boundary. There is a sense in which the local structure of the bulk is the same as that of the boundary (to each boundary subregion we associate its corresponding entanglement wedge --- this is the content of `subregion duality'~\cite{Czech:2012bh,Almheiri:2014lwa,Dong:2016eik,Faulkner:2017vdd}). However, in the bulk classical limit, the local structure respects a gravitational gauge symmetry, whereas in the boundary classical limit it does not. Thus, we have two classical theories, dual to each other in the sense that they are both limits of the same quantum theory, but without the same kind of gauge symmetry. The mechanism we have described provides a natural way for this  (and for this kind of duality more generally) to happen.
    \item One of the key features of gravity is that the topology of spacetime can vary from state to state. Physically, this means that in different states there are different sets of non-local degrees of freedom. The mechanism we have described provides a way for this to happen: the set of non-local degrees of freedom is determined by the structure of entanglement in the underlying quantum state. For different quantum states there are different entanglement structures, and so different sets of non-local degrees of freedom --- and we are inevitably led to physically interpret this in terms of different spacetime topologies. An example of this is illustrated in Figure~\ref{Figure: emergent wormhole}. Quantum theories with variable spacetime topologies would seem to be more difficult to generally construct using traditional constrained quantisation.
    \item Certain models for holographic theories involve tensor networks~\cite{Happy,Donnelly:2016qqt,BeyondToyModels}. In particular, the toy model proposed in~\cite{Happy} involves a tensor network constructed out of perfect tensors. The reduced density matrices in subregions of such tensor networks are approximately proportional to projection operators. Moreover, if one projects the bulk legs of the tensor network onto bulk classical states, it is not hard to show that the subregion reduced density matrices are approximately orthogonal, when the subregion bulk states are different. Thus, these tensor networks yield a classically resolvable local structure, consistent with what has been described in this paper.
    \item The emergent gauge transformations that we have described are essentially approximate `modular symmetries' of each subsystem -- i.e.\ transformations which do not change the reduced density matrix. This is exactly consistent with gravity, where the modular symmetries of a subregion have been shown to be the symmetries of its edge modes~\cite{Czech:2019vih}.
    \item Much conceptual progress in gravity has come from using a quantum error correction (QEC) interpretation of holography~\cite{Almheiri:2014lwa}. It is interesting that QEC also plays a natural role in the context of this paper. To see this, suppose we take the classical limit of a quantum system with Hilbert space $\mathcal{H}$, obtaining a classical system with a space of physical classical states $\mathcal{N}$ with an emergent gauge symmetry, as we have described. Thus, $\mathcal{N}$ can be obtained by imposing constraints and carrying out gauge reduction on a space of kinematical states $\mathcal{N}^{\text{kin.}}$. Suppose we do a constrained quantisation of the classical theory via these kinematical states. In other words, we come up with a `kinematical' Hilbert space $\mathcal{H}^{\text{kin.}}$ and a set of operator constraints which when imposed yield a `physical' Hilbert space $\mathcal{H}^{\text{phys.}}$, whose classical limit has the space of classical states $\mathcal{N}$. There is then a sense in which $\mathcal{H}$ is embedded in $\mathcal{H}^{\text{kin.}}$ via some map $\mathcal{H}\to\mathcal{H}^{\text{kin.}}$ defined by the common classical limit of the two quantum theories. It is natural to think of $\mathcal{H}$ as a code subspace of $\mathcal{H}^{\text{kin.}}$.\footnote{This may be related to work in~\cite{philipp}}. This explains, for example, the resemblance of the three spin toy model we described in Section~\ref{Section: Example} to the three-qutrit code (a similar resemblance was incidentally observed in the holographic context~\cite{Almheiri:2014lwa}). Based on that model, we expect that in general this code can be interpreted as protecting against erasure of subsystem states.
\end{itemize}

\section{Conclusion}
\label{Section: Conclusion}

In this paper, we have described a rather general mechanism for the emergence of classical gauge symmetry from quantum entanglement. This proceeded from understanding what it means for a subsystem to be classically resolvable. We exhibited a toy model demonstrating the mechanism, and a group-theoretic generalisation of that model. We also conjectured that the mechanism is responsible for diffeomorphism invariance in gravity.

Before ending the paper, let us comment on a few possible future directions.

First, everything we have discussed has been at $0^{\text{th}}$ order in the classical $\chi\to 0$ limit. However, to get the full classical picture, one must also investigate the subleading corrections in $\chi$. Indeed, the Poisson bracket of two classical observables $A(x),B(x)$ with corresponding quantum operators $\hat{A},\hat{B}$ is given by the Dirac formula
\begin{equation}
    \pb*{A}{B}(x) = \lim_{\chi\to 0}\frac{1}{i\chi} \mel*{x}{\comm*{\hat{A}}{\hat{B}}}{x}.
\end{equation}
Thus, the Poisson bracket is determined by the leading order corrections to the commutativity of the algebra of classical observables. It would be very worthwhile to figure out how this formula applies to classical limits with emergent gauge symmetry, so that we can understand the phase space structure of the classical theory. In particular, we would like to understand how the symplectic form of the full system decomposes into the symplectic forms of the subsystems. The results of~\cite{Kirklin:2019ror} suggest that Uhlmann holonomy may play a key role here.

Next, in this paper we considered complete classical limits. However, in gravity, the bulk spacetime emerges in a wider regime: a \emph{semi}classical limit $G\to 0$, in which the gravitational degrees of freedom behave classically, but the other fields still behave quantum mechanically. It would thus be useful to have a semiclassical version of the mechanism we have described.

Also, we assumed in this paper that there was no pre-existing gauge symmetry at the quantum level, just because it simplified the analysis. If there were to be a pre-existing gauge symmetry, then it would be good to know how this would interact with the emergent entanglement-based gauge symmetry. The case of a pre-existing gauge symmetry would seem to be more physically relevant, for example in the gravitational setting. In AdS/CFT, the quantum boundary theory typically has something like an $\mathrm{SU}(N)$ gauge symmetry, and in the real world gravitational gauge symmetry coexists with the gauge symmetry of the Standard Model.

On a related note, we have only addressed finite-dimensional Hilbert spaces, but the Hilbert spaces relevant to realistic models are typically infinite-dimensional, so the analysis presented here needs to be extended to the infinite-dimensional setting. The main reason this is non-trivial is that in such a case we are not guaranteed that the quantum states of subsystems have a description in terms of density matrices~\cite{Witten:2018zxz}. Thus, we would need some different way of characterising the classical resolvability of such subsystems. 

A general algebraic approach, accounting for subsystems specified by von Neumann algebras of all Types, as well as those with non-trivial center, would simultaneously address the issues of pre-existing gauge symmetry and infinite-dimensional Hilbert spaces. It is likely that such an approach can be established using the modular framework of Tomita and Takesaki~\cite{Summers:2003tf,Witten:2018zxz}. It is also possible that a semiclassical version of the mechanism described here can involve the emergence of von Neumann algebras of Types that are not part of the full quantum theory, \`a la~\cite{Leutheusser:2021qhd,Leutheusser:2021frk,Witten:2021unn}.

The classical resolvability of subsystems implies strong constraints on the entanglement structure of the full system, as we have described. It would be interesting to ask what other consequences these constraints have, besides leading to emergent gauge symmetry. For example, how much of the holographic entropy cone~\cite{cone} is a consequence of classical resolvability?

Finally, the reader may have noticed the conspicuous absence of any meaningful discussion in this paper of the \emph{dynamical} nature of the classical limits we are considering. The dynamics of a system is usually responsible for the physical relevance of a given classical limit. In particular, the time evolution of a quantum system must map directly onto the time evolution of the classical system, so that the classical picture remains valid at all times. Many interesting phenomena play a role here, such as chaos and decoherence. It would be good to try to understand this better.

\section*{Acknowledgements}
\phantomsection
\addcontentsline{toc}{section}{\protect\numberline{}Acknowledgements}
Thank you to 
Sylvain Carrozza, Stefan Eccles, Philipp H\"ohn, Leon Loveridge, Kelley Kirklin, Isha Kotecha, Slava Lysov, Fabio Mele and Yasha Neiman
for helpful discussions and comments. This work was supported by funding from the Okinawa Institute of Science and Technology. The front image was generated by DALL$\cdot$E 2.

\appendix

\section{Schur's lemma}

We use Schur's lemma several times throughout the paper. Schur's lemma says that a group $G$ acts irreducibly if and only if the only operators which commute with all elements of the group are proportional to the identity. Let us describe some relevant examples of the implications of this. These examples, and generalisations of them, should be sufficient to explain the usage of Schur's lemma in the main body of the paper.

Suppose $G$ is a group with a unitary irreducible representation $U$ on a Hilbert space $\mathcal{H}$, and let $\mu$ be a left-invariant measure on $G$. Let us define an operator $\hat{O}$ acting on $\mathcal{H}$ by
\begin{equation}
    \hat{O} = \int_G \dd{\mu(g)} U(g)\ket{\psi}\bra{\psi} U(g)^\dagger,
\end{equation}
where $\ket{\psi}\in\mathcal{H}$. By the left-invariance of the measure, this operator commutes with all operators of the form $U(g')$, where $g'\in G$. Thus, by Schur's lemma, it must be proportional to the identity. 

Suppose $G_1,G_2$ are groups with unitary irreducible representations $U_1,U_2$ on Hilbert spaces $\mathcal{H}_1,\mathcal{H}_2$, and with left-invariant measures $\mu_1,\mu_2$. Let us define an operator $\hat{O}_{12}$ acting on $\mathcal{H}_1\otimes\mathcal{H}_2$ by
\begin{equation}
    \hat{O}_{12} = \int_{G_1} \dd{\mu_1(g_1)} \int_{G_2} \dd{\mu_2(g_2)} \big(U_1(g_1)\otimes U_2(g_2)\big)\ket{\psi_{12}}\bra{\psi_{12}}\big(U_1(g_1)\otimes U_2(g_2)\big)^\dagger,
\end{equation}
where $\ket{\psi_{12}}\in\mathcal{H}_1\otimes\mathcal{H}_2$. This commutes with all operators of the form $U_1(g'_1)\otimes U_2(g'_2)$, where $g'_1\in G_1$ and $g'_2\in G_2$. By Schur's lemma and linearity, it must therefore by proportional to $\mathds{1}_1\otimes\mathds{1}_2$, so it must be proportional to the identity.

For the last example, let us assume $\mathcal{H}_2$ is finite-dimensional, $G_2$ is compact, and consider the operator
\begin{equation}
    \hat{P} = \int_{G_2} \dd{\mu_2(g_2)} \big(U_1(g_1)\otimes U_2(g_2)\big)\ket{\psi_{12}}\bra{\psi_{12}}\big(U_1(g_1)\otimes U_2(g_2)\big)^\dagger.
    \label{Equation: P}
\end{equation}
This commutes with all operators of the form $\mathds{1}_1\otimes U_2(g_2)$, where $g_2\in G_2$, so by Schur's lemma and linearity we must have $\hat{P}=\hat{P}_1\otimes\mathds{1}_2$ for some $\hat{P}_1$ acting on $\mathcal{H}_1$. By taking the partial trace over $\mathcal{H}_2$ of both sides of~\eqref{Equation: P}, we can deduce that $\hat{P}_1$ must be proportional to
\begin{equation}
    \rho_1 = \tr(\big(U_1(g_1)\otimes U_2(g_2)\big)\ket{\psi_{12}}\bra{\psi_{12}}\big(U_1(g_1)\otimes U_2(g_2)\big)^\dagger),
\end{equation}
which is the reduced density matrix in $\mathcal{H}_1$ of the state $\big(U_1(g_1)\otimes U_2(g_2)\big)\ket{\psi_{12}}$.

\printbibliography

\end{document}